\documentclass[aps,10pt,superscriptaddress,twocolumn,longbibliography]{revtex4-1}

\usepackage[utf8]{inputenc}
\usepackage{lmodern}
\usepackage{amsmath}
\usepackage{graphicx}
\usepackage{float}
\usepackage{amsfonts}
\usepackage{bbold}
\usepackage{braket}

\begin{document}

\title{Error-tolerant quantum convolutional neural networks for symmetry-protected topological phases}

\author{Petr Zapletal}
\affiliation{Department of Physics, Friedrich-Alexander University Erlangen-N\" urnberg (FAU), Erlangen, Germany}
\affiliation{Department of Physics, University of Basel, Klingelbergstrasse 82, 4056 Basel, Switzerland}

\author{Nathan A. McMahon}
\affiliation{Department of Physics, Friedrich-Alexander University Erlangen-N\" urnberg (FAU), Erlangen, Germany}

\author{Michael J. Hartmann}
\affiliation{Department of Physics, Friedrich-Alexander University Erlangen-N\" urnberg (FAU), Erlangen, Germany}

\begin{abstract}
The analysis of noisy quantum states prepared on current quantum computers is getting beyond the capabilities of classical computing. 
Quantum neural networks based on parametrized quantum circuits, measurements and feed-forward can process large amounts of quantum data to reduce measurement and computational costs of detecting non-local quantum correlations. 
The tolerance of errors due to decoherence and gate infidelities is a key requirement for the application of quantum neural networks on near-term quantum computers. Here we construct quantum convolutional neural networks (QCNNs) that can, in the presence of incoherent errors, recognize different symmetry-protected topological phases of generalized cluster-Ising Hamiltonians from one another as well as from topologically trivial phases. Using matrix product state simulations, we show that the QCNN output is robust against symmetry-breaking errors below a threshold error probability and against symmetry-preserving errors provided the error channel is invertible. This is in contrast to string order parameters and the output of previously designed QCNNs, which vanish in the presence of any symmetry-breaking errors. To facilitate the implementation of the QCNNs on near-term quantum computers, the QCNN circuits can be shortened from logarithmic to constant depth in system size by performing a large part of the computation in classical post-processing. These constant-depth QCNNs reduce sample complexity exponentially with system size in comparison to the direct sampling using local Pauli measurements.
\end{abstract}

\maketitle

\section{introduction}

Existing noisy intermediate-scale quantum (NISQ) computers can perform computations that are challenging for classical computers \cite{arute2019}. However, quantum computing hardware and quantum algorithms need to be further developed to enable the exploitation of quantum computers in areas such as the simulation of many-body systems \cite{feynman1982,cao2019} and machine learning \cite{biamonte2017}. One of the major challenges in developing scalable quantum computers is the characterization of noisy quantum data produced by near-term quantum hardware. With increasing system size, standard characterization techniques using direct measurements and classical post-processing become prohibitively demanding due to large measurement counts and computational efforts. While many local properties can be efficiently determined using randomized measurements \cite{huang2020}, global properties of quantum states are typically hard to estimate.

Quantum machine learning techniques based on the direct processing of quantum data on quantum processors can substantially reduce the measurement costs, including quantum principle component analysis \cite{lloyd2014}, quantum autoencoders \cite{romero2017,bondarenko2020,zhang2021}, certification of Hamiltonian dynamics \cite{wiebe2014,gentile021}, quantum reservoir processing \cite{ghosh2019}. Moreover, quantum neural networks based on parametrized quantum circuits, measurements and feed-forward can process large amounts of quantum data, to detect non-local quantum correlations with reduced measurement and computational efforts compared to standard characterization techniques \cite{farhi2018,cong2019,beer2020,kottmann2021,gong2023}. A key requirement for employing quantum neural networks to characterize noisy quantum data produced by near-term quantum hardware is the tolerance to errors due to decoherence and gate infidelities.

The characterization of non-local correlations in quantum states is of key importance to condensed matter physics. It is required for the classification of topological quantum phases of matter \cite{pollmann2010,chen2011,deleseleuc2019,semeghini2021} and for understanding new strongly correlated materials \cite{sachdev2011} such as high-temperature superconductors \cite{wang2016}. Classical machine learning tools for the recognition of topological phases of matter have recently been studied, uncovering phase diagrams from data produced by numerical simulations \cite{carrasquilla2017,vannieuwenburg2017,greplova2020} and measured in experiments \cite{rem2019,bohrdt2021,kaming2021,miles2023}. Moreover, quantum many-body states belonging to topological quantum phases have been prepared on quantum computers using exact matrix product state representations \cite{smith2022}, unitary quantum circuits \cite{satzinger2021}, and measurement and feed-forward \cite{iqbal2023}. Properties of topological phases have been probed on quantum computers by measuring characteristic quantities \cite{smith2022,azses2020} such as string order parameters (SOPs) \cite{perez-garcia2008,pollmann2012}. The detection of topological phases can be enhanced via the processing of measurement data on a classical computer \cite{cong2022}.
Classical machine learning algorithms have been shown to classify topological quantum phases from classical shadows formed by randomized measurements \cite{huang2022}.
However, the rapidly increasing sample complexity with system size remains an outstanding problem for such approaches.

In Ref.~\cite{cong2019}, quantum convolutional neural networks (QCNNs) have been proposed to recognize symmetry-protected topological (SPT) phases \cite{pollmann2010,chen2011} with reduced sample complexity compared to the direct measurement of SOPs. Such QCNNs can be trained to identify characteristics of SPT phases from training data \cite{cong2019,caro2022,liu2023}. Alternatively, QCNNs can be analytically constructed to mimic renormalization-group flow \cite{cong2019,lake2022}, a method for classifying quantum phases \cite{sachdev2011}. A shallow QCNN has been implemented on a 7-qubit superconducting quantum processor in Ref.~\cite{herrmann2022}. This QCNN has exhibited robustness against incoherent errors on the NISQ device which allowed for the recognition of a SPT phase with a higher fidelity than the direct measurement of SOPs. However, the propagation of errors leads to a rapid growth of error density in deeper QCNNs due to the reduction of qubit number from one QCNN layer to the next, which represents a central problem.

Here we overcome this problem by designing QCNNs for generalized cluster-Ising models that can tolerate incoherent errors. The QCNN circuits are constructed by alternating layers, which mimic renormalization-group flow, and new layers, which correct incoherent errors. Due to the tolerance to errors, the QCNNs recognize SPT phases of exact ground states provided access to only noisy states, which approximate the former on NISQ devices. Apart from distinguishing SPT phases from topologically trivial phases as previously shown in Refs.~\cite{cong2019,herrmann2022,lake2022,liu2023}, we newly demonstrate that QCNNs constructed here can recognize two SPT phases from one another.
 
Using matrix product state (MPS) simulations, we show that the QCNN output is robust against symmetry-breaking errors below a threshold error probability. This enables new quantum phase recognition capabilities for QCNNs in scenarios where SOPs and previous QCNN designs \cite{cong2019} are impractical. SOPs rapidly vanish with an increasing length for any probability of symmetry-breaking errors \cite{groot2022}, whereas the QCNN proposed in Ref.~\cite{cong2019} rapidly concentrates symmetry-breaking errors with increasing depth leading to a vanishing output for any error probability. 

 In addition to the tolerance to symmetry-breaking errors, the QCNNs constructed here tolerate symmetry-preserving errors if the error channel is invertible. The error tolerance is limited close to phase boundaries as the threshold error probability decreases with diverging correlation lengths. Nonetheless, a sharp change in the QCNN output at the phase boundaries allows us to precisely determine critical values of Hamiltonian parameters.
 
 To facilitate the implementation of QCNNs on near-term quantum computers, we show that the QCNN circuits constructed here can be shortened from logarithmic to constant depth in system size by efficiently performing a large part of the computation in classical post-processing. The output of the QCNNs corresponds to the expectation value of a multiscale SOP, which is a sum of products of individual SOPs. The multiscale SOP can, in principle, be determined using direct Pauli measurements on the input state without using any quantum circuit. However, the constant-depth QCNN circuits, we derive here, reduce the sample complexity of measuring the multiscale SOP exponentially with system size in comparison to direct Pauli measurements.

The remainder of this manuscript is structured as follows. In Sec.~\ref{sec:model}, we introduce the generalized cluster-Ising model we consider before describing the construction of the QCNNs to analyze it in Sec.~\ref{sec:qcnn}. We investigate the robustness of the QCNN output against incoherent symmetry-preserving errors in Sec.~\ref{sec:preserving} and show how to design QCNNs that tolerate symmetry-breaking errors in Sec.~\ref{sec:breaking}. We investigate the phase transition between two SPT phases in Sec.~\ref{sec:2spt} and study the tolerance to incoherent errors close to phase boundaries in Sec.~\ref{sec:pb}. In Sec.~\ref{sec:complexity}, we compare the sample complexity of QCNNs to the direct Pauli measurement of the input state before presenting concluding remarks and possible applications of error-tolerant QCNNs in Sec.~\ref{sec:con}.

\begin{figure*}[t]
\includegraphics[width = \linewidth]{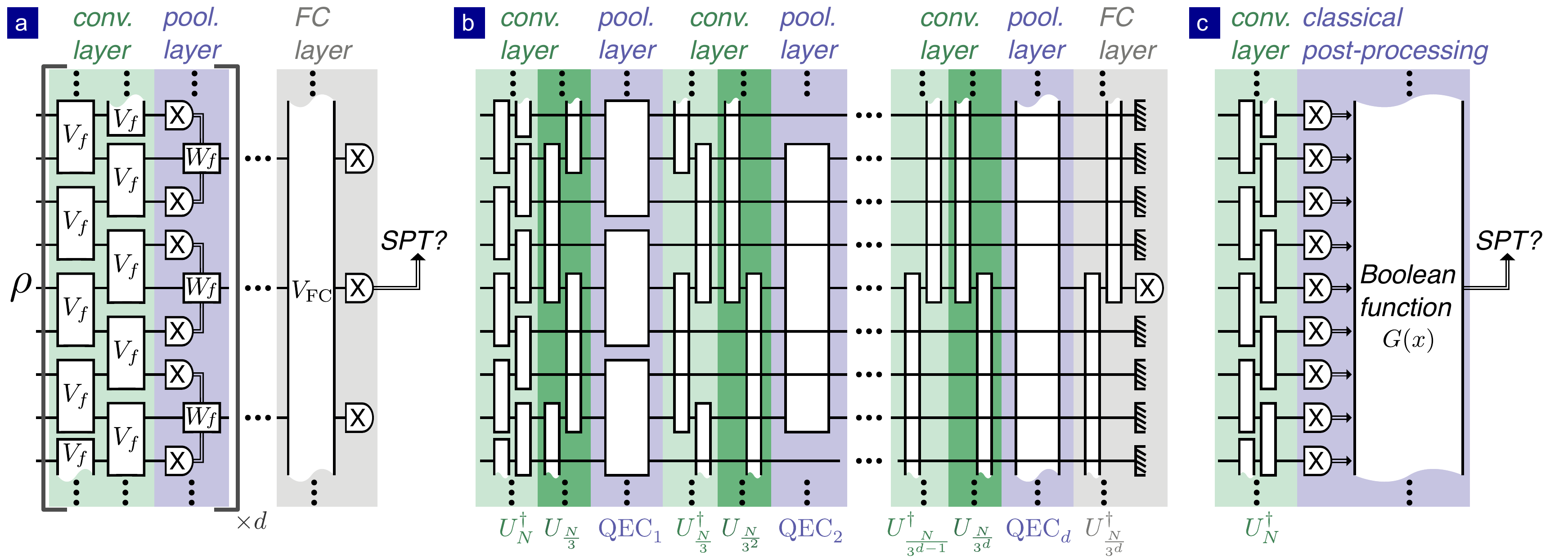}
\caption{Quantum convolutional neural network (QCNN). (a) QCNN quantum circuit consisting of $d$ convolutional layers, $d$ pooling layers and a final fully connected (FC) layer. The measurement of the output qubits labels whether the input state $\rho$ belongs to a given SPT phase. (b) QCNN circuit mimicking renormalization-group flow. In each convolutional layer $f=1,2,...,d$, a disentangling unitary $U_{N/3^{f-1}}^{\dagger}$ and an entangling unitary $U_{N/3^f}$ are applied on sublattices with $N/3^{f-1}$ qubits and $N/3^{f}$ qubits, respectively. In each pooling layer $f$, a quantum-error-correction unitary QEC$_{f}$ is performed on a sublattice with $N/3^{f-1}$ qubits. In the fully connected layer, a disentangling unitary $U_{N/3^d}^{\dagger}$ is applied. At the end, $\lfloor N/3^d\rfloor$ qubits are measured in the $X$ bases. (c) QCNN circuit equivalent to (b) consisting of a constant-depth quantum circuit $U^{\dagger}_N$, the measurement of all qubits in the $X$ basis and classical post-processing. The label of the quantum phase is determined as a Boolean function $G(x)$ of the measured bit strings $x$.}\label{fig:qcnn}
\end{figure*}

\section{Generalized cluster-Ising model in the presence of incoherent errors}\label{sec:model}
We consider a one-dimensional chain of $N$ qubits with open boundary conditions described by the generalized cluster-Ising Hamiltonian
\begin{align}
H =& -J_1\sum_{j=2}^{N-1}\,C_j - J_2\sum_{j=3}^{N-2} D_j  \nonumber\\
&-h_1\sum_{j=1}^N  \,X_j-h_2\sum_{j=1}^{N-1} \,X_j X_{j+1}, \label{eq:ham}
\end{align}
where $C_j = Z_{j-1}X_{j}Z_{j+1}$, $D_j = Z_{j-2}X_{j-1}X_jX_{j+1}Z_{j+2}$, and $X_j$ as well as $Z_j$ are Pauli operators on qubit $j$. The Hamiltonian exhibits $\mathbb{Z}_2\times\mathbb{Z}_2$ symmetry generated by $P_{e/o} = \prod_{j=1}^{N/2} X_{2j/2j-1}$ as well as time-reversal symmetry (complex conjugation). The ground states $\ket{\psi}$ of the Hamiltonian belong to one of four phases: a paramagnetic phase, an antiferromagnetic phase, a `$ZXZ$' SPT phase and a `$ZXXXZ$' SPT phase \cite{verresen2017}. The `$ZXZ$' (`$ZXXXZ$') SPT phase contains the `$ZXZ$' (`$ZXXXZ$') cluster state, which is a stabilizer state with stabilizer elements $C_j$ ($D_j$) and thus the ground state for  $J_2=h_1=h_2=0$ ($J_1=h_1=h_2=0$). SPT phases are characterized by SOPs \cite{perez-garcia2008,pollmann2012}. In particular, the SOPs
\begin{align}
S_{jk} &=Z_j\left(\prod_{i = 1 }^{(k-j)/2} X_{ j + 2i -1}\right) Z_{k},\\
T_{jk} &=Z_jX_{j+1}Y_{j+2}\left(\prod_{i = 2 }^{(k-j)/2-2} X_{ j + 2i}\right) Y_{k-2}X_{k-1}Z_{k},
\end{align}
attain non-vanishing values in the `$ZXZ$' SPT phase and the `$ZXXXZ$' SPT phase, respectively. 

NISQ computers operate in the presence of noise due to decoherence and gate infidelities. To simulate errors that occur during the preparation of the many-body ground states $|\psi\rangle$ of the Hamiltonian \eqref{eq:ham} on NISQ devices, we consider an error channel
\begin{equation}\label{eq:channel}
	\rho = \mathcal{E}(|\psi\rangle\langle \psi |) = \sum_{l=1}^{4^N} K_l |\psi\rangle\langle \psi | K_l^{\dagger},
\end{equation}	
where $K_l\in\{\sqrt{p_{\mathbb{1}}}\mathbb{1},\sqrt{p_X} X, \sqrt{p_Y} Y, \sqrt{p_Z} Z\}^{\otimes N}$ are Kraus operators, $p_E$ are probabilities of Pauli errors $E=X,Y,Z$ and $p_{\mathbb{1}} + p_X + p_Y+p_Z = 1$. For $p_X=p_Y=p_Z$, this error channel describes single-qubit depolarizing noise.

We formulate quantum phase recognition on NISQ devices as a task to identify whether the exact ground state $\ket{\psi}$ of the Hamiltonian \eqref{eq:ham} belongs to a given quantum phase provided access only to the noisy state $\rho$, which approximates $\ket{\psi}$.\\

\section{Quantum convolutional neural networks}\label{sec:qcnn}

Our goal is to design QCNNs that detect the SPT phases of the generalized cluster-Ising model via quantum phase recognition. To perform quantum phase recognition, we process the ground states $\rho$ with the QCNN depicted in Fig.~\ref{fig:qcnn}a consisting of $d$ convolutional layers, $d$ pooling layers and a final fully connected layer. In each convolutional layer $f = 1,2,...,d$, a translationally invariant unitary $V_f$ is applied. In a pooling layer, the system size is reduced by measuring a fraction of qubits and applying feed-forward gates $W_f$ conditioned on the measurement outcomes on the remaining qubits. In this work, we consider the reduction of system size by a factor of three in each pooling layer. As a result, the maximal depth  $d=\lfloor \log_3 N \rfloor$ of the QCNN is logarithmic in system size $N$. In the fully connected layer, a general unitary $V_{\rm FC}$ is performed on all remaining qubits and the qubits are read out labeling whether the ground state $\ket{\psi}$ belongs to a given SPT phase or not.

For each SPT phase, we construct the QCNN depicted in Fig.~\ref{fig:qcnn}b by generalizing the procedure proposed in Ref.~\cite{cong2019}. First, we identify a characteristic state belonging to each SPT phase. For the `$ZXZ$' (`$ZXXXZ$') SPT phase this is the `$ZXZ$' (`$ZXXXZ$') cluster state, which can be mapped onto a product state by a disentangling unitary $U^{\dagger}$ consisting of two (four) layers of two-qubit gates between neighboring qubits, see Appendix~\ref{app:circ} for details. The convolutional layers of the QCNN consist of the disentangling unitary $U^{\dagger}_{N/3^{f-1}}$ mapping the corresponding cluster state on $N/3^{f-1}$ qubits onto a product state and the entangling unitary $U_{N/3^{f}}$ mapping the product state on a sublattice with $N/3^{f}$ qubits onto the cluster state. As a result, we obtain the cluster state for a reduced system size after the measurement of the remaining qubits in each pooling layer. By construction, the cluster state is a fixed point of the QCNN circuit.

Next, we make all states belonging to the `$ZXZ$' (`$ZXXXZ$') SPT phase flow towards the `$ZXZ$' (`$ZXXXZ$') cluster state with the increasing depth of the QCNN. To this end, we implement in pooling layers a procedure that is analogous to quantum error correction (QEC), identifying perturbations away from the cluster state as errors. These errors are detected by measurements in the pooling layers and corrected by feed-forward gates $W_f$ on the remaining qubits which are conditioned on the measurement outcomes. A measurement and a feed-forward gate can be replaced by an entangling gate and tracing out of the "measured" qubits. Using this equivalence, we represent the QEC procedure in each pooling layer $f$ as a unitary $\textrm{QEC}_{f}$ as depicted in Fig.~\ref{fig:qcnn}b.  It has been shown in Ref.~\cite{cong2019} that by correcting $X_j$ and $X_jX_{j+1}$ errors one can make all pure ground states of the cluster-Ising Hamiltonian belonging to the `$ZXZ$' SPT phase (for $J_2=0$) flow towards the `$ZXZ$' cluster state. In this way, the QCNN mimics a renormalization-group flow \cite{sachdev2011}.

In the fully connected layer, we measure stabilizer elements, i.e. either $C_j$ or $D_j$ for the `$ZXZ$' phase or the `$ZXXXZ$' SPT phase, respectively. This measurement is performed by applying the disentangling unitary $U^{\dagger}_{N/3^{d}}$ and reading out all remaining qubits in the $X$ basis. For system size $N$ and depth $d$, we have $m =\lfloor N/3^d\rfloor$ output qubits. The QCNN output 
\begin{equation} \label{eq:qcnn-output}
    y = \frac{1}{m}\sum_{j=-(m-1)/2}^{(m-1)/2}\langle X_{\frac{N+1}{2} + j\cdot3^d}\rangle
\end{equation}
is thus the expectation value of $X$ averaged over the $m$ output qubits.

Before discussing the performance of the constructed QCNNs in the presence of noise due to decoherence and gate infidelities on NISQ devices, we make a crucial observation allowing for a substantial shortening of the QCNN circuits. A large part of the QCNN circuits depicted in Fig.~\ref{fig:qcnn}b can be efficiently implemented in classical post-processing if the QEC procedures $\widetilde{\textrm{QEC}}_{f} = U_{N/3^{f}}^{\dagger}\textrm{QEC}_{f}U_{N/3^{f}}$ transformed by the entangling unitaries $U_{N/3^{f}}$ map $X$-basis eigenstates $\ket{x}$ onto other $X$-basis eigenstates $\ket{x'}$
\begin{equation}\label{eq:cond}
\widetilde{\textrm{QEC}}_{f} \ket{x} \propto \ket{x'}.
\end{equation}
In this case, the QCNNs are equivalent to a constant depth quantum circuit consisting of the disentangling unitary $U^{\dagger}_N$, the measurement of all qubits in the $X$ basis and classical post-processing as depicted in Fig.~\ref{fig:qcnn}c. See Appendix~\ref{app:circ} for the derivation of these equivalent QCNN circuits. In these equivalent QCNN circuits, only the first convolutional layer is implemented on a quantum computer. The remaining convolutional layers, all pooling layers and the fully connected layer are implemented after the measurement of all qubits in classical post-processing as a bit-string-valued Boolean function $G(x)=x'$ of the measured bit strings $x = x_1 x_2 \dots x_N$, where $x_j=0,1$ corresponds to measuring $X_j = +1,-1$. Errors perturbing the cluster states lead to flipped measurement outcomes after the disentangling unitary  $U^{\dagger}_N$. These error syndromes are then corrected in classical post-processing.

Note that the QCNN proposed in Ref.~\cite{cong2019} satisfies the condition \eqref{eq:cond} and its equivalent QCNN circuit consisting of a constant-depth quantum circuit, measurement and classical post-processing has been developed and experimentally realized in Ref.~\cite{herrmann2022}. 

In this work, we consider the equivalent QCNN circuits depicted in Fig.~\ref{fig:qcnn}c. First, we numerically obtain the ground states of the Hamiltonian \eqref{eq:ham} in the thermodynamic limit using the infinite density matrix renormalization group (iDMRG) algorithm \cite{hauschild2018}, see Appendix~\ref{app:num} for details. Next, we perform the constant-depth quantum circuit on the infinite MPSs by sequentially applying two-qubit gates between neighboring qubits. Then, we sample $M_{S}$ outcomes of the measurement of $N$ qubits from the infinite MPSs. Finally, we determine the QCNN output $y$ from the measured bit strings $x$ using the Boolean function $G(x)$ as 
\begin{equation} \label{eq:boolean}
    y = \frac{1}{m}\frac{1}{M_{S}}\sum_{j=-(m-1)/2}^{(m-1)/2}\sum_x\left[1 - 2\,G(x)_{\frac{N+1}{2} + j\cdot3^d}\right].
\end{equation}

\begin{figure}[t]
\includegraphics[width = 0.84\linewidth]{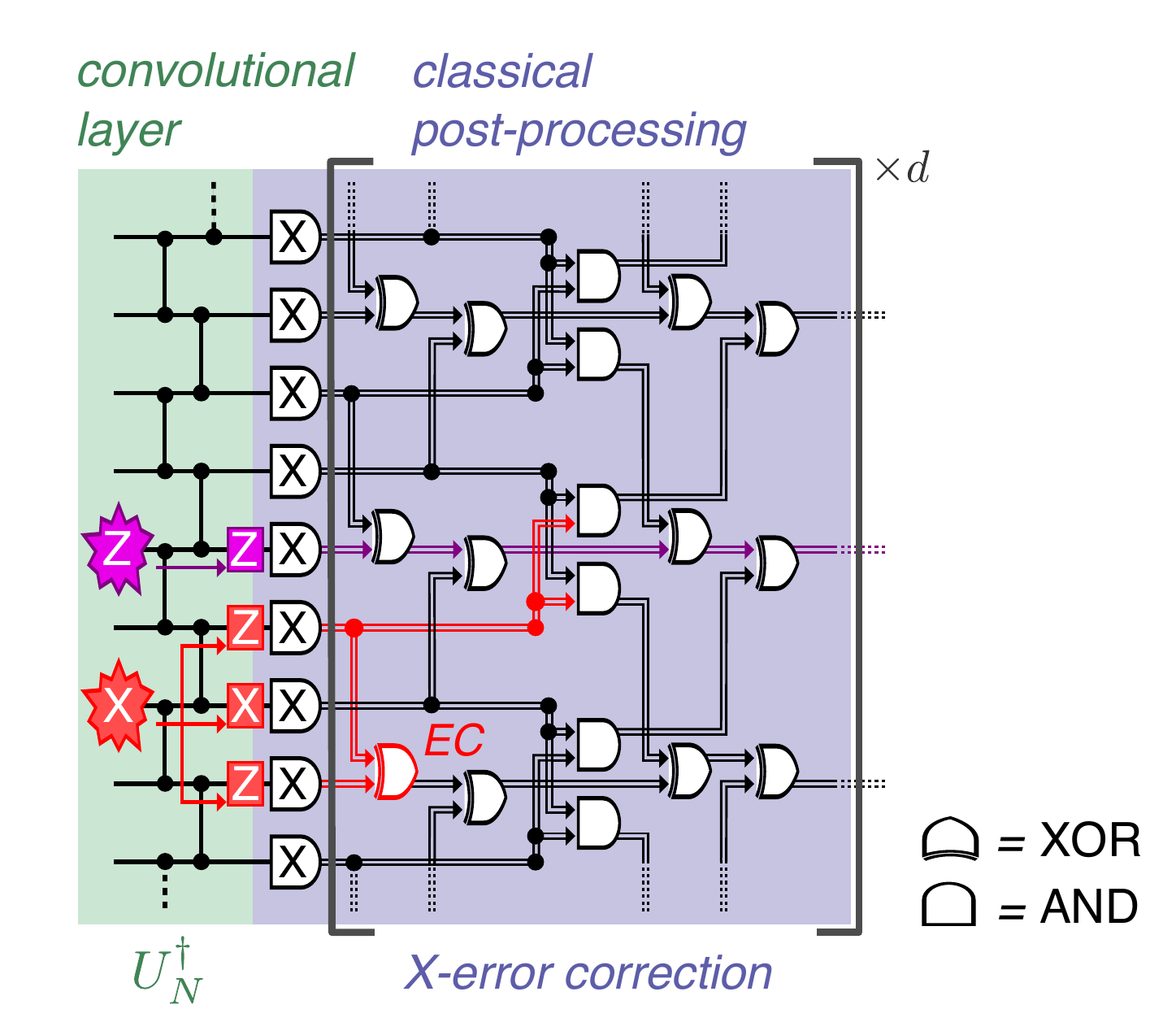}
\caption{QCNN with $X$-error correcting layers detecting the `$ZXZ$' phase. The QCNN circuit consists of a constant-depth quantum circuit, the measurement of all qubits in the $X$ basis and classical post-processing. The quantum circuit performs the disentangling unitary $U_N^{\dagger}$ consisting of controlled $Z$ gates between neighboring qubits. The outcomes $x$ of the measurement in the $X$ basis are processed by the Boolean function $G(x)$, expressed as a logic circuit in terms of AND and XOR gates. The logic circuit is composed of $d$ layers correcting the syndromes of $X$ errors. Red and purple lines show the propagation of $X$ errors and $Z$ errors, respectively, through the QCNN circuit. The $X$-error syndrome is corrected by the XOR gate marked in red.}
\label{fig:qcnnx}
\end{figure}

\section{Tolerance to symmetry-preserving errors}\label{sec:preserving}
NISQ computers operate in the presence of noise due to decoherence and gate infidelities. To enable the exploitation of QCNNs as a characterization tool for NISQ computers, it is thus crucial to investigate the effects of noise on the performance of QCNNs and to construct QCNNs whose output is robust against noise.

We expect that the preparation of typical many-body ground states $|\psi\rangle$ will require substantially deeper quantum circuits than the QCNNs considered in this work which can be implemented in very short constant depth as discussed above.
We thus focus on the robustness of QCNNs against errors that occur during the preparation of many-body ground states $|\psi\rangle$ described by the error channel \eqref{eq:channel} and neglect errors occurring during the QCNN circuits.

\begin{figure}[t]
\centering
\includegraphics[width=0.66\linewidth]{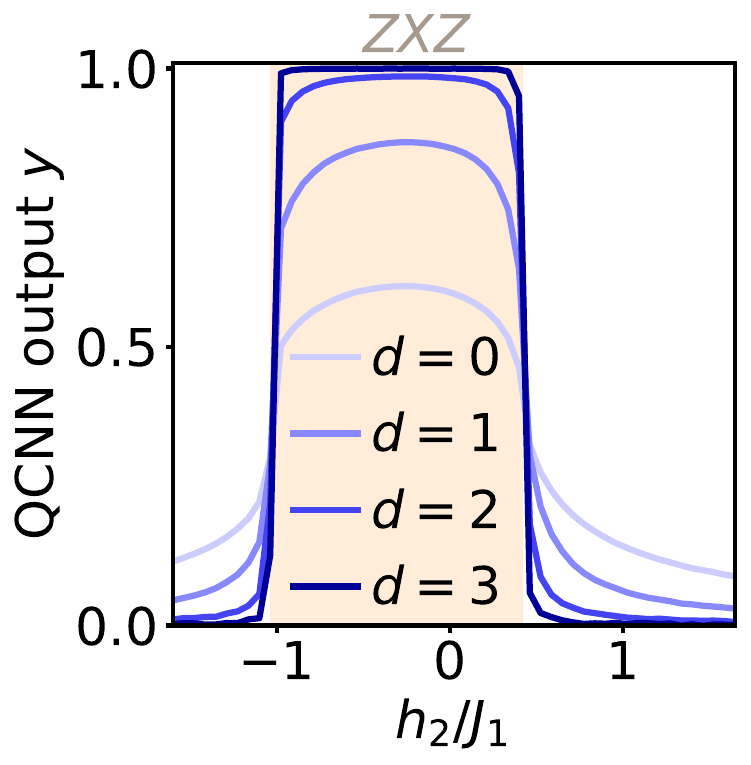}
\caption{QCNN consisting of $X$-error correcting layers for the ground states of the cluster-Ising Hamiltonian \eqref{eq:ham} perturbed by incoherent $X$ errors. The QCNN output as a function of $h_2/J_1$ and different depths $d$ of the QCNN. The orange region denotes the `$ZXZ$' phase. (Parameters: $h_1/J_1 = 0.5$, $J_2 = 0$, $N=1215$, $p_X = 0.1 $, $p_Y=p_Z=0$, $M_{S} = 10^4$)}\label{fig:origx}
\end{figure}

We start by investigating the performance of the QCNN proposed in Ref.~\cite{cong2019} in the presence of incoherent $X$ errors described by the error channel \eqref{eq:channel} with $p_Y=p_Z = 0$. We use the compact implementation as a quantum circuit consisting of the disentangling unitary $U^{\dagger}_N$, the measurement of all qubits in the $X$ basis and classical post-processing. We show the QCNN circuit in Fig.~\ref{fig:qcnnx}. The disentangling unitary $U^{\dagger}_N$ consists of controlled $Z$ gates between neighboring qubits. The outcomes $x$ of the measurement in the $X$ basis are processed by the Boolean function $G(x)$ which is expressed as a logic circuit in terms of AND and XOR gates, see Fig.~\ref{fig:qcnnx}. The key feature of the QCNN is that it identifies and corrects perturbations away from the `$ZXZ$' cluster state. In particular, it corrects coherent $X_j$ and $X_jX_{j+1}$ errors which drive perturbations away from the cluster state to other ground states of the Hamiltonian \eqref{eq:ham} \cite{cong2019}.
The logic circuit is composed of $d$ layers $f=1,2,...,d$, which correspond to the $X$-error correcting $\widetilde{\textrm{QEC}}_f$ procedures transformed by the disentangling unitary $U_N^{\dagger}$. 

We plot in Fig.~\ref{fig:origx} the QCNN output across a cut through the phase diagram as a function of $h_2/J_1$ for fixed $h_1/J_1 = 0.5$ and $J_2=0$ for different depths $d$ of the QCNN. We can see that the QCNN output converges to unity with the increasing depth of the QCNN in the `$ZXZ$' SPT phase but vanishes with the increasing depth outside of the SPT phase. This demonstrates our first observation that QCNNs can tolerate incoherent $X$ errors since their QEC procedures can correct not only coherent perturbations, that transform the cluster state to another ground state in the SPT phase, but also incoherent errors.

The QCNN output converges to ideal noise-free values with increasing depth $d$ for any probability $p_X \neq 0.5$ of incoherent $X$ errors, since the error channel is invertible for these cases, see Appendix~\ref{app:errx} for details. For $p_X=0.5$ the situation is qualitatively different as the error channel \eqref{eq:channel} is not invertible. Invertible symmetry-preserving error channels for $p_X \neq 0.5$ preserve SPT order \cite{groot2022}. In contrast, the non-invertible error channel for $p_X = 0.5$  annihilates SOPs
\begin{equation}\label{eq:sopx}
\mathcal{E}^{\dagger}(S_{jk}) =(1-2p_X)^2 S_{jk} = 0,
\end{equation}
for all $j$ and $k$, where $\mathcal{E}^{\dagger}(\mathcal{O})= \sum_{l=1}^{4^N} K_l^{\dagger} \mathcal{O} K_l$ is the adjoint channel to Eq.~\eqref{eq:channel}. As a result, also the QCNN output vanishes for any input ground state and any depth $d$.

We conclude that QCNNs recognizing the `$ZXZ$' SPT phase can tolerate symmetry-preserving $X$ errors, provided that the error channel is invertible.

\section{Tolerance to symmetry-breaking errors}\label{sec:breaking}

Since noise in NISQ devices typically does not preserve the symmetries of problem Hamiltonians, it is important to investigate the robustness of QCNNs against symmetry-breaking errors.

While coherent and incoherent $X$ errors are tolerated by the QCNN designed in Ref.~\cite{cong2019}, the situation is fundamentally different for incoherent $Z$ errors as they break the $\mathbb{Z}_2\times\mathbb{Z}_2$ symmetry of the Hamiltonian \eqref{eq:ham}. $Z$ errors described by the error channel \eqref{eq:channel} with $p_X=p_Y = 0$ lead to a decrease of the SOPs 
\begin{equation}
\mathcal{E}^{\dagger}(S_{jk}) = (1-2p_Z)^{\frac{L-1}{2}} S_{jk},
\end{equation}
which scales exponentially with their length $L = k - j +1$. As a result, the SOPs rapidly vanish with the increasing length $L$ for any finite (non-unity) $Z$-error probability $p_Z\neq 0,1$. 

Similarly to SOPs, the original design \cite{cong2019} of the QCNN depicted in Fig.~\ref{fig:qcnnx} is substantially affected by $Z$ errors. The syndrome of a $Z_j$ error, i.e., the flipped outcome $x_j$ of the measurement in the $X$ basis, is denoted in Fig.~\ref{fig:qcnnx} by a purple line. In contrast to $X$-error syndromes that are corrected (see red lines in Fig.~\ref{fig:qcnnx}), $Z$-error syndromes (purple line) propagate through the QCNN circuit, see Appendix~\ref{app:err} for more details. As the system size is reduced by a factor of three in each layer, the density of $Z$-error syndromes increases with the increasing depth of the QCNN. As a result, the output of the QCNN rapidly decreases with the depth $d$ both in the `$ZXZ$' SPT phase and outside of the phase for any finite probability $p_Z\neq 0,1$, see Appendix~\ref{app:corx} for details.

\begin{figure}[t]
\includegraphics[width = \linewidth]{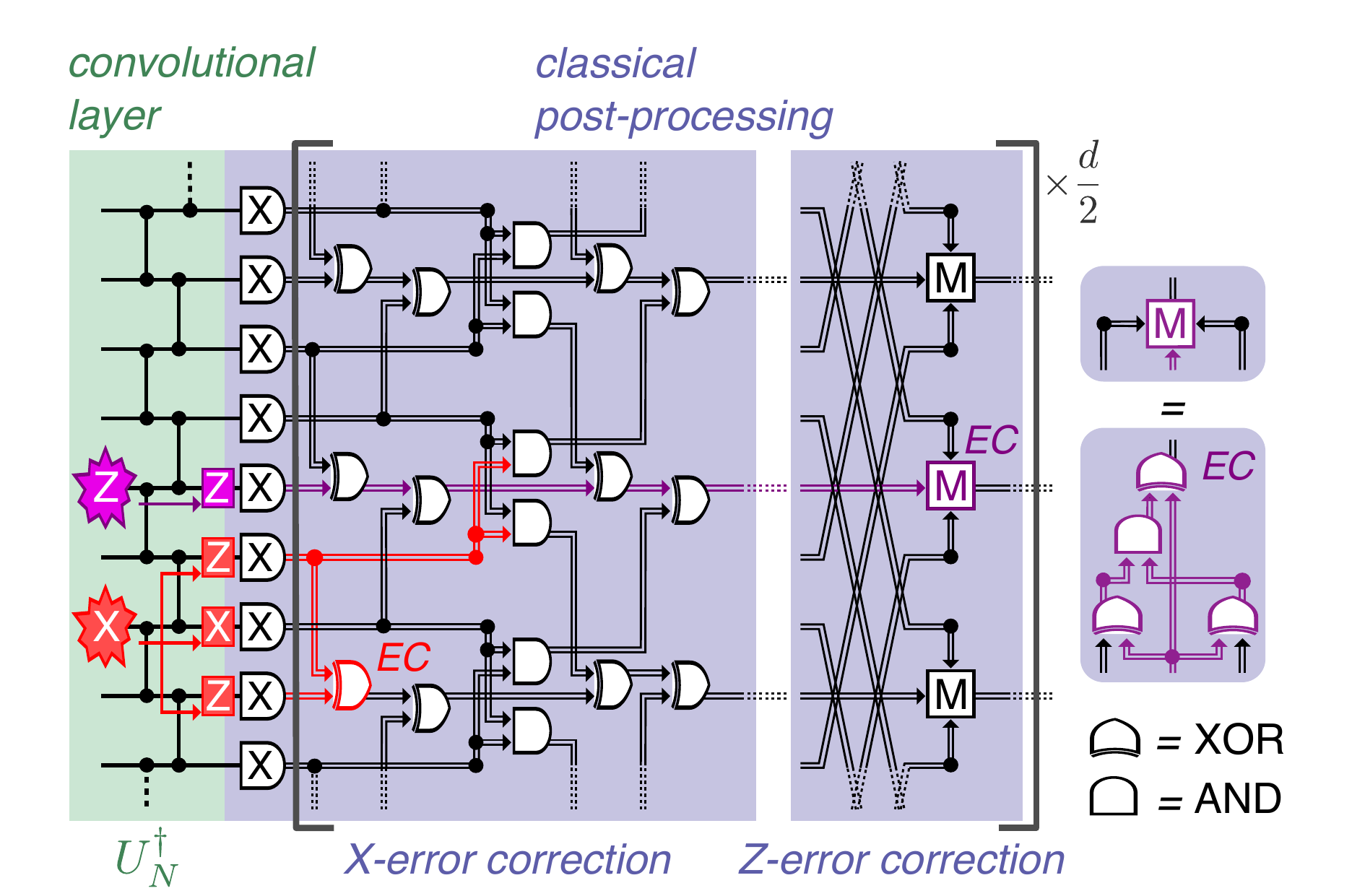}
\caption{QCNN with alternating layers correcting $X$ errors and $Z$ errors for detecting the `$ZXZ$' phase. The QCNN circuit consists of a constant-depth quantum circuit, the measurement of all qubits in the $X$ basis and classical post-processing. The quantum circuit performs the disentangling unitary $U_N^{\dagger}$ consisting of controlled $Z$ gates between neighboring qubits. The outcomes $x$ of the measurement in the $X$ basis are processed by the Boolean function $G(x)$ expressed as a logic circuit in terms of AND and XOR gates as well as of the majority function $M$. The decomposition of the majority function $M$ into AND and XOR gates is shown on the right. Red and purple lines show the propagation of $X$ errors and $Z$ errors, respectively, through the QCNN circuit. The logic circuit consists of  $d$ layers $f=1,2,...,d$. Odd layers $f$ correct the syndromes of $X$ errors (see the XOR gate marked in red) and even layers $f$ correct the syndromes of $Z$ errors (see the majority function $M$ marked in purple).}
\label{fig:qcnnz}
\end{figure}

To perform quantum phase recognition on NISQ devices, QCNNs thus need to be robust against symmetry-breaking errors.
In Ref.~\cite{herrmann2022}, symmetry-breaking errors were corrected in the fully connected layer. While this improved the robustness of the QCNN with depth $d=1$, this approach is impractical for deeper QCNNs due to the rapidly increasing density of $Z$ errors with $d$.
To overcome this issue, we construct a new QCNN depicted in Fig.~\ref{fig:qcnnz} by alternating the original $X$-error correcting layers with new $Z$-error correcting layers. The $Z$-error correcting layer $f$ consists of a new $\textrm{QEC}$ procedure that can be efficiently implemented in classical post-processing as the majority function 
\begin{align}
&M(x_{j-7\cdot3^{f-1}},x_{j},x_{j+7\cdot3^{f-1}}) =\nonumber\\
 &[(x_{j-7\cdot3^{f-1}}\oplus x_j)\wedge (x_j\oplus x_{j+7\cdot3^{f-1}})]\oplus x_j,\label{eq:maj}
\end{align}
where $\wedge$ is the AND gate and $\oplus$ is the XOR gate, see Fig.~\ref{fig:qcnnz}. The majority function $M(x_{j-7\cdot3^{f-1}},x_{j},x_{j+7\cdot3^{f-1}})$ returns the value of the majority of the three bits $x_{j-7\cdot3^{f-1}}$, $x_{j}$, and $x_{j+7\cdot3^{f-1}}$. It thus removes isolated error syndromes, see purple lines in Fig.~\ref{fig:qcnnz} and Appendix~\ref{app:err} for more details. The corresponding QEC unitary is described in Appendix~\ref{app:circ}.

\begin{figure}[t]
\centering
\includegraphics[width=0.66\linewidth]{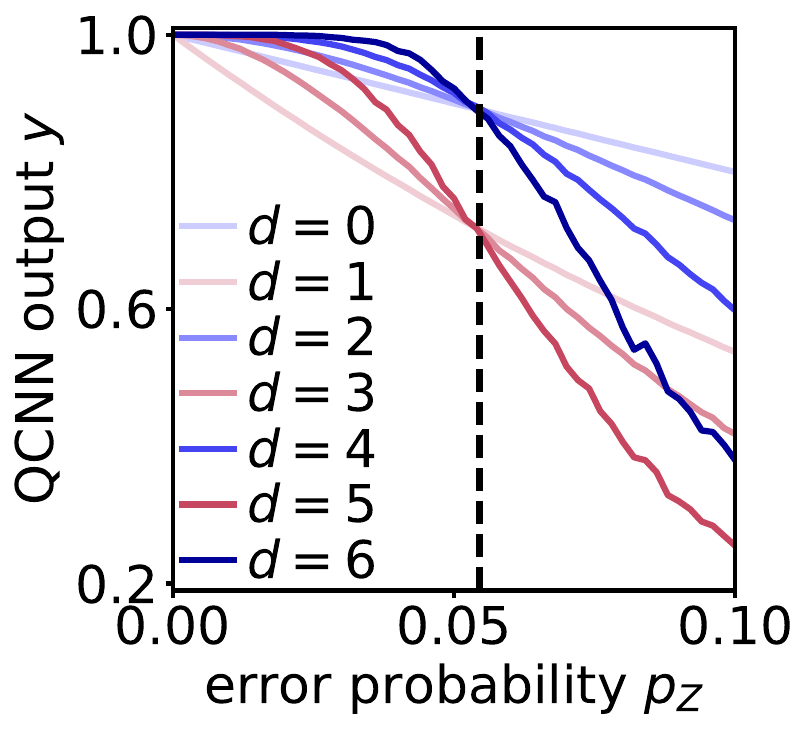}
\caption{QCNN consisting of alternating layers correcting $X$ errors and $Z$ errors for the `$ZXZ$' cluster state perturbed by symmetry-breaking $Z$ errors. The QCNN output as a function of the error probability $p_Z$ of $Z$ errors for different depths $d$. The black dashed line shows the threshold error probability $p_{\rm th} = 0.054$. (Parameters: $N=1215$, $M_{S}  = 10^4$, $p_X=p_Y=0$)}\label{fig:alterz}
\end{figure}

We start by investigating the QCNN with alternating $X$- and $Z$-error correcting layers for the `$ZXZ$' cluster state perturbed by incoherent $Z$ errors as the input state. We plot in Fig.~\ref{fig:alterz} the QCNN output as a function of the $Z$-error probability $p_Z$. We can see an alternating QCNN output after odd and even layers. $Z$ errors propagate through odd, $X$-error correcting layers and, as the system size is reduced by a factor of three, the density of $Z$ errors increases. This error concentration leads to the decrease of the QCNN output after odd layers, compare blue and red lines in Fig.~\ref{fig:alterz}. In contrast, even layers correct $Z$ errors leading to the decrease of their density and the increase in the QCNN output. We find that the QCNN can tolerate $Z$ errors for error probabilities $p_Z$ below a threshold $p_{\rm th} = 0.054$ as the error correction in even layers dominates over the error concentration in odd layers leading to a net increase of the QCNN output after every two layers, see blue lines in Fig.~\ref{fig:alterz}. On the other hand, $Z$ errors cannot be tolerated above the threshold since the error concentration dominates over the error correction leading to a net decrease of the QCNN output. See Appendix~\ref{app:err} for the derivation of the threshold error probability $p_{\rm th}$. This shows that implementing a $Z$-error correcting layer after each $X$-error correcting layer prevents the concentration of symmetry-breaking $Z$ errors below the threshold error probability.

\begin{figure*}[t]
\includegraphics[width = \textwidth]{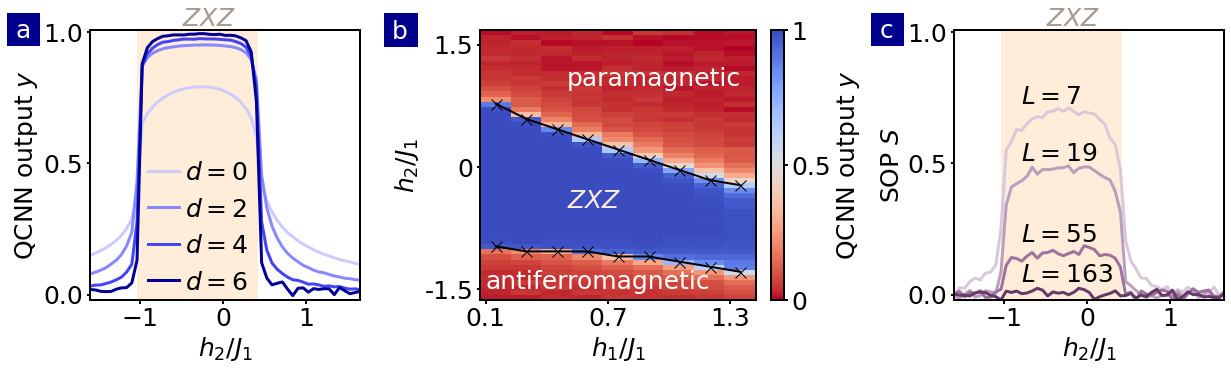}
\caption{Recognition of `$ZXZ$' SPT phase by QCNN consisting of alternating layers correcting $X$ errors and $Z$ errors for ground states of the Hamiltonian \eqref{eq:ham} perturbed by depolarizing noise. (a) The QCNN output $y$ on the cut through the phase diagram as a function of $h_2/J_1$ for different depths $d$  of the QCNN. (b) The QCNN output $y$ as a function of $h_1/J_1$ and $h_2/J_1$ for the depth $d=4$. Black crosses show the phase boundary identified using iDMRG simulations. (c) String order parameters (SOPs) $S_{jk}$ on the cut through the phase diagram as a function of $h_2/J_1$ for different lengths $L = k-j+1$. The orange regions denote the `$ZXZ$' phase. [Parameters: $p_X=p_Y=p_Z = 0.015$, $M_{S} = 10^4$ , $J_2 = 0$; (a) $N=1215$, $h_1/J_1 = 0.5$; (b) $N=135$; (c) $h_1/J_1 = 0.5$]}\label{fig:qcnn dep}
\end{figure*}

We now study the error tolerance of QCNNs with alternating layers for different ground states of the cluster-Ising Hamiltonian. We consider a depolarizing channel with $p_X=p_Y=p_Z$ describing the presence of $X$ errors, $Z$ errors and their simultaneous appearance $Y = iXZ$, representing a typical situation for NISQ devices. We plot in Fig.~\ref{fig:qcnn dep}a the QCNN output for different ground states as a function of $h_2/J_1$ and different depths $d$ of the QCNN, obtained using infinite MPS simulations. We can see that the QCNN tolerates the incoherent errors as the QCNN output converges to unity with the increasing depth $d$ in the SPT phase and it vanishes outside of the SPT phase \footnote{We will show in Sec.~\ref{sec:pb} that the error tolerance is limited close to phase boundaries as the threshold error probability decreases with diverging correlation lengths.}. 

The two types of layers in the QCNN play complementary roles. The $X$-error correcting layers implement renormalization-group flow with states belonging to the `$ZXZ$'  phase flowing towards the `$ZXZ$' cluster state and states outside of the `$ZXZ$'  phase diverging from it. The $X$-error correcting layers are thus crucial for recognizing the `$ZXZ$' phase. In contrast, the $Z$-error correcting layers reduce the density of error syndromes by removing syndromes due to symmetry-breaking errors. As a result, the $Z$-error correcting layers equip the QCNNs with the tolerance to symmetry-breaking errors, see Fig.~\ref{fig:qcnn dep}a.

We now show that the QCNNs with alternating $X$- and $Z$-error correcting layers can perform phase recognition provided that the probability of errors in the prepared states is below the threshold  $p_{\rm th}$. To this end, we plot the QCNN output for the depth $d=4$ as a function of $h_1/J_1$ and $h_2/J_1$ in Fig.~\ref{fig:qcnn dep}b in the presence of depolarizing noise. We can see that the QCNN output attains near unity value in the `$ZXZ$' phase and vanishing value outside of the `$ZXZ$' phase. The abrupt change of the QCNN output from near unity values to vanishing values coincides with the phase boundary (black crosses) determined by iDMRG simulations, see Appendix~\ref{app:num} for more details.

In contrast to the QCNN, SOPs $S_{jk}$ are significantly suppressed in the presence of incoherent errors. We plot in Fig.~\ref{fig:qcnn dep}c SOPs $S_{jk}$ as a function of $h_2/J_1$ for different lengths $L$ in the presence of depolarizing noise. We can see that the SOPs rapidly vanish both in the `$ZXZ$' SPT phase and outside of the phase with the increasing length $L$.

In conclusion, the QCNN constructed here recognizes the `$ZXZ$' SPT phase in the presence of symmetry-breaking errors below the threshold error probability $p_{\rm th}$.
In contrast, SOPs rapidly vanish with the increasing length for any finite probability of symmetry-breaking errors. The previously considered QCNN of Ref.~\cite{cong2019} cannot tolerate symmetry-breaking errors either as its output decreases with the increasing depth for any error probability. As a result, it cannot recognize the SPT phase in the presence of symmetry-breaking errors. 

\section{`$ZXXXZ$' symmetry-protected topological phase}\label{sec:2spt}
We now discuss the extension of phase recognition capabilities of the error-tolerant QCNNs we introduced to distinguish the `$ZXXXZ$' SPT phase from topologically trivial phases as well as the `$ZXZ$' and `$ZXXXZ$' SPT phases from one another.

Similarly, as for the `$ZXZ$' SPT phase, we construct a QCNN that detects the `$ZXXXZ$' phase from the topologically trivial paramagnetic and antiferromagnetic phases. Now the convolutional layer consists of a disentangling unitary $\tilde{U}^{\dagger}_N$ mapping the `$ZXXXZ$' cluster state onto a product state. The QEC procedures are amended to correct $X_j$ and $X_jX_{j+1}$ errors perturbing the `$ZXXXZ$'  cluster state, see Appendix~\ref{app:zxxxz} for more details about the QCNN for the `$ZXXXZ$' phase. To equip the QCNN with the tolerance to state preparation errors, we follow the same approach as in the previous section. We alternate $X$-error correcting layers and $Z$-error correcting layers. While the $X$-error correcting layers had to be amended for the target `$ZXXXZ$' phase, we can employ the same procedure based on the majority function of Eq. \eqref{eq:maj} to correct $Z_j$ errors. In contrast to the disentangling circuit $U^{\dagger}_N$ for the `$ZXZ$' cluster state, which commutes with $Z_j$ errors, the disentangling unitary $\tilde{U}^{\dagger}_N$ for the `$ZXXXZ$' cluster state maps $Z_j$ errors onto the errors $\tilde{U}^{\dagger}_N Z_j\tilde{U}_N = Y_{j-1}Z_jY_{j+1}$, which flip the measurement outcomes $x_{j-1}$, $x_{j}$ and $x_{j+1}$ on the three qubits $j-1$, $j$ and $j+1$. Due to this multiplication of symmetry-breaking errors, the threshold probability $\tilde{p}_{\rm th} =0.018$ is reduced compared to $p_{\rm th}=0.054$ for the `$ZXZ$' cluster state.

We determine the phase boundary between the `$ZXZ$' phase and the  `$ZXXXZ$' phase to be located at $J_1/J_2 = 0.95$ via iDMRG simulations.

We start with a QCNN recognizing the `$ZXZ$' phase from the `$ZXXXZ$' phase.
Before showing the results, we explain the construction of this QCNN, which requires identifying  the perturbations driving the ground states for non-vanishing $h_2$ and $J_2$ away from the characteristic `$ZXZ$' cluster state. These perturbations include the $X_jX_{j+1}$ interactions and the stabilizer elements $D_j$. The $X_jX_{j+1}$ interactions are corrected by the original $X$-error correcting procedure. The $D_j$ stabilizer elements are mapped by the disentangling unitary onto $U_N^{\dagger}D_jU_N = -Y_{j-1}X_jY_{j+1}$, which lead to the same syndromes after the measurement of all qubits in the $X$ basis (flipped measurement outcomes at qubits $j-1$ and $j+1$) as $X_j$ perturbations, for which $U_N^{\dagger}X_jU_N = Z_{j-1}X_jZ_{j+1}$. As a result, the QCNN depicted in Fig.~\ref{fig:qcnnx} constructed in the previous section for correcting $X_j$ and $X_jX_{j+1}$ perturbations, corrects $D_j$ perturbations as well and can be readily used to recognize the `$ZXZ$' phase from the `$ZXXXZ$' phase. To achieve tolerance to state preparation errors on NISQ devices, we can thus alternate the $X$-error correcting layers with $Z$-error correcting layers in the same way as depicted in Fig.~\ref{fig:qcnnz}.

\begin{figure}[t]
\centering
\includegraphics[width=0.66\linewidth]{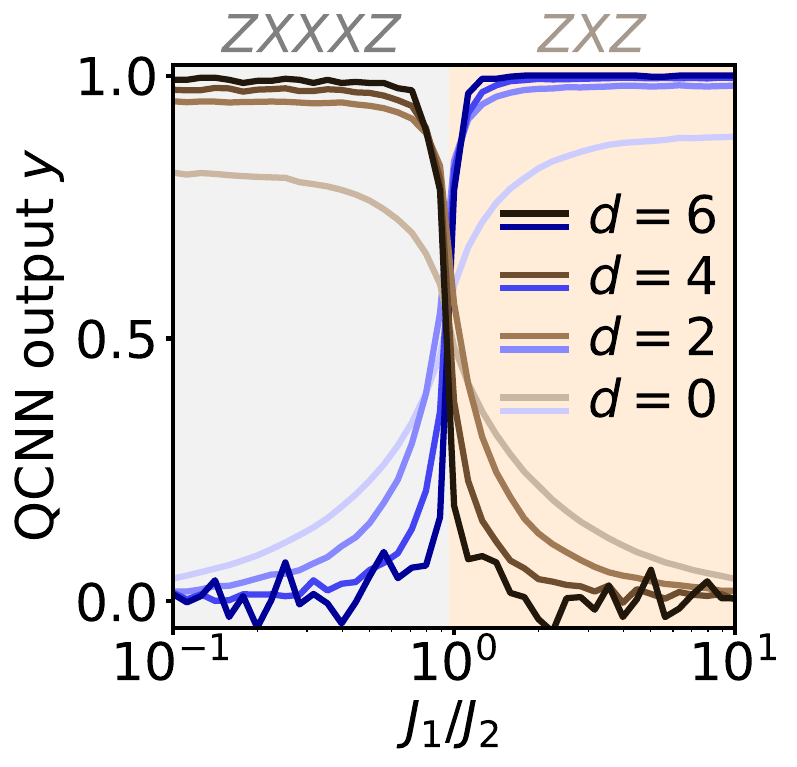}
\caption{Distinguishing the `$ZXZ$’ SPT phase and the `$ZXXXZ$’ SPT phase. The QCNN output as a function of $J_1/J_2$ for different depths $d$ of the QCNN  for ground states perturbed by depolarizing noise. The QCNN with alternating $X$-error correcting layers and $Z$-error correcting layers (blue lines) as well as the QCNN with alternating $C$-error and $Z$-error correcting layers  (brown lines). (Parameters: $N=1215$, $M_{S} = 10^3$, $h_1/J_2=0$, $h_2/J_2=0.1$, $p_X=p_Y=p_Z = 0.01 $)}\label{fig:2top}
\end{figure}

We plot the QCNN output (blue lines) as a function of $J_1/J_2$ in Fig.~\ref{fig:2top} for different depths of the QCNN in the presence of depolarizing noise. We can see that the QCNN detects the `$ZXZ$' phase as its output converges to unity in the phase ($J_1/J_2 > 0.95$) and vanishes in the `$ZXXXZ$'  phase ($J_1/J_2 < 0.95$).

We now discuss the construction of a QCNN recognizing the `$ZXXXZ$' phase from the `$ZXZ$' phase. Here, the stabilizer elements $C_j$ and $X_jX_{j+1}$ interactions play the role of perturbations away from the `$ZXXXZ$' cluster state. The disentangling unitary $\tilde{U}_N^{\dagger}$ for the `$ZXXXZ$' cluster state maps the $C_j$ perturbations onto $\tilde{U}_N^{\dagger}C_j\tilde{U}_N \propto Y_{j-1}X_jY_{j+1}$. The $C_j$ perturbations have different syndromes after the measurement of all qubits in the $X$ basis than $X_j$ and $X_jX_{j+1}$ perturbations, $\tilde{U}_N^{\dagger}X_j\tilde{U}_N = Y_{j-2}X_{j-1}X_jX_{j+1}Y_{j+2}$ and $\tilde{U}_N^{\dagger}X_jX_{j+1}\tilde{U}_N = Y_{j-2}Z_{j-1}Z_{j+2}Y_{j+3}$. As a result, we need to amend the QEC procedures to correct  the $C_j$ perturbations, see Appendix~\ref{app:circ} for more details about this procedure. The $C$-error correcting procedure also corrects $X_jX_{j+1}$ perturbations and $X_j$ perturbations, where the latter now come about only due to noise on NISQ devices, see Appendix~\ref{app:zxxxz} for more details. To achieve tolerance to incoherent $Z_j$ errors, we alternate the $C$-error correcting layers with the $Z$-error correcting layers based on the majority function. We plot the resulting  QCNN output (brown lines) as a function of $J_1/J_2$ in Fig.~\ref{fig:2top} for different depths of the QCNN in the presence of depolarizing noise. We can see that the QCNN detects the `$ZXXXZ$' phase as its output converges to unity in the phase $J_1/J_2 < 0.95$ and vanishes in the `$ZXZ$'  phase $J_1/J_2 > 0.95$.

We have thus demonstrated that the error-tolerant QCNNs we introduced can distinguish not only topological phases from topologically trivial phases but also two topological phases from one another.  To this end, the QCNN for the `$ZXXXZ$' phase needed to be amended to correct $C_j$ perturbations whereas the original QCNN for the `$ZXZ$'  phase was already capable of correcting $D_j$ perturbations. While the QCNNs for the `$ZXZ$' phase and the `$ZXXXZ$' phase are designed to correct different symmetry-preserving errors, the tolerance to symmetry-breaking errors is achieved in a universal manner in all QCNNs by employing the same $Z$-error correcting layer based on the majority function.

\begin{figure*}[t]
\includegraphics[width=0.92\linewidth]{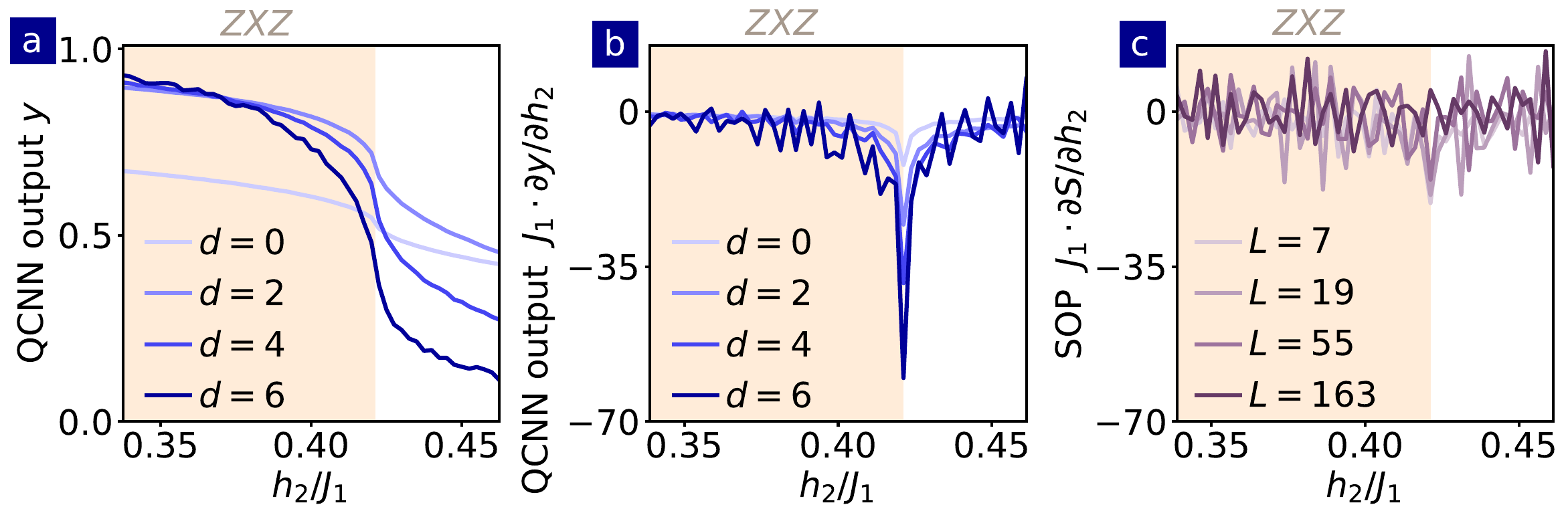}
\caption{Detecting the phase boundary between the `$ZXZ$' phase and the paramagnetic phase in the presence of depolarizing noise. (a) The output of the QCNN with alternating layers correcting $X$ errors and $Z$ errors close to the phase boundary as a function of $h_2/J_1$ for different depths $d$. (b) The slope of the QCNN output $\partial y/\partial h_2$ with respect to the Hamiltonian parameter $h_2$. (c) The slope of SOPs $\partial S_{jk}/\partial h_2$ with respect to the Hamiltonian parameter $h_2$ close to the phase boundary as a function of $h_2/J_1$ for different lengths $L=k-j+1$. The orange regions denote the `$ZXZ$' phase. [Parameters: $M_{S}  = 10^4$, $h_1/J_1 = 0.5$, $J_2 = 0$, $p_X=p_Y=p_Z = 0.015 $; (a) and (b) $N=1215$]}\label{fig:slope}
\end{figure*}

\section{Phase boundary}\label{sec:pb}
So far, we have shown that the QCNNs we consider can recognize SPT phases in the presence of incoherent errors. We now investigate the tolerance of incoherent errors close to phase boundaries. Precisely detecting phase boundaries is one of the major challenges of many-body physics due to diverging correlation lengths and the rapid growth of entanglement in their vicinity \cite{sachdev2011,eisert2010}.

In Fig.~\ref{fig:slope}a we plot the output of the QCNN for the `$ZXZ$' phase close to a phase boundary between the `$ZXZ$' phase and the paramagnetic phase in the presence of depolarizing noise. We can see that the QCNN tolerates incoherent errors well in the SPT phase as its output converges to unity with the increasing depth $d$. On the other hand, close to the phase boundary the QCNN does not tolerate incoherent errors as its output decreases with the increasing depth $d$. We thus observe that while symmetry-preserving $X$ errors can be tolerated for any ground state, the tolerance to symmetry-breaking errors is limited close to phase boundaries.

\begin{figure}[t]
\includegraphics[width=0.66\linewidth]{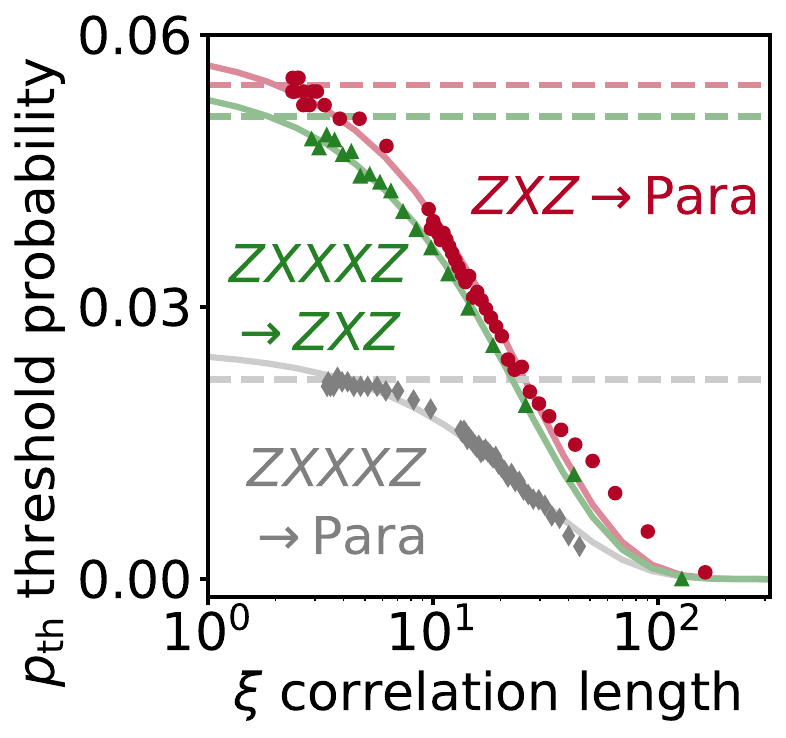}
\caption{Tolerance to incoherent errors close to phase boundaries. The plot shows the threshold probability $p_{\rm th}$ of symmetry-breaking errors as a function of the correlation length $\xi$ in the vicinity of phase boundaries between the `$ZXZ$' phase and the paramagnetic phase (red dots),  the `$ZXXXZ$' phase and the paramagnetic phase (gray diamonds), and  the `$ZXXXZ$' phase and the `$ZXZ$' phase (green triangles). Solid lines show fitted exponential functions and dashed lines show threshold error probabilities of corresponding cluster states. [Parameters: $N=1215$, $M_{S}  = 10^5$, $h_1/J_1 = 0.5$ and $J_2 = 0$ (red dots), $h_1/J_2 = 0.5$ and $J_1 = 0$ (gray diamonds), $h_1 = 0$ and $h_2/J_2 = 0.1$ (green triangles)]}\label{fig:pb}
\end{figure}

We now further quantify the behavior close to the phase boundary between the `$ZXZ$' phase and the paramagnetic phase. We investigate which probabilities $p_Z$ of symmetry-breaking $Z$ errors can be tolerated for each ground state belonging to the `$ZXZ$' phase. To do so, we determine the threshold error probability $p_{\rm th}$ for various ground states below which the QCNN output converges to unity with an increasing depth $d$. Above $p_{\rm th}$ the output decreases with $d$. We plot in Fig.~\ref{fig:pb} the threshold error probability $p_{\rm th}$ (red dots) as a function of the correlation length $\xi$ of the ground states \footnote{The correlation length corresponds to a characteristic length scale at which quantum correlation functions exponentially decay. For matrix product states with the unique largest eigenvalue $|\eta_1| =1$ of the corresponding transfer matrix, the correlation length $\xi \propto - \frac{1}{\log|\eta_2|}$ is determined by the second largest eigenvalue $\eta_2$}. We can see that the threshold error probability decreases with the correlation length. We fit this decrease by the exponential function $p_{\rm th} = p_{\rm th}^0 \exp(- \xi/\bar{\xi})$ with the fitted parameters $p_{\rm th}^0$ and $\bar{\xi}$ in Tab.~\ref{tab}. We can see in Fig.~\ref{fig:pb} a similar exponential decrease of the threshold probability $p_{\rm th}$ close to the phase boundaries between the `$ZXXXZ$' phase and the paramagnetic state (gray diamonds) as well as between the `$ZXXXZ$' phase and the `$ZXZ$' phase (green triangles). The error tolerance is thus strongly suppressed close to all phase boundaries when the correlation length exceeds the characteristic value $\bar{\xi} \approx 26$, c.f.  Tab.~\ref{tab}.

To identify the phase boundary between the `$ZXZ$' phase and the paramagnetic phase in the presence of symmetry-breaking errors, we plot in Fig.~\ref{fig:slope}b the slope $\partial y/\partial h_2$ of the QCNN output $y$ with respect to the Hamiltonian parameter $h_2$ for different depths $d$ of the QCNN. We can see a sharp dip in the slope of the QCNN output precisely located at the phase boundary. The dip becomes more pronounced with the increasing depth $d$ of the QCNN. This shows that while the QCNN output decreases with the increasing depth $d$ close to the phase boundary, see Fig.~\ref{fig:slope}a, we can still precisely identify the phase boundary as a sharp dip in the slope of the QCNN output. We can see in Fig.~\ref{fig:qcnn dep}b that the abrupt change of the QCNN output coincides with the phase boundary (black crosses) determined using iDMRG in the entire phase diagram ($J_2 = 0$). We also observe that the slope of the QCNN output exhibits a sharp dip (or peak) at all other phase boundaries of the generalized cluster-Ising model also for $J_2 \neq 0$ (not shown here). 

In stark contrast to these characteristics, individual SOPs and their slopes are largely suppressed for any finite probability of symmetry-breaking errors. We plot in Fig.~\ref{fig:slope}c the slope $\partial S_{jk}/\partial h_2$ of the SOPs with respect to the Hamiltonian parameter $h_2$ for different lengths $L=k-j+1$. We can see that the slope of the SOPs cannot be distinguished from sampling noise for the number of samples $M_{S}=10^4$. Crucially, the slope of the SOPs does not become more pronounced with increasing length $L$. As a result, one cannot use the SOPs to determine the phase boundary in the presence of symmetry-breaking noise.

The sharp dip in the slope of the QCNN output at critical values of Hamiltonian parameters is a unique feature of phase boundaries and it does not appear at the threshold values of probabilities for incoherent errors, see also Appendix~\ref{app:err}. This feature distinguishes phase boundaries from threshold probabilities of incoherent errors. It thus provides further evidence that the QCNNs we construct perform phase recognition for the error-free ground states while processing only noisy states, which approximate the former.

In conclusion, the tolerance of the considered QCNNs to symmetry-breaking errors is limited close to phase boundaries due to diverging correlation lengths. Nonetheless, we can precisely determine critical values of Hamiltonian parameters as a dip in the slope of the QCNN output. This is in stark contrast to SOPs and their slopes which rapidly vanish for any finite probability of symmetry-breaking errors and thus cannot be used to identify phase boundaries in the presence of symmetry-breaking errors.

\begin{table}[t]
\begin{tabular}{|l|c|c|}
\hline
Phase boundary 			&  $p_{\rm th}^0$ 	& $\bar{\xi}$\\
\hline
`$ZXZ$'$\rightarrow$Paramagnetic		& $0.0588\pm0.0005$					& $26.1\pm0.6$ \\
`$ZXXXZ$'$\rightarrow$Paramagnetic 	    & $0.0254\pm0.0002$ 					& $28.1\pm0.6$ \\
`$ZXXXZ$'$\rightarrow$`$ZXZ$' 	        & $0.0550\pm0.0004$					& $24.6\pm0.6$ \\ 
\hline
\end{tabular}
\caption{Fitted parameters of the threshold probability $p_{\rm th} = p_{\rm th}^0 \exp(- \xi/\bar{\xi})$ of symmetry-breaking errors as a function of the correlation length $\xi$ close to different phase boundaries.}\label{tab}
\end{table}

\section{Sample complexity}\label{sec:complexity}

We now compare QCNNs to the direct measurement of the input state. We focus on sample complexity which quantifies the number of projective measurements required to identify to which quantum phase the input state belongs. In the absence of noise, QCNNs substantially reduce sample complexity compared to the direct measurement of SOPs \cite{cong2019}. In the presence of symmetry-breaking noise, SOPs vanish and thus they cannot detect the SPT phase. Instead, we focus on the observable measured by QCNNs. This observable is different from any single SOP and it is robust against symmetry-breaking noise. In this section, we discuss this observable for the QCNN that detects the `$ZXZ$' phase and compare the cost of directly sampling it from the input state to sampling the QCNN output.

To determine the observable measured by the QCNN with alternating X- and Z-error correcting layers, we represent the QCNN circuit as a unitary $U_{\rm QCNN}$, see Appendix~\ref{app:msop} for details. The measurement of the Pauli $X_{\frac{N+1}{2}} $ operator at the end of the QCNN circuit corresponds to the measurement of a multiscale SOP
\begin{align}	
	S_{\rm M} &= U_{\rm QCNN}^{\dagger} X_{\frac{N+1}{2}} U_{\rm QCNN}\nonumber\\
	&=\sum_{ij}\eta^{(1)}_{ij}S_{ij}+\sum_{ijkl}\eta^{(2)}_{ijkl}S_{ij}S_{kl} + ...,\label{eq:msop}
\end{align}
on the input state. The multiscale SOP $S_{\text{M}}$ is a sum of products of SOPs $S_{ij}$ at different lengths $L = j-i +1$. The length of the SOPs, $L\sim 3^d$, increases exponentially with the depth $d$ of the QCNN. Compared to the QCNN of Ref.~\cite{cong2019}, the change in the coefficients $\eta^{(n)}$ due to our construction equips the QCNN with error tolerance.  The multiscale SOP in Eq.~\eqref{eq:msop} involves at least $2^{3^{d-2}}$ products of SOPs.

As an alternative to executing the QCNN, we can determine the expectation value of the multiscale SOP using the direct measurements of the input state without performing any quantum circuit. Assuming that only measurements in local Pauli bases can be directly performed, which is the case for most devices, we show in Appendix~\ref{app:comp} that the multiscale SOP involves at least $3^{3^{d-4}}$ products of SOPs, which cannot be simultaneously measured via local Pauli measurements as they require sampling in mutually incompatible Pauli bases. As a result, the sample complexity of the direct Pauli measurement scales double exponentially with the depth of the QCNN, corresponding to an exponential scaling with system size for the maximal depth $d=\lfloor \log_3 N\rfloor$. Instead of the direct Pauli measurement, one could estimate the products of SOPs via classical shadow tomography \cite{huang2020}. However, the sample complexity of classical shadow tomography also scales exponentially with the length $L$ of the SOPs. In contrast, all products of SOPs are sampled simultaneously in the QCNN after performing the disentangling unitary. The QCNN thus determines the expectation value of the multiscale SOP with a constant sample complexity in system size (and depth of the QCNN), which exponentially reduces the sample complexity compared to direct Pauli measurements. 

Importantly, the equivalent QCNN circuit depicted in Fig.~\ref{fig:qcnnz}, which is based on a constant-depth quantum circuit, measurement and classical post-processing, measures the multiscale SOP with the same sample complexity as the full quantum QCNN circuit. The constant-depth quantum circuit allows us to simultaneously measure all stabilizer elements $C_j$. From measured bit strings, we then determine the expectation value of the multiscale SOP in classical post-processing with the same sample complexity as for the full quantum QCNN circuit.

QCNNs detecting the `$ZXXXZ$' phase also measure multiscale SOPs which are sums of double exponentially many products of SOPs $T_{jk}$. Similarly to the QCNN for the `$ZXZ$' phase, these QCNNs also reduce sample complexity  exponentially compared to direct local Pauli measurements.

\section{conclusions}\label{sec:con}
We constructed QCNNs that tolerate incoherent errors due to decoherence and gate infidelities during the preparation of their input states. These QCNNs tolerate symmetry-breaking errors below a threshold error probability in contrast to previous QCNN designs and SOPs, which are significantly suppressed for any non-vanishing error probability. Moreover, their output is robust against invertible symmetry-preserving error channels. The error tolerance is limited close to phase boundaries as the threshold error probability decreases with diverging correlation lengths. However, a steep gradient of the QCNN output at phase boundaries between SPT phases and topologically trivial phases as well as between two SPT phases allows us to precisely determine critical values of the Hamiltonian parameters. 

The QCNN quantum circuits constructed here can be shortened from logarithmic depth in input size to short, constant depth by performing a large part of computation in classical post-processing after the measurement of all qubits. This substantially improves the performance of QCNNs under NISQ conditions by reducing the number of finite-fidelity quantum gates. The classical post-processing part of QCNNs consists of  logic circuits with at most logarithmic depth in input size. The QCNNs we constructed reduce sample complexity exponentially in input size in comparison to the direct sampling of the QCNN output using local Pauli measurements.  

Our work provides new insights into SPT order in open quantum systems, which are subject to decoherence and dissipation. Apart from NISQ computers, the error channel we consider, see Eq. \eqref{eq:channel}, describes typical open quantum systems \cite{groot2022}. On the one hand, SOPs rapidly vanish with an increasing length for any symmetry-breaking error channel as shown in Ref.~\cite{groot2022}. On the other hand, our results show that SPT order is not completely washed out for probabilities of symmetry-breaking errors below a finite threshold. This distinction emerges because the multiscale SOPs, that are efficiently measured by the QCNNs we introduce, exploit information about SPT order at different length scales to detect SPT phases in the presence of symmetry-breaking noise.

Due to the tolerance of errors and the short depth of their quantum circuits, the QCNNs constructed here can be readily realized on current NISQ computers to efficiently measure characteristic non-local observables of SPT phases. This will facilitate the investigation of topological quantum phases of matter on quantum computers.

As a next step, the construction of QCNN circuits mimicking renormalization-group flow depicted in Fig.~\ref{fig:qcnn}b could be extended for two- and higher-dimensional systems, which can, in addition to SPT phases, feature intrinsic topological order \cite{kitaev2003} and symmetry enriched topological phases \cite{mesaros2013}. Using preparation circuits for characteristic ground states belonging to SPT phases \cite{ge2016} as well as symmetry enriched topological phases \cite{liu2022}, these ground states can be considered as fixed points of QCNNs and the QEC unitaries in pooling layers can be constructed in analogy to renormalization-group decoders \cite{duclos-cianci2010}.

QCNN circuits for two- and higher-dimensional systems can be equipped with the tolerance to incoherent errors by using our approach. To this end, the QCNN layers mimicking renormalization-group flow are alternated with layers correcting symmetry-breaking errors. While the layers mimicking renormalization-group flow need to be specifically designed for each target quantum phase and given Hamiltonian perturbations, the correction of symmetry-breaking errors can be implemented in a universal manner using the majority function. In this way, all errors perturbing the characteristic state are corrected provided that the error density is small enough, as we demonstrated for different SPT phases of the generalized cluster-Ising model. The majority function provides a universal tool to correct symmetry-breaking errors in QCNNs for two- and higher-dimensional systems.

Other interesting future directions include detecting less understood topological phases such as anyonic chains \cite{feiguin2007} and quantum spin liquids \cite{savary2016}. Another promising direction is the training of QCNNs based on parametrized quantum circuits to identify non-local observables characterizing topological phases from training data \cite{cong2019,pesah2021,caro2022,liu2023}.

The QCNNs we constructed open the way for efficiently characterizing noisy quantum data produced by near-term quantum hardware. In addition to the recognition of topological phases, reducing the sample complexity of non-local observables will substantially speed up other quantum algorithms. A prominent example is the variational quantum eigensolver for quantum chemistry problems which involves many repetitions of demanding measurements of molecular Hamiltonians  \cite{peruzzo2014,mcclean2016}.

\section{Acknowledgements}
We thank R. Mansuroglu for insightful discussions.
This work was supported by the EU program H2020-FETOPEN project 828826 Quromorphic as well as the EU program HORIZON-MSCA-2022-PF project 101108476 HyNNet NISQ and is part of the Munich Quantum Valley, which is supported by the Bavarian state government with funds from the Hightech Agenda Bayern Plus. NAM is funded by the Alexander von Humboldt Foundation.

\appendix

\section{Numerical simulations}\label{app:num}
Our main results are based on MPS simulations implemented using the Python library TeNPy \cite{hauschild2018}. Using iDMRG with the maximal bond dimension $\chi=150$, we obtain numerically exact ground states $\ket{\psi}$ of the Hamiltonian \eqref{eq:ham} in the thermodynamic limit to avoid finite size effects. First, we identify phase boundaries as sharp peaks in the second derivative of the ground state energy with respect to $h_2/J_1$ for constant $h_1/J_1$ and $J_2=0$ in Figs.~\ref{fig:origx}, \ref{fig:qcnn dep}, and \ref{fig:slope}, with respect to $J_1/J_2$ for constant $h_1=0$ and $h_2/J_2 = 0.1$ in Fig.~\ref{fig:2top}, as well as with respect to $h_2/J_2$ for constant $h_1/J_2$ and $J_1=0$ in Fig.
~\ref{fig:zxxxz}.

We implement the QCNN circuits depicted in Fig.~\ref{fig:qcnn}c consisting of a constant-depth quantum circuit, the measurement of all qubits in the $X$ basis and classical post-processing. The constant-depth quantum circuit performs the disentangling unitary $U^{\dagger}_N$ consisting of nearest-neighbor two-qubit gates which can be efficiently applied on the MPSs $|\psi\rangle$ obtained using iDMRG. We simulate the measurement outcomes of $N$ qubits by sampling spin configurations $x$ in the $X$ basis from their probability distribution $P_x = \textrm{Tr}[|x\rangle\langle x|U^{\dagger}_N|\psi\rangle \langle \psi|U_N]$
corresponding to the MPS $U^{\dagger}_N|\psi\rangle$ after having performed the disentangling unitary $U_N^{\dagger}$. QCNN outputs are determined from the sampled bit strings $x$ as a Boolean function which is expressed as a logic circuit, see Figs.~\ref{fig:qcnnx} and \ref{fig:qcnnz}.

To explore incoherent errors using the error channel \eqref{eq:channel}, we implement the error channel $\mathcal{E}$ by sampling error events $E_l = K_l/\sqrt{p_l}$ from their probability distribution $p_l = {\rm Tr}[K_l^{\dagger}K_l]/2^N$. The error events $E_l$ are products of Pauli operators, which can be efficiently implemented on the MPSs $|\psi\rangle$. We then sample bit strings $x$ from the joint probability distribution 
\begin{align}
P_x &= p_l\textrm{Tr}[\ket{x}\bra{x}U^{\dagger}_N E_l\ket{\psi}\bra{\psi}E_l^{\dagger}U_N]
\end{align}
which correspond to the measurement outcomes for the noisy state $\mathcal{E}(\ket{\psi}\bra{\psi})$ after having performed the disentangling unitary $U_N^{\dagger}$.

Increasing the bond dimension to $\chi=200$ does not lead to a visible change in our findings showing that the MPSs accurately describe the ground states of the Hamiltonian \eqref{eq:ham} and their processing with the QCNNs. 

\section{QCNN circuits}\label{app:circ}
\begin{figure*}[t]
\includegraphics[width = \linewidth]{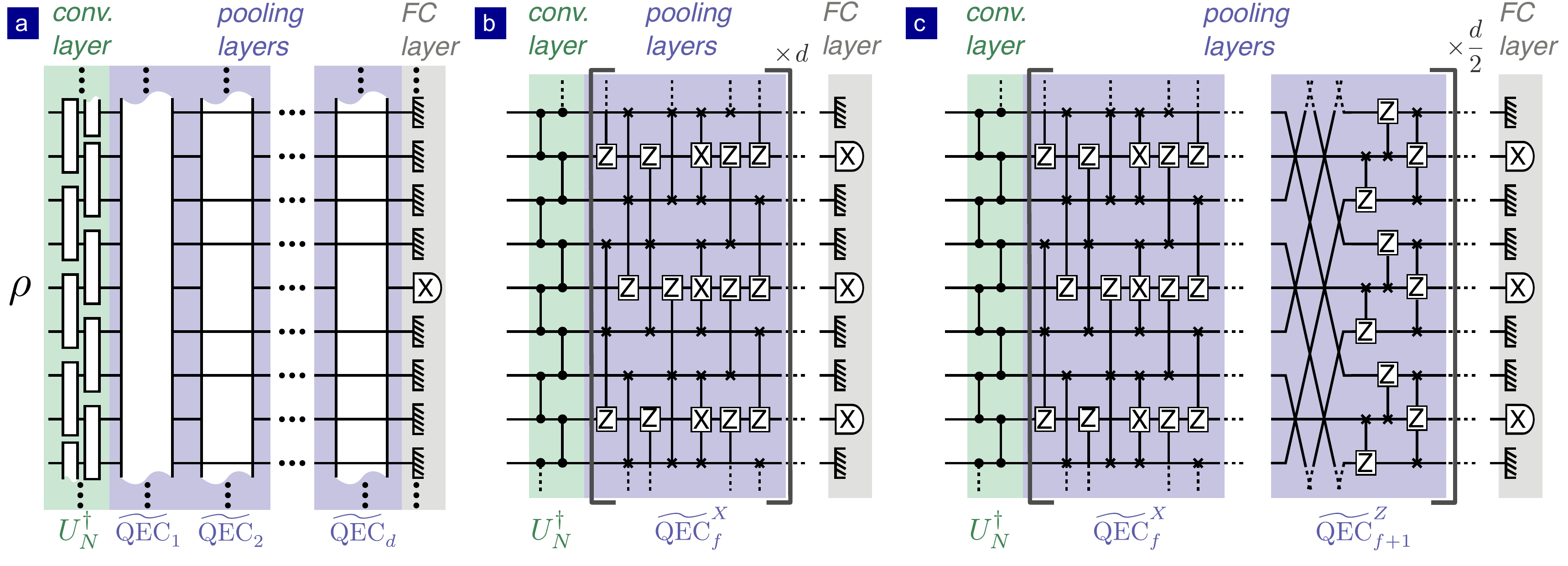}
\caption{QCNN quantum circuits. (a) QCNN circuit equivalent to Fig.~\ref{fig:qcnn}b consisting of a single convolutional layer, $d$ pooling layers and a fully connected (FC) layer. The convolutional layer performs the disentangling unitary $U_N^{\dagger}$. Each pooling layer $f=1,2,...,d$ performs the $\widetilde{\textrm{QEC}}_{f}$ procedure transformed by the entangling unitary $U_{N/3^{f}}$ on a sublattice with $N/3^{f-1}$ qubits. The fully connected layer involves the measurement of $\lfloor N/3^d\rfloor$ qubits in the $X$ basis. (b) QCNN detecting the `$ZXZ$' phase with $X$-error correcting procedures $\widetilde{\textrm{QEC}}_{f}^{X}$. (c) QCNN detecting the `$ZXZ$' phase with alternating procedures $\widetilde{\textrm{QEC}}_{f}^X$ and $\widetilde{\textrm{QEC}}_{f+1}^Z$ correcting $X$ errors and $Z$ errors, respectively. The disentangling unitary $U_N^{\dagger}$ involves controlled $Z$ gates with controls in the computational basis. The $\widetilde{\textrm{QEC}}_{f}^X$ procedure consists of controlled-controlled Z gates C$_x$C$_x$Z, controlled Z gates C$_x$Z and controlled-controlled NOT gates C$_x$C$_x$NOT with all controls in the $X$ basis. The $\widetilde{\textrm{QEC}}_{f+1}^Z$ procedure consists of SWAP, C$_x$C$_x$Z and C$_x$Z gates.}\label{fig:qcnntr}
\end{figure*}
In this appendix, we describe in detail the QCNN circuits used in this work. We first discuss the QCNN detecting the `$ZXZ$' phase. Then we describe the QCNN detecting the `$ZXXXZ$' phase. Finally, we discuss the QCNNs distinguishing the two SPT phases from one another.  

All QCNNs considered in this work consist of $d$ convolutional layers, $d$ pooling layers and a fully connected layer as depicted in Fig.~\ref{fig:qcnn}a. Each convolutional layer $f=1,2,...,d$ consists of a disentangling unitary $U^{\dagger}_{N/3^{f-1}}$ on $N/3^{f-1}$ qubits followed by an entangling unitary $U_{N/3^{f}}$ on a sublattice with $N/3^{f}$ qubits as depicted in Fig.~\ref{fig:qcnn}b. Each pooling layer involves a QEC procedure $\textrm{QEC}_{f}$. In the fully connected layer, the disentangling unitary $U^{\dagger}_{N/3^d}$ is applied and all $\lfloor N/3^d\rfloor$ remaining qubits are measured in the $X$ basis. Note that each $\textrm{QEC}_{f}$ procedure is preceded by the entangling unitary $U_{N/3^{f}}$, which is implemented in the preceding convolutional layer, and followed by the disentangling unitary $U^{\dagger}_{N/3^{f}}$, which is implemented in the following convolutional layer for $f<d$ and in the fully connected layer for $f=d$, see Fig.~\ref{fig:qcnn}b. The QCNN circuit is thus equivalent to a single convolutional layer followed by $d$ pooling layers as depicted in Fig.~\ref{fig:qcnntr}a. The convolutional layer performs the disentangling unitary $U^{\dagger}_N$. The pooling layer $f$ involves the QEC unitary $\widetilde{\textrm{QEC}}_{f} = U_{N/3^{f}}^{\dagger}\textrm{QEC}_{f}U_{N/3^{f}}$ transformed by the entangling unitary $U_{N/3^{f}}$. Note that in this equivalent quantum circuit, convolutional layers for $f>1$ are absorbed into the $\widetilde{\textrm{QEC}}_{f}$ unitaries in the pooling layers. The disentangling unitary $U_{N/3^{d}}^{\dagger}$ from the fully connected layer is also absorbed into the $\widetilde{\textrm{QEC}}_{d}$ unitary. The fully connected layer in this equivalent quantum circuit thus consists only of the measurement of remaining qubits in the $X$ basis.

For conciseness, we focus here on the equivalent quantum circuits depicted in Fig.~\ref{fig:qcnntr}a as the $\widetilde{\textrm{QEC}}_{f}$ procedures transformed by the entangling unitary consist of fewer gates than the bare QEC$_f$ procedures.

We first discuss the QCNN detecting the `$ZXZ$' phase. We start with the QCNN consisting of $X$-error correcting layers depicted in Fig.~\ref{fig:qcnntr}b. The disentangling unitary $U^{\dagger}_{N}$ consists of controlled $Z$ gates between neighboring qubits. The transformed QEC procedure $\widetilde{\textrm{QEC}}^X_{f}$ consists of controlled-controlled Z gates C$_x$C$_x$Z, controlled Z gates C$_x$Z and controlled-controlled NOT gates C$_x$C$_x$NOT with all controls in the $X$ basis. The error-tolerant design of the QCNN depicted in Fig.~\ref{fig:qcnntr}c consists of alternating layers correcting $X$ errors and $Z$ errors. The new $Z$-error correcting procedure $\widetilde{\textrm{QEC}}^Z_{f}$ involves SWAP, C$_x$Z and C$_x$C$_x$Z gates.

Since all gates are controlled in the $X$ basis and they implement either Pauli $X$ or Pauli $Z$ operations on the target qubit, the $\widetilde{\textrm{QEC}}_{f}^X$ and $\widetilde{\textrm{QEC}}_{f}^Z$ procedures map $X$-basis eigenstates $\ket{x}$ onto other $X$-basis eigenstates $\pm\ket{g_f(x)}$, where $g_f:\{0,1\}^N \to \{0,1\}^N$ is a Boolean function. As we also measure in the $X$ basis in the fully-connected layer, the processing of a quantum state $\rho$ by the QCNN can be implemented in classical post-processing as a Boolean function $G(x) =  (g_d\circ g_{d-1}\circ...\circ g_1)(x)$ after measuring all qubits in the $X$ basis. In particular, the output 
\begin{equation}
\langle X_j\rangle = \textrm{Tr}\left[X_jQ_dU^{\dagger}_N\rho U_N Q_d^{\dagger}\right] =\sum_{x}P_x (1 -2G(x)_j),
\end{equation}
of qubit $j$ measured in the fully-connected layer of the full quantum QCNN circuit (Fig.~\ref{fig:qcnntr}a) can be determined from bit strings $x$ measured after the convolutional layer by using the $j$th element of the output of the Boolean function $ G(x)$, where 
$P_x = \textrm{Tr}[|x\rangle\langle x|U^{\dagger}_N\rho U_N]$ is the probability of measuring a bit string $x$ after the first convolutional layer and
\begin{equation}
Q_d = \widetilde{\textrm{QEC}}_{d}...\widetilde{\textrm{QEC}}_{2}\widetilde{\textrm{QEC}}_{1}.
\end{equation}

We thus only need to apply the single disentangling unitary $U_N^{\dagger}$ on a quantum computer, measure all qubits in the $X$ basis and determine the QCNN output in classical post-processing from the measured bit strings $x$, see Fig.~\ref{fig:qcnn}c. The QCNN quantum circuits depicted in Figs.~\ref{fig:qcnntr}b and \ref{fig:qcnntr}c can thus be implemented as equivalent circuits depicted in Figs.~\ref{fig:qcnnx} and \ref{fig:qcnnz}, respectively. The Boolean function $g_f$ corresponding to the pooling layer $f$ performing the $\widetilde{\textrm{QEC}}_{f}$ procedure can be expressed as a logic circuit with a constant depth in system size, see Figs.~\ref{fig:qcnnx} and \ref{fig:qcnnz}. As a result, the Boolean function $G$ corresponding to the QCNN with $d$ pooling layers can be implemented as a logic circuit with a depth proportional to $d$ which can be at most $d = \lfloor\log_3 N\rfloor$ logarithmic in system size $N$.

We now describe QCNNs detecting the `$ZXXXZ$' phase. The QCNN consisting of alternating layers correcting $X$ errors and $Z$ errors is depicted in Fig.~\ref{fig:zxxxztr}a. The disentangling unitary $\tilde{U}^{\dagger}$ consists of controlled $Z$ gates CZ between all neighboring qubits controlled in the computational basis, controlled $Y$ gates C$_y$Y between all neighboring qubits controlled in the $Y$ basis and $Z$ gates. The $X$-error correcting procedure $\widetilde{\textrm{QEC}}_{f}^X$ consists of controlled $Y$ gates C$_x$Y and controlled-controlled $Y$ gates C$_x$C$_x$Y with controls in the $X$ basis. The $Z$-error correcting procedure $\widetilde{\textrm{QEC}}_{f}^Z$ is the same as for the `$ZXZ$' phase involving SWAP, C$_x$Z and C$_x$C$_x$Z gates. The QCNN consisting of alternating layers correcting $C$ errors and $Z$ errors is depicted in Fig.~\ref{fig:zxxxztr}b.  The $C$-error correcting procedure $\widetilde{\textrm{QEC}}_{f}^C$ consists of C$_x$Y and C$_x$C$_x$Y gates. 

As all $\widetilde{\textrm{QEC}}_{f}$ procedures consist of gates controlled in the $X$ basis implementing the Pauli $Y$ operation or the Pauli $Z$ operation on the target qubit, they satisfy the condition \eqref{eq:cond} and thus they can be implemented in classical post-processing as a Boolean function $g_f(x)=x'$. The output 
\begin{equation}
\langle X_j\rangle = \textrm{Tr}\left[X_jQ_d\tilde{U}^{\dagger}_N\rho \tilde{U}_N Q_d^{\dagger}\right] =\sum_{x}\tilde{P}_x (1-2G(x)_j),
\end{equation}
of qubit $j$ measured in the fully-connected layer of the full quantum QCNN circuit can be determined from bit strings $x$ measured after the first convolutional layer $\tilde{U}^{\dagger}_N$ by using the $j$th element of the output of the Boolean function $G(x)$, where 
$\tilde{P}_x = \textrm{Tr}[|x\rangle\langle x|\tilde{U}^{\dagger}_N\rho \tilde{U}_N]$.

Note that the Boolean function corresponding to the $C$-error correcting procedure for the `$ZXXXZ$' phase is the same as the Boolean function performing the $X$-error correcting procedure for the `$ZXZ$' phase depicted in Fig.~\ref{fig:qcnnx}. 

\section{Propagation of symmetry-preserving errors in QCNN circuits}\label{app:errx}

In this appendix, we discuss the propagation of symmetry-preserving $X$ errors in the QCNN detecting the `$ZXZ$' phase and consisting of $X$-error correcting layers. We exploit that the QCNN can be implemented as the constant-depth quantum circuit $U^{\dagger}_N$, measurements in the $X$ basis and classical post-processing. We focus on the propagation of errors in the constant-depth quantum circuit $U^{\dagger}_N$ and the logic circuit implemented in classical post-processing.

The `$ZXZ$' cluster state $\ket{C}$ is mapped by the disentangling unitary $U^{\dagger}_N$ onto the product state $\ket{+}^{\otimes N}$, where $\ket{+}$ is the $+ 1$ eigenstate of the Pauli $X$ operator.  The subsequent measurement thus deterministically yields the outcome $x_j = 0$ corresponding to $X_j = +1$ for all qubits $j$. A single $X_j$ error perturbing the cluster state $X_j\ket{C}$ is mapped onto $U^{\dagger}_NX_jU_N = Z_{j-1}X_jZ_{j+1}$ by the disentangling unitary $U^{\dagger}_N$ leading to the flip of two measurement outcomes $x_{j\pm1} = 1$, see red lines in Fig.~\ref{fig:qcnnx}. This $X$-error syndrome is corrected by the $X$-error correcting procedure such that $g(x)_k = 0$ for all $N/3$ classical bits $k$ propagating to the next layer, see Fig.~\ref{fig:qcnnx}. The other $2\cdot N/3$ bits are discarded.

We now investigate the cluster state $K\ket{C}$ perturbed by an arbitrary number $\Upsilon = \sum_{j=1}^{N}\upsilon^{(0)}_j$ of $X$ errors as the input state, where $K = \prod_{j=1}^{N}X_j^{\upsilon^{(0)}_j}$ and $\upsilon^{(0)}_j = 0,1$. The perturbed cluster state is mapped by the disentangling unitary $U^{\dagger}_N$ onto the product state $\bigotimes_{j=1}^N \ket{\pm}_j$, where qubit $j$ is in the state $\ket{+}$ if $\upsilon^{(0)}_{j-1}\oplus\upsilon^{(0)}_{j+1} = 0$ and in the state $\ket{-}$ otherwise. The subsequent measurement in the Pauli $X$ basis thus yields the deterministic outcome $x_j = \upsilon^{(0)}_{j-1}\oplus\upsilon^{(0)}_{j+1}$. The measured bit string $x$ is processed in the logic circuit depicted in Fig.~\ref{fig:qcnnx} consisting of $X$-error correcting layers. A bit $x_j$ at the output of each $X$-error correcting layer $f$ depends on five bits $x_{j-4\cdot3^{f-1}}$, $x_{j-2\cdot3^{f-1}}$, $x_{j}$, $x_{j+2\cdot3^{f-1}}$, and $x_{j+4\cdot3^{f-1}}$ at the input of this layer. We express the bit string $x_j = \upsilon^{(f)}_{j-3^f}\oplus\upsilon^{(f)}_{j+3^f}$ at every layer $f>0$ in terms of $\upsilon^{(f)}_{j}=0,1$ similarly to the bit string $x_j = \upsilon^{(0)}_{j-1}\oplus\upsilon^{(0)}_{j+1}$ at layer $f=0$, i.e., directly after the measurement. We can interpret $\upsilon^{(f)}_{j}$ as the syndrome of the $X_j$ error on the sublattice with $N/3^f$ qubits. Each $X$-error correcting layer implements the majority function 
\begin{equation}\label{eq:majx}
    \upsilon^{(f)}_{j} = M(\upsilon^{(f-1)}_{j-2\cdot3^{f-1}},\upsilon^{(f-1)}_{j},\upsilon^{(f-1)}_{j+2\cdot3^{f-1}})
\end{equation}
for the $X$-error syndromes. As a result, all $X$ error syndromes are corrected for a small initial number $\Upsilon\ll N$ of $X$ errors and a sufficiently large depth $d$. The QCNN output thus converges to unity with the increasing depth $d$.

As shown in Fig.~\ref{fig:origx}, the QCNN output converges to unity for all states in the `$ZXZ$' phase. The phase boundary coincides with a threshold density of coherent $X$ errors perturbing the cluster state \cite{cong2019,lake2022}. Above the threshold density, $X$-error syndromes are concentrated in the QCNN circuit and the QCNN output vanishes with increasing depth $d$.

We now discuss the propagation of incoherent $X$ errors in the QCNN for the cluster state $\mathcal{E}(|C\rangle\langle C|)$ with the probability $p_X$ of $X$ errors described by the error channel \eqref{eq:channel} with $p_Y=p_Z = 0$. At layer $f=0$, the probability $p_0 = p_X$ of the $X$ error is uniform at every qubit and the probabilities of $X$ errors at different qubits $j$ and $k$ are not correlated. We will now describe the probability $p_f$ of $X$-error syndromes $\upsilon_j^{(f)}$ at each layer $f$. The majority function \eqref{eq:majx} dictates that the probabilities of error syndromes at different qubits remain uniform and uncorrelated. According to the majority function, the error probabilities follow the recursion relation
\begin{equation}\label{eq:rx}
p_f = p_{f-1}^2(3-2p_{f-1}).
\end{equation}
After $d$ $X$-error correcting layers, we obtain $x_j = \upsilon^{(d)}_{j-3^d}\oplus\upsilon^{(d)}_{j+3^d} = 1$ with the probability $2p_d(1-p_d)$.

We can exactly determine the QCNN output 
\begin{equation}
    y = 1- 4 p_d(1-p_d)    
\end{equation}
for the initial error probability $p_0$ by iterating the recursion relation \eqref{eq:rx}. The probability $p_f$ after each $X$-error correcting layer decreases, i.e., $p_f<p_{f-1}$, for $0<p_{f-1}<0.5$ and it increases, i.e., $p_f>p_{f-1}$, for $0.5<p_{f-1}<1$. As a result, all $X$-error syndromes are corrected for $p_0<0.5$ and a sufficiently large depth $d$. The QCNN output converges to unity with increasing depth. For $p_0>0.5$, the error probability $p_d$ monotonously increases with increasing depth $d$. For large depths $d$, we obtain $p_d \approx 1$ and thus the QCNN also attains a near unity output. The situation is qualitatively different for $p_0 = 0.5$. This value of the error probability is the fixed point $p_f=p_{f-1} = 0.5$ of the recursion relation \eqref{eq:rx} and the QCNN output vanishes for every depth $d$.

The action of the error channel $\mathcal{E}$ on the cluster state is qualitatively different for $p_0 = p_X \neq 0.5$ and $p_X = 0.5$. For $p_X \neq 0.5$, the error channel is invertible and it thus preserves SPT order \cite{groot2022}. In the presence of incoherent errors, the SOPs $S_{jk}$ decrease by the factor $(1-2p_X)^2$ but they retain non-vanishing values, see Eq.~\eqref{eq:sopx}. In this case, the QCNN is able to correct the incoherent errors and it attains ideal noise-free values for a sufficiently large depth $d$. In contrast, the error channel is not invertible for $p_X = 0.5$. The SOPs are annihilated by the error channel (see Eq.~\eqref{eq:sopx}) and SPT order is completely washed out \cite{groot2022}. In this case, the QCNN is not able to correct the incoherent errors and it attains a vanishing output for every depth $d$ as discussed above.

\section{Propagation of symmetry-breaking errors in QCNN circuits}\label{app:err}
In this appendix, we discuss the propagation of symmetry-breaking $Z$ errors in QCNNs detecting the `$ZXZ$' phase.

The situation is more complicated for $Z$ errors than for symmetry-preserving $X$ errors discussed in Appendix~\ref{app:errx}. A single $Z_j$ error perturbing the cluster state $Z_j\ket{C}$ leads to the flip of a single measurement outcome $x_{j} = 1$ as the error $Z_j$ commutes with the disentangling unitary $U^{\dagger}_N$, see Fig.~\ref{fig:qcnnx}. This syndrome of the $Z_j$ error propagates through the $X$-error correcting layer such that $g(x)_k = 1$ for bits $k$ on the sublattice with $N/3$ bits in the next layer if $k=j-2$, $k=j$, or $k=j+2$. As $2\cdot N/3$ bits are discarded in each layer, the density of error syndromes increases. This leads to a decreasing QCNN output for any probability $p_Z\neq0$ or $p_Z\neq 1$ of $Z$ errors. To correct $Z$ errors, we construct a new $Z$-error correcting $\widetilde{\textrm{QEC}}^Z_f$ procedure, depicted in Fig.~\ref{fig:qcnntr}c, with a corresponding logic circuit depicted in Fig.~\ref{fig:qcnnz}. This logic circuit consists of the majority function, see Eq. \eqref{eq:maj}. The majority function $M(x_{j-7\cdot3^{f-1}},x_{j},x_{j+7\cdot3^{f-1}})$ in layer $f$ returns the value of the majority of the three bits $x_{j-7\cdot3^{f-1}}$, $x_{j}$, and $x_{j+7\cdot3^{f-1}}$. It thus removes isolated syndromes $x_{j}=1$ of $Z_j$ errors, see Fig.~\ref{fig:qcnnz}. Provided that the initial density of $Z_j$ errors is small enough, the majority vote further decreases the density of error syndromes, preventing their concentration in the QCNN circuit.

We now investigate the propagation of $Z$ errors in the QCNN with alternating $X$-error and $Z$-error correcting layers for the cluster state $\rho = \mathcal{E}(\ket{C}\bra{C})$ with the probability $p_Z$ of $Z$ errors described by the error channel \eqref{eq:channel} with $p_X=p_Y=0$. As $Z$ errors commute with the disentangling unitary $U_N^{\dagger}$, we measure $x_j = 1 $ with the identical probability $p_Z$ at all qubits $j$. Moreover, the probabilities of measuring the values $x_j = 1$ and $x_k=1$ on different qubits $j$ and $k$ are not correlated.  We will now describe how the probability $p_f$ of $Z$-error syndromes evolves in each layer $f$ of the QCNN. We first note that the probability $p_f$ remains identical in each layer, i.e. the same for all qubits $j$, as the QCNN circuit is translationally invariant. We also assume that the probability of $Z$-error syndromes on different qubits remains uncorrelated. This assumption is well justified by the agreement with our numerical simulations showing that correlations between error syndromes that build up in the logic circuit are negligible.

We start with the probability $p_0 = p_Z$ of measuring $x_j=1$ at each qubit $j$ after the disentangling unitary $U_N^{\dagger}$. The measured bit strings $x$ are now processed in the $X$-error correcting layers for $f$ odd and in the $Z$-error correcting layers for $f$ even. A bit $x_j$ at the output of the $X$-error correcting layer $f$ depends on five bits $x_{j-4\cdot3^{f-1}}$, $x_{j-2\cdot3^{f-1}}$, $x_{j}$, $x_{j+2\cdot3^{f-1}}$, and $x_{j+4\cdot3^{f-1}}$ at the input of this layer, see Fig.~\ref{fig:qcnnz}, each of which has the value $1$ with the probability $p_{f-1}$. Using a truth table for the output of the $X$-error correcting layer, we determine that the output value $x_j=1$ occurs with the probability
\begin{align}
p_f &= f_X(p_{f-1}) \nonumber\\
&= p_{f-1}^3 +p_{f-1}(1-p_{f-1})^2\left(3-2p_{f-1}+4p_{f-1}^2\right).\label{eq:fx}
\end{align}
The probability $p_f$ after each $X$-error correcting layer increases, i.e., $p_f>p_{f-1}$ for $0<p_{f-1}<0.5$ resulting in a decreased QCNN output. This can also be seen in Fig.~\ref{fig:alterz}, where after each $X$-error correcting layer, the QCNN output decreases, compare red and blue lines.

\begin{figure}[t]
\centering
\includegraphics[width=0.66\linewidth]{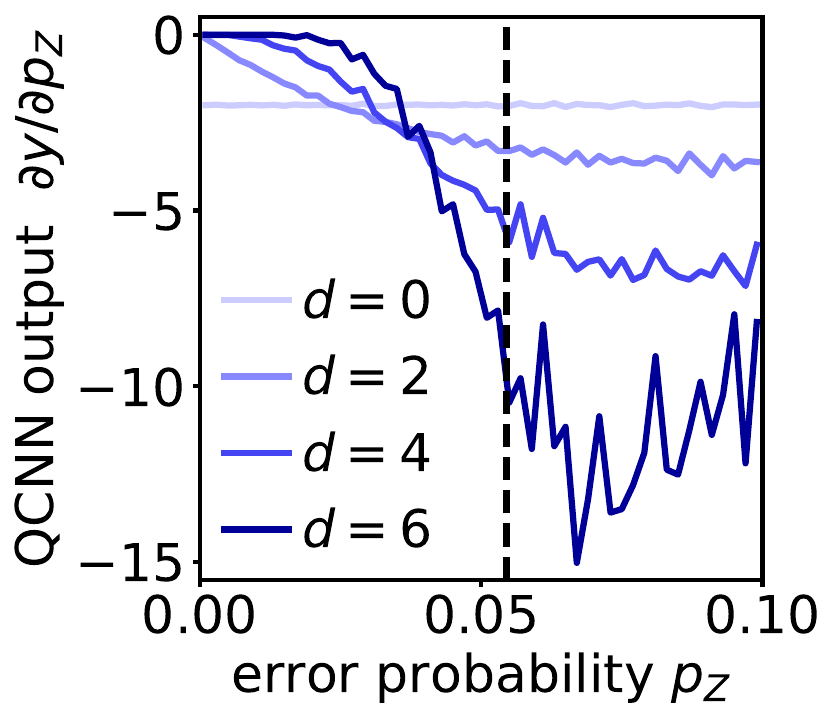}
\caption{QCNN consisting of alternating layers correcting $X$ errors and $Z$ errors for the `$ZXZ$' cluster state perturbed by symmetry-breaking $Z$ errors. The slope $\partial y/\partial p_z$ of the QCNN output with respect to the probability $p_Z$ of $Z$ errors as a function of the probability $p_Z$ for different depths $d$. The black dashed line shows the threshold error probability $p_{\rm th} = 0.054$. (Parameters: $N=1215$, $M_{S}  = 10^5$, $p_X=p_Y=0$)}\label{fig:slopealterz}
\end{figure}

\begin{figure*}[t]
\centering
\includegraphics[width=\linewidth]{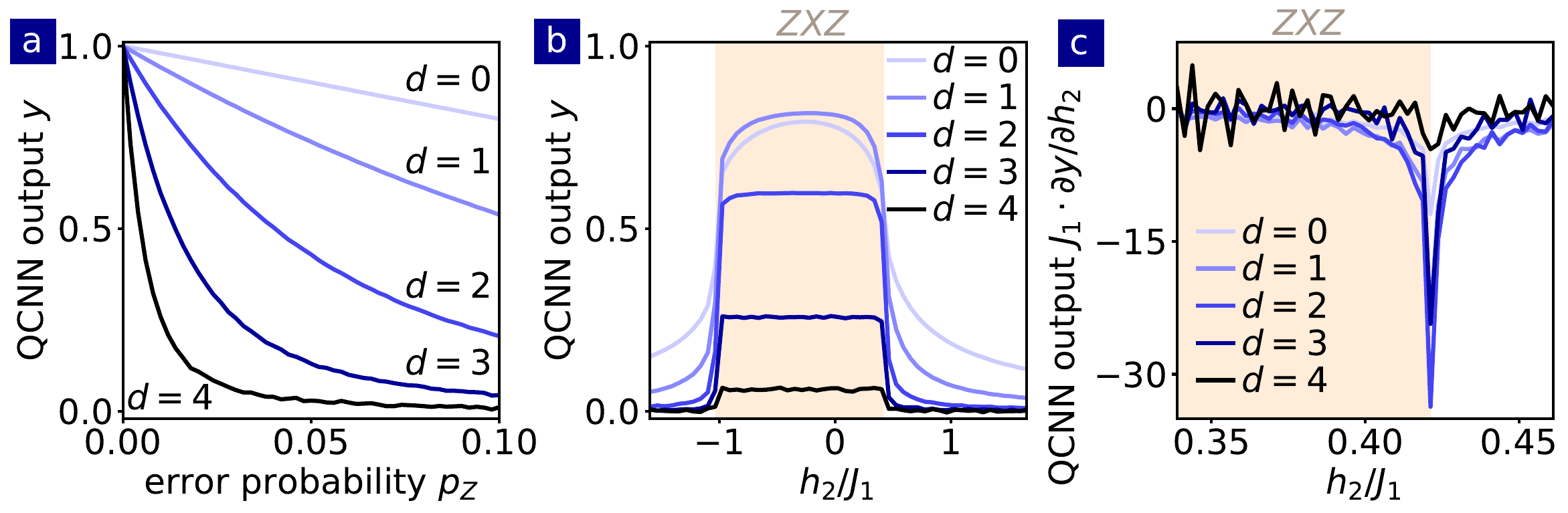}
\caption{QCNN detecting the `$ZXZ$' SPT phase consisting of $X$-error correcting layers for the ground states of the cluster-Ising Hamiltonian \eqref{eq:ham} perturbed by incoherent symmetry-breaking errors. (a) The QCNN output for the `$ZXZ$' cluster state perturbed by $Z$ errors as a function of the error probability $p_Z$ for different depths $d$ of the QCNN. (b) The QCNN output as a function of $h_2/J_1$ and different depths $d$  in the presence of depolarizing noise. (c) The slope of the QCNN output $\partial y/\partial h_2$ with respect to the Hamiltonian parameter $h_2$ close to the phase boundary between the `$ZXZ$' phase and the paramagnetic phase as a function of $h_2/J_1$ for different depths $d$. The orange regions denote the `$ZXZ$' phase. [Parameters: $N=1215$, $M_{S} = 10^4$; (a) $p_X=p_Y=0$; (b) and (c) $h_1/J_1 = 0.5$, $J_2 = 0$, $p_X=p_Y=p_Z = 0.02 $]}\label{fig:origz}
\end{figure*}

A bit $x_j$ at the output of the $Z$-error correcting layer $f$ depends on three bits $x_{j-7\cdot3^{f-1}}$, $x_{j}$, and $x_{j+7\cdot3^{f-1}}$ at the input of this layer, see Fig.~\ref{fig:qcnnz}, each of which has the value $1$ with the probability $p_{f-1}$. Using a truth table for the output of the $Z$-error correcting layer, we determine that the output value $x_j=1$ occurs with the probability
\begin{equation}
p_f = f_Z(p_{f-1}) = p_{f-1}^2(3 - 2p_{f-1}).\label{eq:fz}
\end{equation}
The probability $p_f$ after each $Z$-error correcting layer decreases, i.e., $p_f<p_{f-1}$ for $0<p_{f-1}<0.5$ resulting in an increased QCNN output. This can also be seen in Fig.~\ref{fig:alterz}, where after each $Z$ error correcting layer the QCNN output increases, compare red and blue lines.

We identify two distinct regimes depending on the initial error probability $p_0 = p_Z$. For error probabilities below the threshold $p_Z < p_{\rm th}$, error correction in even ($Z$-error correcting) layers dominates over error concentration in odd ($X$-error correcting) layers resulting in a net reduction of errors after two subsequent layers. For error probabilities above the threshold $p_Z>p_{\rm th}$, error concentration in odd layers dominates over error correction in even layers resulting in a net concentration of errors after two subsequent layers. We determine the threshold probability $p_{\rm th} = 0.054$ as the fixed point of the recursion relation $p_f = f_Z(f_X(p_{f-2}))$.

We now study the slope of the QCNN output with respect to the error probability $p_Z$. We plot the slope as a function of the probability $p_Z$ in Fig.~\ref{fig:slopealterz} for different depths $d$ of the QCNN. We can see a key difference compared to the slope of the QCNN output with respect to the Hamiltonian parameter $h_2$ depicted in Fig.~\ref{fig:slope}b. The slope with respect to $h_2$ exhibits a sharp dip precisely located at the phase boundary between the `$ZXZ$' SPT phase and the paramagnetic phase. In contrast, the slope with respect to the probability $p_Z$ keeps decreasing with the increasing probability $p_Z$ above the threshold value $p_{\rm th} = 0.054$. We conclude that the sharp dip in the slope of the QCNN output with respect to the Hamiltonian parameters is a unique feature of phase boundaries, which distinguishes them from the behavior of the QCNN output at the threshold error probability $p_{\rm th}$.

\section{QCNN consisting of $X$-error correcting layers}\label{app:corx}

In this appendix, we discuss the QCNN recognizing the `$ZXZ$' SPT phase consisting of $X$-error correcting layers and how it is affected by symmetry-breaking errors. The compact implementation of the QCNN as a quantum circuit consisting of the disentangling unitary $U^{\dagger}_N$, the measurement of all qubits in the $X$ basis and classical post-processing is depicted in Fig.~\ref{fig:qcnnx}.

\begin{figure*}[t]
\includegraphics[width = \linewidth]{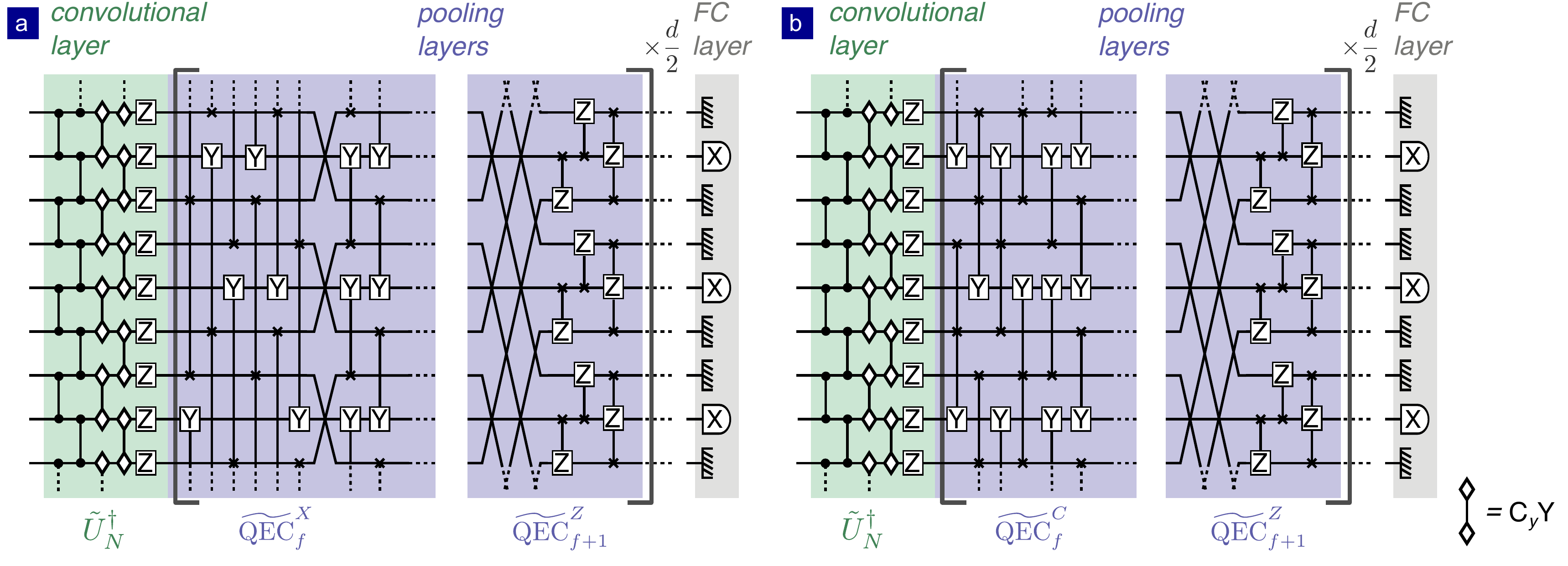}
\caption{QCNN quantum circuit detecting the `$ZXXXZ$' phase. (a) QCNN with alternating procedures $\widetilde{\textrm{QEC}}_{f}^X$ and $\widetilde{\textrm{QEC}}_{f+1}^Z$ correcting $X$ errors and $Z$ errors, respectively, for the recognition of the `$ZXXXZ$' phase from topologically trivial phases. (b) QCNN with alternating procedures $\widetilde{\textrm{QEC}}_{f}^C$ and $\widetilde{\textrm{QEC}}_{f+1}^Z$ correcting $C$ errors and $Z$ errors, respectively, for the recognition of the `$ZXXXZ$' phase from the `$ZXZ$' phase. The disentangling unitary $\tilde{U}^{\dagger}$ involves controlled $Z$ gates CZ with controls in the computational basis, controlled $Y$ gates C$_y$Y with controls in the $Y$ basis and $Z$ gates. The $\widetilde{\textrm{QEC}}_{f}^X$ procedure consists of SWAP gates as well as controlled-controlled Y gates C$_x$C$_x$Y and controlled Y gates C$_x$Y with all controls in the $X$ basis. The $\widetilde{\textrm{QEC}}_{f+1}^Z$ procedure consists of SWAP, C$_x$C$_x$Z and C$_x$Z gates. The $\widetilde{\textrm{QEC}}_{f}^C$ procedure consists of C$_x$C$_x$Y and C$_x$Y gates.}\label{fig:zxxxztr}
\end{figure*}

We start by investigating the QCNN for the `$ZXZ$' cluster state perturbed by incoherent $Z$ errors as the input state. We plot in Fig.~\ref{fig:origz}a the QCNN output as a function of the $Z$-error probability $p_Z$. We can see that the QCNN output decreases with increasing depth for all error probabilities. $Z$ errors propagate through the $X$-error correcting layers and, as the system size is reduced by a factor of three in each layer, the density of $Z$ errors increases. As a result, the probability of $Z$ errors increases with increasing depth $d$ according to the recursion relation \eqref{eq:fx}. This error concentration leads to the decrease of the QCNN output for any initial error probability $p_Z$. This is in contrast to the QCNN consisting of alternating layers correcting $X$ errors and $Z$ errors that tolerates symmetry-breaking $Z$ errors below the threshold error probability $p_{\rm th} = 0.054$.

We now study the QCNN consisting of $X$-error correcting layers for different ground states of the cluster-Ising Hamiltonian in the presence of depolarizing noise. We plot in Fig.~\ref{fig:origz}b the QCNN output as a function of $h_2/J_1$ and different depths $d$ of the QCNN. We can see that the QCNN output vanishes with increasing depth $d$ both in the SPT phase (orange region) and outside of the SPT phase (white regions). This is in contrast to the QCNN consisting of alternating layers correcting $X$ errors and $Z$ errors that tolerates symmetry-breaking errors and thus attains near unity values in the SPT phase for large depths $d$.

We plot in Fig.~\ref{fig:origz}c the slope $\partial y/\partial h_2$ of the QCNN output $y$ with respect to the Hamiltonian parameter $h_2$ for different depths $d$ of the QCNN close to the phase boundary between the `$ZXZ$' phase and the paramagnetic phase. We can see a dip in the slope of the QCNN output located at the phase boundary. However, the dip vanishes for large depths $d\geq4$ of the QCNN. This is again in contrast to the QCNN consisting of alternating layers correcting $X$ errors and $Z$ errors. For this network, the dip in the slope becomes more pronounced with increasing depth due to the tolerance of errors.

We conclude that the QCNN consisting of $X$-error correcting layers is largely affected by symmetry-breaking errors. Due to the concentration of symmetry-breaking errors in the QCNN, the QCNN output rapidly vanishes with the increasing depth $d$. Also the slope of the QCNN output at phase boundaries vanishes with the increasing depth. This is in contrast to the QCNN consisting of alternating layers correcting $X$ errors and $Z$ errors that tolerates symmetry-breaking errors and thus attains near unity values in the SPT phase for large depths. Moreover, the dip in the slope of the QCNN output at phase boundaries becomes more pronounced with increasing depth.

\section{`$ZXXXZ$' SPT phase}\label{app:zxxxz}
\begin{figure*}[t]
\centering
\includegraphics[width=\linewidth]{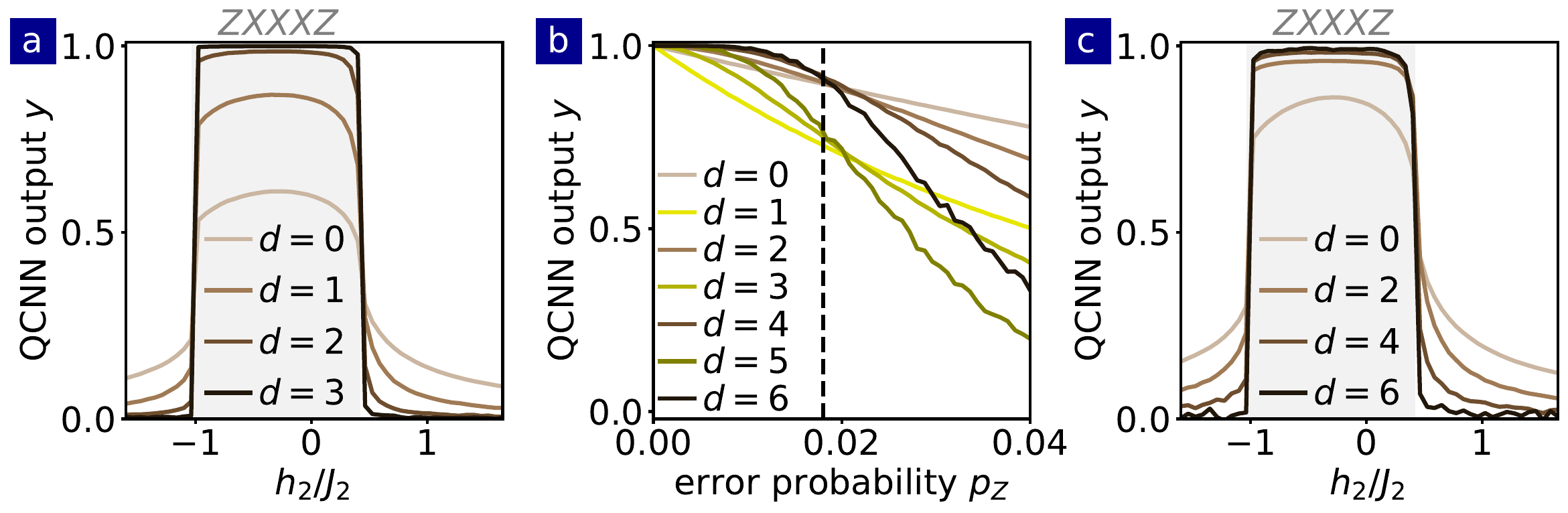}
\caption{QCNN recognizing the `$ZXXXZ$' phase from the paramagnetic phase and the anti-ferromagnetic phase. (a) The output of the QCNN consisting of $X$-error correcting layers for ground states of the cluster-Ising Hamiltonian \eqref{eq:ham} perturbed by incoherent $X$ errors as a function of $h_2/J_2$ for different depths $d$ of the QCNN. (b) The output of the QCNN consisting of alternating layers correcting $X$ errors and $Z$ errors as a function of the $Z$-error probability $p_Z$ for the `$ZXXXZ$' cluster state for different depths $d$. The black dashed line shows the threshold error probability $\tilde{p}_{\rm th} = 0.018$. (c) The output of the QCNN consisting of alternating layers correcting $X$ errors and $Z$ errors for ground states of the cluster-Ising Hamiltonian \eqref{eq:ham} perturbed by depolarizing noise as a function of $h_2/J_2$ for different depths $d$. The gray regions denote the `$ZXXXZ$' phase. [Parameters: $N=1215$, $M_{S} = 10^4$; (a) $p_X = 0.1 $, $p_Y=p_Z=0$, $h_1/J_2=0.5$, $J_1=0$; (b) $p_X=p_Y=0$; (c) $p_X=p_Y=p_Z = 0.005 $, $h_1/J_2=0.5$, $J_1=0$]}\label{fig:zxxxz}
\end{figure*}

In this appendix, we discuss QCNNs recognizing the `$ZXXXZ$' SPT phase and their tolerance to different types of errors. Similarly, as for the `$ZXZ$' SPT phase, we construct a QCNN to recognize the `$ZXXXZ$' SPT phase from the paramagnetic phase and the antiferromagnetic phase. The $X$-error correcting procedure $\widetilde{\textrm{QEC}}^X_f$ is depicted in Fig.~\ref{fig:zxxxztr}a.

The `$ZXXXZ$' cluster state $\ket{D}$ is mapped by the disentangling unitary $\tilde{U}^{\dagger}_N$ in the first convolutional layer onto the product state $\ket{+}^{\otimes N}$.  The subsequent measurement thus deterministically yields the outcome $x_j = 0$ for all qubits $j$. A single $X_j$ error perturbing the cluster state $X_j\ket{D}$ is mapped onto $\tilde{U}_N^{\dagger}X_j\tilde{U}_N = Y_{j-2}X_{j-1}X_jX_{j+1}Y_{j+2}$ by the disentangling unitary $\tilde{U}^{\dagger}_N$ leading to the flip of two measurement outcomes $x_{j\pm2} = 1$. This error is corrected by the $X$-error correcting QEC procedure such that $g(x)_k = 0$ for all $N/3$ classical bits $k$ propagating to the next layer. Similarly, the syndrome $x_{j\pm2}=x_{j-1}=x_{j+3}=1$ of a $X_jX_{j+1}$ error is corrected by the $X$-error correcting procedure.

We now investigate the QCNN with $X$-error correcting layers. We plot in Fig.~\ref{fig:zxxxz}a the QCNN output across a cut through the phase diagram as a function of $h_2/J_2$ for $J_1=0$ and different depths $d$ of the QCNN in the presence of incoherent $X$ errors. We can see that the QCNN output converges to unity with the increasing depth of the QCNN in the `$ZXXXZ$' SPT phase (gray region). On the other hand, the QCNN output vanishes with the increasing depth in the topologically trivial phases (white regions). This shows that the QCNN can recognize the `$ZXXXZ$' SPT phase from topologically trivial phases in the presence of incoherent $X$ errors. Incoherent $X$ errors can be tolerated for any probability $p_X\neq0.5$.

To equip the QCNN with the tolerance to symmetry-breaking $Z$ errors, we alternate $X$-error correcting layers with $Z$-error correcting layers, see Fig.~\ref{fig:zxxxztr}a. The $\widetilde{\textrm{QEC}}^Z_f$ procedure, correcting $Z$ errors, is the same as that for the `$ZXZ$' phase, c.f. Fig.~\ref{fig:qcnntr}c, which can be implemented in classical post-processing as the majority function \eqref{eq:maj}, depicted in Fig.~\ref{fig:qcnnz}. We start by investigating the QCNN with alternating layers for the `$ZXXXZ$' cluster state perturbed by incoherent $Z$ errors as the input state. We plot in Fig.~\ref{fig:zxxxz}b the QCNN output as a function of the $Z$-error probability. We can see an alternating QCNN output after odd and even layers. $Z$ errors propagate through odd $X$-error correcting layers and, as the system size is reduced by a factor of three, the density of $Z$ errors increases. This error concentration leads to the decrease of the QCNN output after odd layers, compare brown and yellow lines in Fig.~\ref{fig:zxxxz}b. In contrast, even layers correct $Z$ errors leading to the decrease of their density and the increase of the QCNN output. The QCNN can tolerate $Z$ errors below the threshold error probability $\tilde{p}_{\rm th} =0.018$ as the error correction in even layers dominates over the error concentration in odd layers. This shows that implementing the $Z$-error correcting layer after each $X$-error correcting layer prevents the concentration of symmetry-breaking $Z$ errors for small error probabilities.

The threshold probability $\tilde{p}_{\rm th} = 0.018$ for the `$ZXXXZ$' phase is smaller than the threshold probability $p_{\rm th}=0.054$ for the `$ZXZ$' phase. This decrease in the tolerated error probabilities can be understood by investigating the propagation of $Z$ errors in the QCNN circuit. In contrast to the disentangling circuit $U^{\dagger}_N$ for the `$ZXZ$' phase, which commutes with $Z_j$ errors, the disentangling unitary $\tilde{U}^{\dagger}$ for the `$ZXXXZ$' phase maps $Z_j$ errors onto three errors $\tilde{U}^{\dagger}_N Z_j\tilde{U}_N = Y_{j-1}Z_jY_{j+1}$, which flip the measurement outcomes $x_{j}=x_{j\pm1}=1$ on qubits $j-1$, $j$ and $j+1$. This error syndrome is corrected by the $Z$-error correcting procedure provided that the $Z_j$ error is isolated. However, the threshold error probability $\tilde{p}_{\rm th}$ is smaller than for the `$ZXZ$' phase due to the multiplication of symmetry-breaking errors by the disentangling unitary.

We now study the error tolerance of the QCNN with alternating layers for different ground states of the cluster-Ising Hamiltonian \eqref{eq:ham} perturbed by depolarizing noise. We plot in Fig.~\ref{fig:zxxxz}c the QCNN output as a function of $h_2/J_2$ for $J_1=0$ and different depths $d$ of the QCNN. We can see that the QCNN tolerates the incoherent errors due to depolarizing noise as its output converges to unity with the increasing depth $d$ in the `$ZXXXZ$' phase (gray region) and vanishes in the topologically trivial phases (white regions).

In conclusion, we constructed a QCNN for the `$ZXXXZ$' phase that tolerates symmetry-preserving $X$ errors if the error channel is invertible and symmetry-breaking $Z$ errors for small error probabilities. The QCNN is constructed similarly as for the `$ZXZ$' phase by amending the QEC procedures to correct $X_j$ and $X_jX_{j+1}$ errors perturbing the `$ZXXXZ$'  cluster state. As the disentangling unitary $\tilde{U}^{\dagger}_N$ for the `$ZXXXZ$' phase maps $Z_j$ errors onto three errors, $\tilde{U}^{\dagger}_N Z_j\tilde{U}_N = Y_{j-1}Z_jY_{j+1}$, the threshold probability $\tilde{p}_{\rm th}=0.018$ of $Z$ errors is reduced compared to the QCNN for the `$ZXZ$' phase.

We finally discuss the propagation of errors in the QCNN detecting the `$ZXXXZ$' phase from the `$ZXZ$' phase. This QCNN consists of alternating $C$-error and $Z$-error correcting layers as depicted in Fig.~\ref{fig:zxxxztr}b and discussed in Sec.~\ref{sec:2spt} of the main text. The $C$-error correcting layers are essential for the detection of the `$ZXXXZ$' phase while the $Z$-error correcting layers equip the QCNN with error tolerance. A single $C_j$ error is transformed by the disentangling unitary $\tilde{U}^{\dagger}_N$ as $\tilde{U}_N^{\dagger}C_j\tilde{U}_N \propto Y_{j-1}X_jY_{j+1}$ and thus leads to the error syndrome $x_{j\pm1}=1$. This error syndrome is corrected by the $C$-error correcting procedure. $X_j$ and $X_jX_{j+1}$ errors are mapped onto $\tilde{U}_N^{\dagger}X_j\tilde{U}_N = Y_{j-2}X_{j-1}X_jX_{j+1}Y_{j+2}$ and $\tilde{U}_N^{\dagger}X_jX_{j+1}\tilde{U}_N = Y_{j-2}Z_{j-1}Z_{j+2}Y_{j+3}$ with the corresponding error syndromes $x_{j\pm2}=1$ and $x_{j\pm2}=x_{j-1}=x_{j+3}=1$, respectively. The syndrome $x_{j\pm2}=1$ of the $X_j$ error is corrected by the $C$-error correcting layer only if bit $j$ propagates to the next layer. If bit $j$ is discarded, the $X_j$-error syndrome is transformed into either $x_{j-2}=x_{j+4} = 1$ or into $x_{j-4}=x_{j+2} = 1$. On the sublattice with $N/3$ bits in the next layer, this corresponds in both cases to the $C_k$-error syndrome $x_{k\pm3}=1$. In the subsequent $Z$-error correcting layer, we take the majority value $M(x_{k-7\cdot3},x_k,x_{k+7\cdot3})$ of every triple of qubits $x_{k-7\cdot3}$, $x_k$, and $x_{k+7\cdot3}$. As these bits are separated by the distance $7\cdot3$, the single $C_k$-error syndrome does not change any of the majority values and it is thus removed. Similarly, also $X_jX_{j+1}$ error syndromes are removed by two subsequent layers. The QCNN can thus distinguish the `$ZXXXZ$' phase from the `$ZXZ$' phase, see Fig.~\ref{fig:2top}.

Note that the QCNN consisting of only $C$-error correcting layers also corrects $X_j$ and $X_jX_{j+1}$ errors. As we discussed above, the syndrome of the $X_j$ error is transformed by the $C$-error correcting layer into the $C_k$-error syndrome on the sublattice with $N/3$ bits. This $C_k$-error syndrome is corrected in the subsequent $C$-error correcting layer. Similarly, the syndrome of a single $X_jX_{j+1}$ error is also corrected by two subsequent $C$-error correcting layers.

\section{Multiscale string order parameter}\label{app:msop}
In this appendix, we describe the multiscale SOP $S_{\text{M}}$, see Eq. \eqref{eq:msop}, that is measured by the QCNNs considered in this work. First, we show that $S_{\text{M}}$ is a sum of products of SOPs $S_{jk}$. Then, we demonstrate that the length of the SOPs involved in $S_{\text{M}}$ increases exponentially with the depth $d$ of the QCNN. Finally, we determine a lower bound for the number of products of SOPs that are summed together to construct $S_{\text{M}}$.

\begin{table*}[t]
\begin{tabular}{c|c|c|c}
$f$	&  Measured Observable	& \# of products & \\
\hline
$d$ 		& $G^{(d)}_{ii}=X_{i}$ & 1 & $\widetilde{\textrm{QEC}}_d^{X}$ 		\\
$d-1$ 	& $\frac{1}{4}\sum_{\alpha'\beta'} G^{(d-1)}_{\left(i - \alpha'\right)\left(i+\beta'\right)}-\frac{1}{4}\sum_{\alpha' } G^{(d-1)}_{\left(i - \alpha'\right)i}G^{(d-1)}_{\left(i+\gamma'\right)\left(i+\gamma'\right)} -\frac{1}{4} \sum_{\alpha' } G^{(d-1)}_{\left(i - \gamma'\right)\left(i - \gamma'\right)}G^{(d-1)}_{i\left(i+\alpha'\right)}\hdots$ & 16 & $\widetilde{\textrm{QEC}}_{d-1}^{Z}$ 	\\
$d-2$ 	&  $\frac{1}{8}\sum_{\kappa}G^{(d-2)}_{(i+\kappa)(i+\kappa)}-\frac{1}{16}\sum_{\lambda}G^{(d-2)}_{(i+\lambda-3^{d-2})(i+\lambda+3^{d-2})} + \frac{1}{32}\sum_{\lambda'}H^{(d-2)}_{(i+\lambda'-3^{d-2})(i+\lambda'+3^{d-2})} \hdots $ & 2500 & $\widetilde{\textrm{QEC}}_{d-2}^{X}$ 	\\
& \vdots &  &		\\
$0$ 	& $\sum_{ij}\eta^{(1)}_{ij}G^{(0)}_{(i+1)(j-1)}+\sum_{ijkl}\eta^{(2)}_{ijkl}G^{(0)}_{(i+1)(j-1)}G^{(0)}_{(k+1)(l-1)} + \hdots$ &$\Omega(2^{3^{d-2}})$ & $U^{\dagger}_N$ 	\\
\hline
input	&$S_{\rm M} =\sum_{ij}\eta^{(1)}_{ij}S_{ij}+\sum_{ijkl}\eta^{(2)}_{ijkl}S_{ij}S_{kl} + \hdots$ &$\Omega(2^{3^{d-2}})$ & --- \\	
\end{tabular}
\caption{Backpropagation of the measured observable through the QCNN circuit with the depth $d$. The table displays the measured observable backpropagated to layer $f$, where the ellipsis denotes products of two and more products of Pauli strings. The table also displays the number of products of Pauli strings $G^{(f)}_{jk}$ involved in the measured observable and the unitary performed at layer $f$. The multiscale SOP $S_{\rm M}$ measured on the input state and the number of products of SOPs involved in $S_{\rm M}$ are stated in the last row. We use $ H_{jk}^{(d-2)}=G^{(d-2)}_{jk}G^{(d-2)}_{\left(j+3^{d-2}\right)\left(k- 3^{d-2}\right)}$, $i = \frac{N+1}{2}$, $\alpha',\beta' \in \left\{0, 2\cdot 3^{d-1}, 4\cdot 3^{d-1} \right\}$, $\gamma' = 4\cdot 3^{d-1}$, $\kappa\in\{-7\cdot3^{d-2},0,7\cdot3^{d-2}\}$, $\lambda\in\{-6\cdot3^{d-2},6\cdot3^{d-2}\}$, and $\lambda'\in\{-6\cdot3^{d-2},0,6\cdot3^{d-2}\}$.}\label{tab:msop}
\end{table*}

We focus here on the QCNN detecting the `$ZXZ$' phase, consisting of alternating $X$-error and $Z$-error correcting layers. We consider the form of the QCNN depicted in Fig.~\ref{fig:qcnntr}c with all convolutional layers for $f>1$ and the fully connected layer absorbed into the $\widetilde{\textrm{QEC}}_{f}$ procedures in pooling layers. The QCNN circuit thus performs the unitary 
\begin{equation}
U_{\rm QCNN} =\widetilde{\textrm{QEC}}_d^X \widetilde{\textrm{QEC}}_{d-1}^Z...\widetilde{\textrm{QEC}}_2^Z\widetilde{\textrm{QEC}}^X_1U_N^{\dagger}
\end{equation}
consisting of the disentangling unitary $U_N^{\dagger}$ and $d$ pooling layers $\widetilde{\textrm{QEC}}_f$ where $f=1,2,...,d$. For odd (even) $f$, the pooling layers perform the $X$-error (Z-error) correcting procedure $\widetilde{\textrm{QEC}}_{f}^X$ ($\widetilde{\textrm{QEC}}_{f}^Z$), see Fig.~\ref{fig:qcnntr}c. We also assume that the QCNN has an odd number $d$ of layers.

\textbf{Sum of products of string order parameters.} We first show that the observable measured by the QCNN corresponds to the multiscale SOP $S_{\text{M}}$ which is a sum of products of SOPs, c.f. Eq. \eqref{eq:msop}.

The measurement of the Pauli operator
\begin{align}
\langle X_{\frac{N+1}{2}} \rangle =& \textrm{Tr}[X_{\frac{N+1}{2}}U_{\rm QCNN}\rho U_{\rm QCNN}^{\dagger}]\nonumber\\ 
						 =& \textrm{Tr}[U_{\rm QCNN}^{\dagger}X_{\frac{N+1}{2}}U_{\rm QCNN}\rho ]\label{eq:msop2}
\end{align}
at the end of the QCNN circuit corresponds to the measurement of the observable $U_{\rm QCNN}^{\dagger}X_{\frac{N+1}{2}}U_{\rm QCNN}$ on the input state $\rho$. We used the cyclic property of the trace in the second equality in Eq.~\eqref{eq:msop2}. We backpropagate the measured observable through the QCNN circuit to the input state (zeroth layer). To this end, we use the recursion relations
\begin{align}
  	&\widetilde{\textrm{QEC}}_f^{X\,\dagger}G^{(f)}_{jk}\widetilde{\textrm{QEC}}_f^{X} = \frac{1}{4}\Biggl[\sum_{\alpha\beta} G^{(f-1)}_{\left(j - \alpha\right)\left(k+\beta\right)}\nonumber\\
	&-\sum_{\alpha } G^{(f-1)}_{\left(j - \alpha\right)k}G^{(f-1)}_{\left(k+\gamma\right)\left(k+\gamma\right)}-\sum_{\alpha} G^{(f-1)}_{\left(j - \gamma\right)\left(j - \gamma\right)}G^{(f-1)}_{j\left(k+\alpha\right)}\nonumber\\
	&+G^{(f-1)}_{\left(j - \gamma\right)\left(j - \gamma\right)}G^{(f-1)}_{jk}G^{(f-1)}_{\left(k+\gamma\right)\left(k+\gamma\right)}\Biggr],\label{eq:gx}\\
  	&\widetilde{\textrm{QEC}}_f^{Z\,\dagger}G^{(f)}_{jk}\widetilde{\textrm{QEC}}_f^{Z} \nonumber\\
	&=\frac{1}{2^{l_f}} \prod_{\delta}\left[ X_{\delta-\epsilon}+X_{\delta}+X_{\delta+\epsilon}- X_{\delta-\epsilon}X_{\delta}X_{\delta+\epsilon}\right],\label{eq:gz}
\end{align}
for $f$ odd and $f$ even, respectively, where $\alpha,\beta \in \left\{0, 2\cdot 3^{f-1}, 4\cdot 3^{f-1} \right\}$, $\gamma = 4\cdot 3^{f-1}$, $\delta \in \left\{j,j+2\cdot3^f,...,k\right\}$ and $\epsilon = 7\cdot3^{f-1}$. These relations describe the backpropagation of Pauli strings
\begin{align}
	G^{(f)}_{jk} =&X_j X_{j+2\cdot3^{f}}...X_k.\label{eq:string}
\end{align}
The length of the Pauli strings $G^{(f)}_{jk}$ is defined as $l_f = (k - j)/(2\cdot3^f) + 1$.

The backpropagation of the measured observable is summarized in Tab.~\ref{tab:msop}. The Pauli operator $X_{\frac{N+1}{2}} $ measured at the end of the QCNN circuit corresponds to the Pauli string $G^{(d)}_{\frac{N+1}{2}\frac{N+1}{2}} = X_{\frac{N+1}{2}} $ with the minimal length $l_d=1$. The recursion relation \eqref{eq:gx} dictates that backpropagating this operator through the $X$-error correcting layer $\widetilde{\textrm{QEC}}_d^X$ gives rise to a sum of 16 terms including nine Pauli strings $G^{(d-1)}_{jk}$, six products of two Pauli strings $G^{(d-1)}_{jk}$ and a single product of three Pauli strings $G^{(d-1)}_{jk}$ at layer $f=d-1$, see Tab.~\ref{tab:msop}. 

Next, we backpropagate these Pauli strings and the products of Pauli strings through the layer $\widetilde{\textrm{QEC}}^Z_{d-1}$ performing the $Z$-error correcting procedure. Due to the linearity of the unitary $\widetilde{\textrm{QEC}}^Z_{d-1}$, we can separately backpropagate each product in the sum. Each Pauli string $G^{(d-1)}_{jk}$ gives rise to $4^{l_{d-1}}$ products of Pauli $X_i$ operators at layer $f=d-2$, see Eq.~ \eqref{eq:gz}. These products can be expressed in terms of Pauli strings $G^{(d-2)}_{jk}$ by using Eq.~\eqref{eq:string}. We thus again obtain a sum of products of Pauli strings $G_{jk}^{(d-2)}$ at layer $f=d-2$, see Tab.~\ref{tab:msop}.

We continue backpropagating these products of Pauli strings towards the input state at layer $f=0$. Backpropagating the Pauli string $G^{(f)}_{jk}$ through the $X$-error correcting layer $\widetilde{\textrm{QEC}}^X_f$ gives rise to $16$ products of Pauli strings $G^{(f-1)}_{jk}$, see Eq.~\eqref{eq:gx}. In every $X$-error correcting layer as well as in every $Z$-error correcting layer, we again obtain a sum of products of Pauli strings $G_{jk}^{(f)}$. At layer $f=0$, Pauli strings $G^{(0)}_{jk}$ are mapped by the disentangling unitary $U_N^{\dagger}$ onto SOPs,
\begin{equation}\label{eq:ugu}
U_N G^{(0)}_{jk}U^{\dagger}_N = S_{(j-1)(k+1)},
\end{equation}
see Tab.~\ref{tab:msop}.
As a result, we measure on the input state a sum of products of SOPs, i.e., the multiscale SOP $S_{\text{M}}$ of Eq. \eqref{eq:msop}.

\textbf{Length of string order parameters.} The backpropagation of all Pauli strings and their products is intractable due to their rapidly increasing number with the depth of the QCNN, see Tab.~\ref{tab:msop}. However, we now show that the multiscale SOP involves a SOP whose length increases exponentially with the depth $d$ of the QCNN. 

To this end, we focus on the product
\begin{equation}\label{eq:gg}
H_{jk}^{(f)}=\mathcal{L}_j^{(f)} G^{(f)}_{jk}G^{(f)}_{\left(j+3^{f}\right)\left(k- 3^{f}\right)} \mathcal{R}_k^{(f)}
\end{equation}
of Pauli strings $G^{(f)}_{jk}$, $G^{(f)}_{\left(j- 3^{f}\right)\left(k+ 3^{f}\right)}$, $\mathcal{L}^{(f)}_j$ and $\mathcal{R}^{(f)}_k $. The Pauli strings $\mathcal{L}^{(d-2)}_j = \mathcal{R}^{(d-2)}_k = \mathbb{1}$ reduce to the identity operator at layer $f=d-2$ and they are defined recursively for $f<d-2$ by relations 
\begin{align}
\mathcal{L}^{(f)}_j = \mathcal{L}^{(f+1)}_{j+2\cdot3^f},\label{eq:lo}\\
\mathcal{R}^{(f)}_k = \mathcal{R}^{(f+1)}_{k-2\cdot3^f},\label{eq:ro}
\end{align}
for $f$ being even and 
\begin{align}
\mathcal{L}^{(f)}_j = \mathcal{L}^{(f+1)}_{j-5\cdot3^f}X_{j-6\cdot3^f}X_{j-3\cdot3^f},\label{eq:le}\\
\mathcal{R}^{(f)}_k = X_{k+3\cdot3^f}X_{k+6\cdot3^f}\mathcal{R}^{(f+1)}_{k+5\cdot3^f},\label{eq:re}
\end{align}
for $f$ being odd. 

We show  in Appendix~\ref{app:hstring} that the product $H^{(f)}_{jk}$ appears at every layer $f\leq d-2$.
The backpropagation of the products $H^{(f)}_{jk}$ is summarized in Tab.~\ref{tab:prod}. The first product $H^{(d-2)}_{(\frac{N+1}{2}-3^{d-2})(\frac{N+1}{2}+3^{d-2})}$ appears at layer $f=d-2$, see Tab.~\ref{tab:msop}. The product $H^{(f)}_{jk}$ recursively appears at every layer $f<d-2$. After the disentangling unitary $U_N^{\dagger}$ at layer $f=0$, the product $H^{(0)}_{jk}$ gives rise to a product of SOPs, see Tab.~\ref{tab:prod}. 

\begin{table}[t]
\begin{tabular}{c|c|c|c}
$f$	&  Product of Pauli strings	& $l_f$ & \\
\hline
$d-2$ 	& $H^{(d-2)}_{(i -3^{d-2})(i + 3^{d-2})}$ & 1 & $\widetilde{\textrm{QEC}}_{d-2}^{X}$ 	\\
	 	& 		\vdots		&  & \\
$f$ even 	& $H^{(f)}_{(i -l_{f}\cdot3^{f})(i +l_{f}\cdot3^{f})}$ & $\frac{3^{d-f}+13}{8}$ & $\widetilde{\textrm{QEC}}_{f}^{Z}$ 	\\
$f$ odd	& $H^{(f)}_{(i -l_{f}\cdot3^{f})(i +l_{f}\cdot3^{f})}$ & $\frac{3^{d-f}-1}{8}$ & $\widetilde{\textrm{QEC}}_{f}^{X}$ 	\\
	 	& 		\vdots		&  & \\
$0$ 	& $H^{(0)}_{(i -l_{0})(i +l_{0})}$&$\frac{3^{d}+13}{8}$ & $U^{\dagger}_N$ 	\\
\hline
input		& $\tilde{\mathcal{L}}S_{(i-l_0-1)(i+l_0+1)}S_{(i-l_0)(i+l_0)} \tilde{\mathcal{R}}$ &$L=\frac{3^{d}+17}{4}$ & 	---			\\	
\end{tabular}
\caption{Backpropagation of products $H^{(f)}_{jk}$ of Pauli strings through the QCNN circuit with the depth $d$. We focus on a single product at each layer $f=1,2,...,d-2$. The table displays the length $l_f$ of the Pauli string $G^{(f)}_{(j+3^{f})(k-3^f)}$ in the product $H^{(f)}_{jk}$ and the unitary performed at layer $f$. The corresponding product of SOPs measured on the input state and the length $L$ of the SOP $S_{(i-l_0)(i+l_0)}$ are displayed in the last row. Operators $\tilde{\mathcal{L}}$ and $\tilde{\mathcal{R}}$ are defined in Appendix~\ref{app:hstring} and $i=\frac{N+1}{2}$.}\label{tab:prod}
\end{table}

The Pauli sting $G^{(0)}_{(\frac{N+1}{2} -l_{0}+1)(\frac{N+1}{2} +l_{0}-1)}$ in the product $H^{(0)}_{(\frac{N+1}{2} -l_{0})(\frac{N+1}{2} +l_{0})}$ at layer $f=0$ attains the length $l_0 = \frac{3^{d}+13}{8}$, see Tab.~\ref{tab:prod}. The Pauli string $G^{(0)}_{(\frac{N+1}{2} -l_{0}+1)(\frac{N+1}{2} +l_{0}-1)}$ is mapped by the disentangling unitary $U_N^{\dagger}$ onto the SOP $S_{(\frac{N+1}{2} -l_{0})(\frac{N+1}{2} +l_{0})}$ with the length 
\begin{equation}
L = 2l_0 + 1 = \frac{3^{d}+17}{4}\sim3^d.
\end{equation}
This shows that the multiscale SOP \eqref{eq:msop} involves a SOP whose length increases exponentially for large depths $d$ of the QCNN.
For the depth $d = \log_3 N$, this SOP exhibits the length $L\approx N/4$ comparable to system size $N$. By extending the analysis presented here, it can be shown that the multiscale SOP $S_{\rm M}$ involves also other SOPs with exponentially increasing lengths $L\sim3^d$ as well as SOPs at all length scales between $L=1$ and $L\sim3^d$.

\begin{table}[t]
\begin{tabular}{c|c|c|c}
$f$	&  Products of Pauli strings	& $l_f$ & \\
\hline
$2$ 	& $H^{(2)}_{(i -9l_{2})(i +9l_{2})}$ & $\frac{3^{d-2}+13}{8}$ & $\widetilde{\textrm{QEC}}_{2}^{Z}$ 	\\
$1$	& $H^{(2)}_{(i -9l_{2})(i +9l_{2})}$ & --- & $\widetilde{\textrm{QEC}}_{1}^{X}U^{\dagger}_N$\\
\hline
input		& $\bar{\mathcal{L}}\left(\prod_{\zeta}A_{\zeta}C_{\zeta}B_{\zeta}\right)\bar{\mathcal{R}}$ & --- & 	---			\\	
\end{tabular}
\caption{Backpropagation of products of Pauli strings through the QCNN circuit with the depth $d$. We focus on a single product of Pauli strings $H^{(2)}_{jk}$ at layers $f=1,2$ and display all products of SOPs measured on the input state that emerge by backpropagating the product $H^{(2)}_{jk}$ through $\widetilde{\textrm{QEC}}_{1}^{X}U^{\dagger}_N$. The table displays the length $l_f$ of the Pauli string $G^{(f)}_{(j+3^{f})(k-3^f)}$ in the product $H^{(f)}_{jk}$ and the unitary performed at layer $f$. $C_{\zeta}=S_{(\zeta-1)(\zeta+1)}$ are stabilizer elements as defined in the main text, $A_{\zeta}$ and $B_{\zeta}$ are defined in Eqs.~\eqref{eq:a} and \eqref{eq:b}, respectively, operators  $\bar{\mathcal{L}}$ and $\bar{\mathcal{R}}$ are defined in the Appendix~\ref{app:hstring}, $\zeta\in\{\frac{N+1}{2} -9l_{2},\frac{N+1}{2} -9(l_{2}-1),...,\frac{N+1}{2} +9l_{2}\}$, and $i=\frac{N+1}{2}$.}\label{tab:number}
\end{table}

\textbf{Number of products of string order parameters.} Finally, we determine a lower bound for the number of products of SOPs in the multiscale SOP $S_{\text{M}}$. To this end, we focus on products of Pauli strings displayed in Tab.~\ref{tab:number}. 

We start with the product $H^{(2)}_{(\frac{N+1}{2} -9l_{2})(\frac{N+1}{2} +9l_{2})}$ which appears at layer $f=2$, see Tab.~\ref{tab:prod}. The recursion relation \eqref{eq:gz} dictates that this product appears at layer $f=1$ as well. In contrast to the discussion above, we now focus on the product $H^{(2)}_{(\frac{N+1}{2} -9l_{2})(\frac{N+1}{2} +9l_{2})}$ at layer $f=1$. The Pauli strings $G^{(2)}_{[\frac{N+1}{2} -9(l_{2}-1)][\frac{N+1}{2} +9(l_{2}-1)]}$ and $G^{(2)}_{(\frac{N+1}{2} -9l_{2})(\frac{N+1}{2} +9l_{2})}$ in this product have lengths $l_2=\frac{3^{d-2}+13}{8}$ and $l_2+1$. We backpropagate this product through the $X$-error correcting layer $\widetilde{\textrm{QEC}}_1^X$ and the disentangling unitary $U_{N}^{\dagger}$
\begin{align}\label{eq:gg0}
&U_N \widetilde{\textrm{QEC}}_1^{X\,\dagger} H^{(2)}_{(i -9l_{2})(i +9l_{2})}
\widetilde{\textrm{QEC}}_1^{X} U_N^{\dagger}\nonumber\\
&= \bar{\mathcal{L}}\left(\prod_{\zeta}A_{\zeta}C_{\zeta}B_{\zeta}\right)\bar{\mathcal{R}},
\end{align}
where $\zeta\in\{\frac{N+1}{2} -9l_{2},\frac{N+1}{2} -9(l_{2}-1),...,\frac{N+1}{2} +9l_{2}\}$, $C_{\zeta} = S_{(\zeta-1)(\zeta+1)}$ are stabilizer elements as defined in the main text, and 
\begin{align}
A_{\zeta} =& \frac{1}{2}\left(C_{\zeta-4}C_{\zeta-2} - C_{\zeta-4} +C_{\zeta-2} + \mathbb{1}\right),\label{eq:a}\\
B_{\zeta} =& \frac{1}{2}\left(\mathbb{1} +  C_{\zeta+2} - C_{\zeta+4} + C_{\zeta+2}C_{\zeta+4}\right).\label{eq:b}
\end{align}
In Eq.~\eqref{eq:gg0}, we used the recursion relation \eqref{eq:gx} as well as Eq.~\eqref{eq:ugu}, see Appendix~\ref{app:hstring} for details. The product \eqref{eq:gg0} involves $2l_2+1=\frac{3^{d-2}+17}{4}>\frac{3^{d-2}}{4}$ terms $A_{\zeta}$ and $B_{\zeta}$. By distributing the parentheses in all terms $A_{\zeta}$ and $B_{\zeta}$, we obtain a sum of $16^{2l_2+1}>2^{3^{d-2}}$ products of SOPs $S_{jk}$. Note that these products of SOPs emerge from only the single product $H^{(2)}_{(\frac{N+1}{2} -9l_{2})(\frac{N+1}{2} +9l_{2})}$ at layer $f=1$. This places the lower bound $2^{3^{d-2}}$ on the total number of products of SOPs in the multiscale SOP $S_{\text{M}}$, which involves also many other products of SOPs.

In summary, we showed in this appendix that the QCNN with alternating $X$-error and $Z$-error correcting layers detecting the `$ZXZ$' phase measures the multiscale SOP $S_{\text{M}}$, see Eq. \eqref{eq:msop}. This multiscale SOP is a sum of products of SOPs $S_{jk}$ whose length $L\sim 3^{d}$ increases exponentially with the depth $d$ of the QCNN. The lower bound for the number of products of SOPs in the sum is $2^{3^{d-2}}$.

\section{Sample complexity of the multiscale string order parameter}\label{app:comp}
In this appendix, we discuss the sample complexity of directly sampling the multiscale SOP of Eq.  \eqref{eq:msop} from the input state via local Pauli measurements without using any quantum circuit. 

A local Pauli measurement consists of simultaneously reading out all qubits $j$ in the basis of Pauli operators $\sigma_j=X_j,Y_j,Z_j$. We thus measure in the basis of the tensor product $\mathcal{B} = \bigotimes_{j=1}^N \sigma_j$ of the Pauli operators $\sigma_j$. A product of SOPs can be sampled via the local Pauli measurement in any Basis $\mathcal{B}$ in which it is diagonal. Several products of SOPs that are diagonal in the same tensor product basis can be simultaneously sampled in this basis. We can express the multiscale SOP 
\begin{equation}\label{eq:om}
S_{\rm M} = \sum_{m=1}^b O_m
\end{equation}
in terms of operators $O_m$ which are sums of products of SOPs that are diagonal in the tensor product basis $\mathcal{B}_m$. To determine the expectation value of the multiscale SOP, we can individually measure each operator $O_m$ using local Pauli measurements. While the sample complexity of this measurement depends on the variance $\langle O_m^2 \rangle - \langle O_m \rangle^2$ of the operators $O_m$, the number $b$ of different bases $\mathcal{B}_m$ in which we need to measure places a lower bound on the sample complexity. Note that the decomposition \eqref{eq:om} of the multiscale SOP is not unique as a product of SOPs can be diagonal in several bases $\mathcal{B}_m$. To determine the lower bound for the sample complexity of the multiscale SOP, we now investigate the minimal number $B$ of bases in which we need to measure.

To this end, we focus on the products \eqref{eq:gg0} of SOPs. We rewrite the products \eqref{eq:gg0} as 
\begin{align}\label{eq:ck}
&\bar{\mathcal{L}}\left(\prod_{\zeta}A_{\zeta}C_{\zeta}B_{\zeta}\right)\bar{\mathcal{R}} \nonumber\\
&=\bar{\mathcal{L}}A_{\frac{N+1}{2} -9l_{2}}\left(\prod_{\zeta}C_{\zeta}\right)\left(\prod_{\zeta'}K_{\zeta'}\right) B_{\frac{N+1}{2} +9l_{2}} \bar{\mathcal{R}},
\end{align}
where $\zeta'\in\{\frac{N+1}{2} -9l_{2}+4,\frac{N+1}{2} -9l_{2} + 13,...,\frac{N+1}{2} +9l_{2}-5\}$ and 
\begin{align}
K_{\zeta'} = & \frac{1}{4}\left[\left(C_{\zeta'-2} + \mathbb{1}\right)\left(C_{\zeta'+3} + \mathbb{1}\right)\right.\label{eq:ii}\\
& +\left(C_{\zeta'-2} - \mathbb{1}\right)C_{\zeta'}\left(C_{\zeta'+3} + \mathbb{1}\right)\label{eq:xz}\\
& +\left(C_{\zeta'-2} + \mathbb{1}\right)C_{\zeta'+1}\left(C_{\zeta'+3} - \mathbb{1}\right)\label{eq:zx}\\
& +\left.\left(C_{\zeta'-2} - \mathbb{1}\right)C_{\zeta'}C_{\zeta'+1}\left(C_{\zeta'+3} - \mathbb{1}\right)\right].\label{eq:yy}
\end{align}
Crucially, each operator $K_{\zeta'}$ involves terms that need to be measured in three different tensor product bases. Recalling that $C_{\zeta'} = Z_{\zeta'-1}X_{\zeta'}Z_{\zeta'+1}$, the terms in the line \eqref{eq:xz} are diagonal in the $X$ basis on qubit $\zeta'$ and in the computational basis on qubit $\zeta'+1$. The terms in the line \eqref{eq:zx} are diagonal in the computational basis on qubit $\zeta'$ and in the $X$ basis on qubit $\zeta'+1$. The terms in the line \eqref{eq:yy} are diagonal in the $Y$ basis on qubits $\zeta'$ and $\zeta'+1$. As a result, we need to measure in three different bases $X_{\zeta'}Z_{\zeta'+1}$, $Z_{\zeta'}X_{\zeta'+1}$ and $Y_{\zeta'}Y_{\zeta'+1}$ on qubits $\zeta'$ and $\zeta'+1$. The terms in the line \eqref{eq:ii} are diagonal in any of these three bases as they act as the identity operator on qubits $\zeta'$ and $\zeta'+1$.
  
The product of SOPs \eqref{eq:ck} involves $2\,l_2 = \frac{3^{d-2}+13}{4}>3^{d-4}$ operators $K_{\zeta'}$. Distributing the parentheses in all operators $K_{\zeta'}$ on the right-hand side of Eq.~\eqref{eq:ck} gives rise to products of SOPs with mutually incompatible bases $\bigotimes_{\zeta'}\{X_{\zeta'}Z_{\zeta'+1},Z_{\zeta'}X_{\zeta'+1},Y_{\zeta'}Y_{\zeta'+1} \}$ on qubits $\zeta'$ and $\zeta'+1$. As a result, we need to measure in $3^{2l_2} > 3^{3^{d-4}}$ different tensor product bases $\mathcal{B}_m$. This places the lower bound $3^{3^{d-4}}$ for the sample complexity of measuring the multiscale SOP via local Pauli measurements.

\section{Backpropagation of products (\ref{eq:gg}) of Pauli strings through the QCNN circuit}\label{app:hstring}
In this appendix, we discuss the backpropagation of products $H^{(f)}_{jk}$ of Pauli strings defined in Eq.~(\ref{eq:gg}) through the QCNN circuit. We show that the product $H^{(f)}_{jk}$ recursively appears at every layer $f< d-2$ of the QCNN. We also derive Eq.~\eqref{eq:gg0} describing the backpropagation of the product $H^{(2)}_{jk}$ through the first $X$-error correcting layer $\widetilde{\textrm{QEC}}^X_1$ and the disentangling unitary $U_N^{\dagger}$.

Following the recursion relations \eqref{eq:lo}, \eqref{eq:ro}, \eqref{eq:le} and \eqref{eq:re}, the Pauli strings $\mathcal{L}^{(f)}_j$ and $\mathcal{R}^{(f)}_k $ in the product $H^{(f)}_{jk}$ for $f$ being odd can be explicitly expressed as
\begin{align}
\mathcal{L}^{(f)}_j =& \prod_{g=0}^{(d-4-f)/2}X_{j-\kappa_g-3\cdot3^{f+2g}}X_{j-\kappa_g},\\
\mathcal{R}^{(f)}_k =& \prod_{g=0}^{(d-4-f)/2}X_{k+\kappa_g}X_{k+\kappa_g+3\cdot3^{f+2g}},
\end{align}
where
\begin{equation}
\kappa_g = \frac{23\cdot3^{2g} +1}{8}3^f.
\end{equation}
For $f$ being even, the Pauli strings $\mathcal{L}^{(f)}_j$ and $\mathcal{R}^{(f)}_k $ can be explicitly expressed as
\begin{align}
\mathcal{L}^{(f)}_j =& \prod_{g=0}^{(d-5-f)/2}X_{j-\lambda_g-9\cdot3^{f+2g}}X_{j-\lambda_g},\\
\mathcal{R}^{(f)}_k =& \prod_{g=0}^{(d-5-f)/2}X_{k+\lambda_g}X_{k+\lambda_g+9\cdot3^{f+2g}},
\end{align}
where
\begin{equation}
\lambda_g = \frac{69\cdot3^{2g} -13}{8}3^f.
\end{equation}
The Pauli strings $\mathcal{L}^{(f)}_j$ and $\mathcal{R}^{(f)}_k$ involve $2\lfloor(d-f-2)/2\rfloor$ Pauli $X_l$ operators at layer $f$. 

\textbf{Backpropagation of products (\ref{eq:gg}) of Pauli strings.} We now show that the product $H^{(f)}_{jk}$ appears at every layer $f< d-2$, see Tab.~\ref{tab:prod} in Appendix~\ref{app:msop}. We start with $H^{(d-2)}_{(\frac{N+1}{2}-3^{d-2})(\frac{N+1}{2}+3^{d-2})}$ at layer $f=d-2$. The product $H^{(f)}_{(\frac{N+1}{2}-l_{f}\cdot3^{f})(\frac{N+1}{2}+l_{f}\cdot3^{f})}$ backpropagates through the $X$-error correcting layer $\widetilde{\textrm{QEC}}_f^X$ according to the recursion relation
\begin{align}\label{eq:ggx}
	&\widetilde{\textrm{QEC}}_{f}^{X\,\dagger}H^{(f)}_{(\frac{N+1}{2}-l_{f}\cdot3^{f})(\frac{N+1}{2}+l_{f}\cdot3^{f})}\widetilde{\textrm{QEC}}_{f}^{X}\nonumber\\
	&\rightarrow H^{(f-1)}_{(\frac{N+1}{2}-l_{f-1}\cdot3^{f-1})(\frac{N+1}{2}+l_{f-1}\cdot3^{f-1})},
\end{align}
where $l_{f-1}=3l_f+2$.
Due to the distance $|l-m|\geq3\cdot3^f$ between every pair of Pauli operators $X_l$ and $X_m$ in the strings $\mathcal{L}^{(f)}_{\frac{N+1}{2}-l_f\cdot3^f}$ and $\mathcal{R}^{(f)}_{\frac{N+1}{2}+l_f\cdot3^f} $, each Pauli $X_l$ operator separately backpropagates through the odd $X$-error correcting layer $\widetilde{\textrm{QEC}}_f^X$ according to Eq.~\eqref{eq:gx}. The Pauli strings $G^{(f)}_{(\frac{N+1}{2}-l_{f}\cdot3^{f})(\frac{N+1}{2}+l_{f}\cdot3^{f})}$ and $G^{(f)}_{[\frac{N+1}{2}-(l_{f}-1)\cdot3^{f}][\frac{N+1}{2}+(l_{f}-1)\cdot3^{f}]}$ in the product $H^{(f)}_{(\frac{N+1}{2}-l_{f}\cdot3^{f})(\frac{N+1}{2}+l_{f}\cdot3^{f})}$ also separately backpropagate through the $X$-error correcting layer $\widetilde{\textrm{QEC}}_f^X$ according to Eq.~\eqref{eq:gx}.
As a result, backpropagating the product $H^{(f)}_{(\frac{N+1}{2}-l_{f}\cdot3^{f})(\frac{N+1}{2}+l_{f}\cdot3^{f})}$ through the $X$-error correcting layer $\widetilde{\textrm{QEC}}_f^X$ gives rise to a sum of $16^{2(d-f-1)}$ terms including the product $H^{(f-1)}_{(\frac{N+1}{2}-l_{f-1}\cdot3^{f-1})(\frac{N+1}{2}+l_{f-1}\cdot3^{f-1})}$, where we used Eqs.~\eqref{eq:lo} and \eqref{eq:ro}. We focus on the single product $H^{(f-1)}_{(\frac{N+1}{2}-l_{f-1}\cdot3^{f-1})(\frac{N+1}{2}+l_{f-1}\cdot3^{f-1})}$ as indicated by the recursion relation \eqref{eq:ggx} and displayed in Tab.~\ref{tab:prod}.

The product $H^{(f)}_{(\frac{N+1}{2}-l_{f}\cdot3^{f})(\frac{N+1}{2}+l_{f}\cdot3^{f})}$ backpropagates through the $Z$-error correcting layer $\widetilde{\textrm{QEC}}_f^Z$ according to the recursion relation
\begin{align}\label{eq:ggz}
	&\widetilde{\textrm{QEC}}_{f}^{Z\,\dagger}H^{(f)}_{(\frac{N+1}{2}-l_{f}\cdot3^{f})(\frac{N+1}{2}+l_{f}\cdot3^{f})}\widetilde{\textrm{QEC}}_{f}^{Z}\nonumber\\&\rightarrow H^{(f-1)}_{(\frac{N+1}{2}-l_{f-1}\cdot3^{f-1})(\frac{N+1}{2}+l_{f-1}\cdot3^{f-1})},
\end{align}
where $l_{f-1}=3l_f-5$.
Each Pauli string in the product $H^{(f)}_{(\frac{N+1}{2}-l_{f}\cdot3^{f})(\frac{N+1}{2}+l_{f}\cdot3^{f})}$ backpropagates separately through the $Z$-error correcting layer $\widetilde{\textrm{QEC}}^Z_f$ according to Eq.~\eqref{eq:gz} giving rise to a sum of $4^{2(l_f +d - f) -5}$ terms including the product $H^{(f-1)}_{\frac{N+1}{2}-l_{f-1}\cdot3^{f-1})(\frac{N+1}{2}+l_{f-1}\cdot3^{f-1})}$ as indicated by the recursion relation \eqref{eq:ggz} and stated in Tab.~\ref{tab:prod}. To show this, we distribute the parentheses in Eq.~\eqref{eq:gz} and exploit Eqs.~\eqref{eq:le} and \eqref{eq:re} as well as that
\begin{align}
G^{(f-1)}_{jk}=&X_{j+2\cdot3^{f-1}}X_{j+4\cdot3^{f-1}}X_{j+8\cdot3^{f-1}}\left(\prod_{\delta'}X_{\delta'-\epsilon} X_{\delta'+\epsilon}\right)\nonumber\\
&\times\left(\prod_{\delta''}X_{\delta''}\right)X_{k-8\cdot3^{f-1}} X_{k-4\cdot3^{f-1}}X_{k-2\cdot3^{f-1}},
\end{align}
and
\begin{align}
G^{(f-1)}_{\left(j+ 3^{f-1}\right)\left(k- 3^{f-1}\right)} &= X_{j+3^{f-1}}X_{j+5\cdot3^{f-1}}X_{j+11\cdot3^{f-1}}\nonumber\\
&\times\left(\prod_{\delta'}X_{\delta'}\right)\left(\prod_{\delta''}X_{\delta''-\epsilon} X_{\delta''+\epsilon}\right)\nonumber\\
&\times X_{k-11\cdot3^{f-1}} X_{k-5\cdot3^{f-1}}X_{k-3^{f-1}},
\end{align}
where $\delta' \in \left\{j+7\cdot3^{f-1},j+13\cdot3^{f-1},...,k-7\cdot3^{f-1}\right\}$, $\delta'' \in \left\{j+10\cdot3^{f-1},j+16\cdot3^{f-1}...,k-10\cdot3^{f-1}\right\}$ and  $\epsilon = 7\cdot3^{f-1}$.

To summarize the backpropagation of the products $H^{(f)}_{(\frac{N+1}{2}-l_{f}\cdot3^{f})(\frac{N+1}{2}+l_{f}\cdot3^{f})}$ in the QCNN circuit, we start with the first product $H^{(d-2)}_{(\frac{N+1}{2} -3^{d-2})(\frac{N+1}{2} + 3^{d-2})}$ at layer $f = d-2$, c.f. Tab.~\ref{tab:prod}. Using the recursion relations \eqref{eq:ggx} and \eqref{eq:ggz} for odd layers $\widetilde{\textrm{QEC}}_f^X$ and even layers $\widetilde{\textrm{QEC}}_f^Z$, respectively, we obtain the product $H^{(f)}_{(\frac{N+1}{2}-l_{f}\cdot3^{f})(\frac{N+1}{2}+l_{f}\cdot3^{f})}$ at every layer $f<d-2$. Note that the product $H^{(f)}_{(\frac{N+1}{2}-l_{f}\cdot3^{f})(\frac{N+1}{2}+l_{f}\cdot3^{f})}$ at layer $f$ emerges only by backpropagating the product $H^{(f+1)}_{(\frac{N+1}{2}-l_{f+1}\cdot3^{f+1})(\frac{N+1}{2}+l_{f+1}\cdot3^{f+1})}$ from layer $f+1$. The recursion relations \eqref{eq:gx} and \eqref{eq:gz} dictate that all other products of Pauli strings at layer $f+1$ give rise to products at layer $f$ different to $H^{(f)}_{(\frac{N+1}{2}-l_{f}\cdot3^{f})(\frac{N+1}{2}+l_{f}\cdot3^{f})}$.

We now investigate the length $l_f$ of the Pauli string $G^{(f)}_{[\frac{N+1}{2}-(l_{f}-1)\cdot3^{f}][\frac{N+1}{2}+(l_{f}-1)\cdot3^{f}]}$ in the product $H^{(f)}_{(\frac{N+1}{2}-l_{f}\cdot3^{f})(\frac{N+1}{2}+l_{f}\cdot3^{f})}$ at every layer $f$. We start with $l_{d-2}=1$ at layer $f=d-2$, see Tab.~\ref{tab:prod}. By backpropagating through the $X$-error correcting layer, the length $l_{f-1} = 3l_f+2$ of the Pauli string $G^{(f-1)}_{[\frac{N+1}{2}-(l_{f-1}-1)\cdot3^{f-1}][\frac{N+1}{2}+(l_{f-1}-1)\cdot3^{f-1}]}$ at layer $f-1$ increases by a factor of three compared to the length $l_f$ of the Pauli string $G^{(f)}_{[\frac{N+1}{2}-(l_{f}-1)\cdot3^{f}][\frac{N+1}{2}+(l_{f}-1)\cdot3^{f}]}$ at layer $f$, see Eq.~\eqref{eq:ggx}. By backpropagating through the $Z$-error correcting layer, the length $l_{f-1} = 3l_f-5$ also increases by a factor of three, see Eq.~\eqref{eq:ggz}. After two successive layers --- the first layer correcting $X$ errors and the second layer correcting $Z$ errors --- the length of the Pauli string  increases by a factor of nine from $l_{f}$ to $l_{f-2} = 9l_f +1$. At every odd layer $f$, the length can be expressed as a sum 
\begin{equation}
l_f = \sum_{g=0}^{(d-f-2)/2}9^g=\frac{3^{d-f}-1}{8}.
\end{equation}
At every even layer $f$, the length can be expressed as $l_f =\frac{3^{d-f}+13}{8}$, c.f. Tab.~\ref{tab:prod}.

The product $H^{(0)}_{(\frac{N+1}{2} -l_{0})(\frac{N+1}{2} +l_{0})}$ at layer $f=0$ is mapped by the disentangling unitary $U_N^{\dagger}$ onto the product of SOPs
\begin{align}
  &U_N H^{(0)}_{(\frac{N+1}{2} -l_{0})(\frac{N+1}{2} +l_{0})}U_N^{\dagger}\nonumber\\
  &= \tilde{\mathcal{L}}S_{(i-l_0-1)(i+l_0+1)}S_{(i-l_0)(i+l_0)} \tilde{\mathcal{R}},
\end{align}
see Tab.~\ref{tab:prod} where
 \begin{align}
\tilde{\mathcal{L}}&=U_N \mathcal{L}^{(0)}_{\frac{N+1}{2} -l_{0}}U_N^{\dagger}\nonumber\\
&= \prod_{g=0}^{(d-5)/2}C_{\frac{N+1}{2} -l_{0}-\lambda_g-9\cdot3^{2g}}C_{\frac{N+1}{2} -l_{0}-\lambda_g},\\
\tilde{\mathcal{R}}&=U_N \mathcal{R}^{(0)}_{\frac{N+1}{2} +l_{0}}U_N^{\dagger} \nonumber\\
&= \prod_{g=0}^{(d-5)/2}C_{\frac{N+1}{2} +l_{0}+\lambda_g}C_{\frac{N+1}{2} +l_{0}+\lambda_g+9\cdot3^{2g}}.
\end{align}

\textbf{Derivation of Eq.~\eqref{eq:gg0}.} In Eq.~\eqref{eq:gg0}, we backpropagate the product $H^{(2)}_{(\frac{N+1}{2} -9l_{2})(\frac{N+1}{2} +9l_{2})}$  that appears at layer $f=1$ through the first $X$-error correcting layer $\widetilde{\textrm{QEC}}_1^X$ and the disentangling unitary $U^{\dagger}_N$ using the recursion relation \eqref{eq:gx} and Eq.~\eqref{eq:ugu}, respectively. Note that every Pauli operator $X_{\zeta}$ in the product  $H^{(2)}_{(\frac{N+1}{2} -9l_{2})(\frac{N+1}{2} +9l_{2})}$ separately backpropagates through the $X$-error correcting layer $\widetilde{\textrm{QEC}}_1^X$ due to the distance at least 9 between these Pauli operators. The Pauli strings $G^{(2)}_{[\frac{N+1}{2} -9(l_{2}-1)][\frac{N+1}{2} +9(l_{2}-1)]}$ and $G^{(2)}_{(\frac{N+1}{2} -9l_{2})(\frac{N+1}{2} +9l_{2})}$ in the product $H^{(2)}_{jk}$ involve $l_2=\frac{3^{d-2}+13}{8}$ and $l_2+1$, respectively, Pauli $X_\zeta$ operators. Backpropagating these Pauli strings through the layer $\widetilde{\textrm{QEC}}_1^X$ and the disentangling unitary $U_N^{\dagger}$ gives rise to the product
\begin{align}\label{eq:acb}
&U_N \widetilde{\textrm{QEC}}_1^{X\,\dagger} G^{(2)}_{(\frac{N+1}{2} -9l_{2})(\frac{N+1}{2} +9l_{2})}\nonumber\\
&\times G^{(2)}_{[\frac{N+1}{2} -9(l_{2}-1)][\frac{N+1}{2} +9(l_{2}-1)]}\widetilde{\textrm{QEC}}_1^{X} U_N^{\dagger} = \prod_{\zeta}E_{\zeta}
\end{align}
of $2l_2+1=\frac{3^{d-2}+17}{4}$ terms $E_{\zeta}=A_{\zeta}C_{\zeta}B_{\zeta}$ where $\zeta\in\{\frac{N+1}{2} -9l_{2},\frac{N+1}{2} -9(l_{2}-1),...,\frac{N+1}{2} +9l_{2}\}$ and $C_{\zeta} = S_{(\zeta-1)(\zeta+1)}$. Operators $A_{\zeta}$ and $B_{\zeta}$ are defined in Eqs.~\eqref{eq:a} and \eqref{eq:b}. Backpropagating the Pauli strings $\mathcal{L}^{(2)}_{\frac{N+1}{2} -9l_{2}}$ and $\mathcal{R}^{(2)}_{\frac{N+1}{2} +9l_{2}}$ through the layer $\widetilde{\textrm{QEC}}_1^X$ and the disentangling unitary $U_N^{\dagger}$ gives rise to the products
\begin{align}
\bar{\mathcal{L}} &=U_N \widetilde{\textrm{QEC}}_1^{X\,\dagger} \mathcal{L}^{(2)}_{\frac{N+1}{2} -9l_{2}}\widetilde{\textrm{QEC}}_1^{X} U_N^{\dagger} \nonumber\\
&= \prod_{g=0}^{(d-7)/2}E_{\frac{N+1}{2} -9l_{2}-\lambda_g-81\cdot3^{2g}}E_{\frac{N+1}{2} -9l_{2}-\lambda_g},\label{eq:lbar}\\
\bar{\mathcal{R}} &=U_N \widetilde{\textrm{QEC}}_1^{X\,\dagger} \mathcal{R}^{(2)}_{\frac{N+1}{2} +9l_{2}}\widetilde{\textrm{QEC}}_1^{X} U_N^{\dagger}\nonumber\\
&= \prod_{g=0}^{(d-7)/2}E_{\frac{N+1}{2} +9l_{2}+\lambda_g}E_{\frac{N+1}{2} +9l_{2}+\lambda_g+81\cdot3^{2g}}.\label{eq:rbar}
\end{align}
By combining Eqs.~\eqref{eq:acb}, \eqref{eq:lbar} and \eqref{eq:rbar} we obtain Eq.~\eqref{eq:gg0}.

\bibliography{QCNN_paper}

\begin{thebibliography}{58}%
\makeatletter
\providecommand \@ifxundefined [1]{%
 \@ifx{#1\undefined}
}%
\providecommand \@ifnum [1]{%
 \ifnum #1\expandafter \@firstoftwo
 \else \expandafter \@secondoftwo
 \fi
}%
\providecommand \@ifx [1]{%
 \ifx #1\expandafter \@firstoftwo
 \else \expandafter \@secondoftwo
 \fi
}%
\providecommand \natexlab [1]{#1}%
\providecommand \enquote  [1]{``#1''}%
\providecommand \bibnamefont  [1]{#1}%
\providecommand \bibfnamefont [1]{#1}%
\providecommand \citenamefont [1]{#1}%
\providecommand \href@noop [0]{\@secondoftwo}%
\providecommand \href [0]{\begingroup \@sanitize@url \@href}%
\providecommand \@href[1]{\@@startlink{#1}\@@href}%
\providecommand \@@href[1]{\endgroup#1\@@endlink}%
\providecommand \@sanitize@url [0]{\catcode `\\12\catcode `\$12\catcode
  `\&12\catcode `\#12\catcode `\^12\catcode `\_12\catcode `\%12\relax}%
\providecommand \@@startlink[1]{}%
\providecommand \@@endlink[0]{}%
\providecommand \url  [0]{\begingroup\@sanitize@url \@url }%
\providecommand \@url [1]{\endgroup\@href {#1}{\urlprefix }}%
\providecommand \urlprefix  [0]{URL }%
\providecommand \Eprint [0]{\href }%
\providecommand \doibase [0]{http://dx.doi.org/}%
\providecommand \selectlanguage [0]{\@gobble}%
\providecommand \bibinfo  [0]{\@secondoftwo}%
\providecommand \bibfield  [0]{\@secondoftwo}%
\providecommand \translation [1]{[#1]}%
\providecommand \BibitemOpen [0]{}%
\providecommand \bibitemStop [0]{}%
\providecommand \bibitemNoStop [0]{.\EOS\space}%
\providecommand \EOS [0]{\spacefactor3000\relax}%
\providecommand \BibitemShut  [1]{\csname bibitem#1\endcsname}%
\let\auto@bib@innerbib\@empty
\bibitem [{\citenamefont {Arute}\ and\ \citenamefont
  {others}(2019)\citenamefont {Arute} \emph {et~al.}}]{arute2019}%
  \BibitemOpen
  \bibfield  {author} {\bibinfo {author} {\bibfnamefont {Frank}\ \bibnamefont
  {Arute}} \emph {et~al.},\ }\bibfield  {title} {\enquote {\bibinfo {title}
  {Quantum supremacy using a programmable superconducting processor},}\ }\href
  {\doibase 10.1038/s41586-019-1666-5} {\bibfield  {journal} {\bibinfo
  {journal} {Nature}\ }\textbf {\bibinfo {volume} {574}},\ \bibinfo {pages}
  {505--510} (\bibinfo {year} {2019})}\BibitemShut {NoStop}%
\bibitem [{\citenamefont {Feynman}(1982)}]{feynman1982}%
  \BibitemOpen
  \bibfield  {author} {\bibinfo {author} {\bibfnamefont {Richard~P.}\
  \bibnamefont {Feynman}},\ }\bibfield  {title} {\enquote {\bibinfo {title}
  {Simulating physics with computers},}\ }\href {\doibase 10.1007/BF02650179}
  {\bibfield  {journal} {\bibinfo  {journal} {Int J Theor Phys}\ }\textbf
  {\bibinfo {volume} {21}},\ \bibinfo {pages} {467--488} (\bibinfo {year}
  {1982})}\BibitemShut {NoStop}%
\bibitem [{\citenamefont {Cao}\ \emph {et~al.}(2019)\citenamefont {Cao},
  \citenamefont {Romero}, \citenamefont {Olson}, \citenamefont {Degroote},
  \citenamefont {Johnson}, \citenamefont {Kieferov{\'a}}, \citenamefont
  {Kivlichan}, \citenamefont {Menke}, \citenamefont {Peropadre}, \citenamefont
  {Sawaya}, \citenamefont {Sim}, \citenamefont {Veis},\ and\ \citenamefont
  {{Aspuru-Guzik}}}]{cao2019}%
  \BibitemOpen
  \bibfield  {author} {\bibinfo {author} {\bibfnamefont {Yudong}\ \bibnamefont
  {Cao}}, \bibinfo {author} {\bibfnamefont {Jonathan}\ \bibnamefont {Romero}},
  \bibinfo {author} {\bibfnamefont {Jonathan~P.}\ \bibnamefont {Olson}},
  \bibinfo {author} {\bibfnamefont {Matthias}\ \bibnamefont {Degroote}},
  \bibinfo {author} {\bibfnamefont {Peter~D.}\ \bibnamefont {Johnson}},
  \bibinfo {author} {\bibfnamefont {M{\'a}ria}\ \bibnamefont {Kieferov{\'a}}},
  \bibinfo {author} {\bibfnamefont {Ian~D.}\ \bibnamefont {Kivlichan}},
  \bibinfo {author} {\bibfnamefont {Tim}\ \bibnamefont {Menke}}, \bibinfo
  {author} {\bibfnamefont {Borja}\ \bibnamefont {Peropadre}}, \bibinfo {author}
  {\bibfnamefont {Nicolas P.~D.}\ \bibnamefont {Sawaya}}, \bibinfo {author}
  {\bibfnamefont {Sukin}\ \bibnamefont {Sim}}, \bibinfo {author} {\bibfnamefont
  {Libor}\ \bibnamefont {Veis}}, \ and\ \bibinfo {author} {\bibfnamefont
  {Al{\'a}n}\ \bibnamefont {{Aspuru-Guzik}}},\ }\bibfield  {title} {\enquote
  {\bibinfo {title} {Quantum {{Chemistry}} in the {{Age}} of {{Quantum
  Computing}}},}\ }\href {\doibase 10.1021/acs.chemrev.8b00803} {\bibfield
  {journal} {\bibinfo  {journal} {Chem. Rev.}\ }\textbf {\bibinfo {volume}
  {119}},\ \bibinfo {pages} {10856--10915} (\bibinfo {year}
  {2019})}\BibitemShut {NoStop}%
\bibitem [{\citenamefont {Biamonte}\ \emph {et~al.}(2017)\citenamefont
  {Biamonte}, \citenamefont {Wittek}, \citenamefont {Pancotti}, \citenamefont
  {Rebentrost}, \citenamefont {Wiebe},\ and\ \citenamefont
  {Lloyd}}]{biamonte2017}%
  \BibitemOpen
  \bibfield  {author} {\bibinfo {author} {\bibfnamefont {Jacob}\ \bibnamefont
  {Biamonte}}, \bibinfo {author} {\bibfnamefont {Peter}\ \bibnamefont
  {Wittek}}, \bibinfo {author} {\bibfnamefont {Nicola}\ \bibnamefont
  {Pancotti}}, \bibinfo {author} {\bibfnamefont {Patrick}\ \bibnamefont
  {Rebentrost}}, \bibinfo {author} {\bibfnamefont {Nathan}\ \bibnamefont
  {Wiebe}}, \ and\ \bibinfo {author} {\bibfnamefont {Seth}\ \bibnamefont
  {Lloyd}},\ }\bibfield  {title} {\enquote {\bibinfo {title} {Quantum machine
  learning},}\ }\href {\doibase 10.1038/nature23474} {\bibfield  {journal}
  {\bibinfo  {journal} {Nature}\ }\textbf {\bibinfo {volume} {549}},\ \bibinfo
  {pages} {195--202} (\bibinfo {year} {2017})}\BibitemShut {NoStop}%
\bibitem [{\citenamefont {Huang}\ \emph {et~al.}(2020)\citenamefont {Huang},
  \citenamefont {Kueng},\ and\ \citenamefont {Preskill}}]{huang2020}%
  \BibitemOpen
  \bibfield  {author} {\bibinfo {author} {\bibfnamefont {Hsin-Yuan}\
  \bibnamefont {Huang}}, \bibinfo {author} {\bibfnamefont {Richard}\
  \bibnamefont {Kueng}}, \ and\ \bibinfo {author} {\bibfnamefont {John}\
  \bibnamefont {Preskill}},\ }\bibfield  {title} {\enquote {\bibinfo {title}
  {Predicting many properties of a quantum system from very few
  measurements},}\ }\href {\doibase 10.1038/s41567-020-0932-7} {\bibfield
  {journal} {\bibinfo  {journal} {Nat. Phys.}\ }\textbf {\bibinfo {volume}
  {16}},\ \bibinfo {pages} {1050--1057} (\bibinfo {year} {2020})}\BibitemShut
  {NoStop}%
\bibitem [{\citenamefont {Lloyd}\ \emph {et~al.}(2014)\citenamefont {Lloyd},
  \citenamefont {Mohseni},\ and\ \citenamefont {Rebentrost}}]{lloyd2014}%
  \BibitemOpen
  \bibfield  {author} {\bibinfo {author} {\bibfnamefont {Seth}\ \bibnamefont
  {Lloyd}}, \bibinfo {author} {\bibfnamefont {Masoud}\ \bibnamefont {Mohseni}},
  \ and\ \bibinfo {author} {\bibfnamefont {Patrick}\ \bibnamefont
  {Rebentrost}},\ }\bibfield  {title} {\enquote {\bibinfo {title} {Quantum
  principal component analysis},}\ }\href {\doibase 10.1038/nphys3029}
  {\bibfield  {journal} {\bibinfo  {journal} {Nat. Phys.}\ }\textbf {\bibinfo
  {volume} {10}},\ \bibinfo {pages} {631--633} (\bibinfo {year}
  {2014})}\BibitemShut {NoStop}%
\bibitem [{\citenamefont {Romero}\ \emph {et~al.}(2017)\citenamefont {Romero},
  \citenamefont {Olson},\ and\ \citenamefont {{Aspuru-Guzik}}}]{romero2017}%
  \BibitemOpen
  \bibfield  {author} {\bibinfo {author} {\bibfnamefont {Jonathan}\
  \bibnamefont {Romero}}, \bibinfo {author} {\bibfnamefont {Jonathan~P.}\
  \bibnamefont {Olson}}, \ and\ \bibinfo {author} {\bibfnamefont {Alan}\
  \bibnamefont {{Aspuru-Guzik}}},\ }\bibfield  {title} {\enquote {\bibinfo
  {title} {Quantum autoencoders for efficient compression of quantum data},}\
  }\href {\doibase 10.1088/2058-9565/aa8072} {\bibfield  {journal} {\bibinfo
  {journal} {Quantum Sci. Technol.}\ }\textbf {\bibinfo {volume} {2}},\
  \bibinfo {pages} {045001} (\bibinfo {year} {2017})}\BibitemShut {NoStop}%
\bibitem [{\citenamefont {Bondarenko}\ and\ \citenamefont
  {Feldmann}(2020)}]{bondarenko2020}%
  \BibitemOpen
  \bibfield  {author} {\bibinfo {author} {\bibfnamefont {Dmytro}\ \bibnamefont
  {Bondarenko}}\ and\ \bibinfo {author} {\bibfnamefont {Polina}\ \bibnamefont
  {Feldmann}},\ }\bibfield  {title} {\enquote {\bibinfo {title} {Quantum
  {{Autoencoders}} to {{Denoise Quantum Data}}},}\ }\href {\doibase
  10.1103/PhysRevLett.124.130502} {\bibfield  {journal} {\bibinfo  {journal}
  {Phys. Rev. Lett.}\ }\textbf {\bibinfo {volume} {124}},\ \bibinfo {pages}
  {130502} (\bibinfo {year} {2020})}\BibitemShut {NoStop}%
\bibitem [{\citenamefont {Zhang}\ \emph {et~al.}(2021)\citenamefont {Zhang},
  \citenamefont {Kong}, \citenamefont {Farooq}, \citenamefont {Yung},
  \citenamefont {Guo},\ and\ \citenamefont {Wang}}]{zhang2021}%
  \BibitemOpen
  \bibfield  {author} {\bibinfo {author} {\bibfnamefont {Xiao-Ming}\
  \bibnamefont {Zhang}}, \bibinfo {author} {\bibfnamefont {Weicheng}\
  \bibnamefont {Kong}}, \bibinfo {author} {\bibfnamefont {Muhammad~Usman}\
  \bibnamefont {Farooq}}, \bibinfo {author} {\bibfnamefont {Man-Hong}\
  \bibnamefont {Yung}}, \bibinfo {author} {\bibfnamefont {Guoping}\
  \bibnamefont {Guo}}, \ and\ \bibinfo {author} {\bibfnamefont {Xin}\
  \bibnamefont {Wang}},\ }\bibfield  {title} {\enquote {\bibinfo {title}
  {Generic detection-based error mitigation using quantum autoencoders},}\
  }\href {\doibase 10.1103/PhysRevA.103.L040403} {\bibfield  {journal}
  {\bibinfo  {journal} {Phys. Rev. A}\ }\textbf {\bibinfo {volume} {103}},\
  \bibinfo {pages} {L040403} (\bibinfo {year} {2021})}\BibitemShut {NoStop}%
\bibitem [{\citenamefont {Wiebe}\ \emph {et~al.}(2014)\citenamefont {Wiebe},
  \citenamefont {Granade}, \citenamefont {Ferrie},\ and\ \citenamefont
  {Cory}}]{wiebe2014}%
  \BibitemOpen
  \bibfield  {author} {\bibinfo {author} {\bibfnamefont {Nathan}\ \bibnamefont
  {Wiebe}}, \bibinfo {author} {\bibfnamefont {Christopher}\ \bibnamefont
  {Granade}}, \bibinfo {author} {\bibfnamefont {Christopher}\ \bibnamefont
  {Ferrie}}, \ and\ \bibinfo {author} {\bibfnamefont {D.~G.}\ \bibnamefont
  {Cory}},\ }\bibfield  {title} {\enquote {\bibinfo {title} {Hamiltonian
  {{Learning}} and {{Certification Using Quantum Resources}}},}\ }\href
  {\doibase 10.1103/PhysRevLett.112.190501} {\bibfield  {journal} {\bibinfo
  {journal} {Phys. Rev. Lett.}\ }\textbf {\bibinfo {volume} {112}},\ \bibinfo
  {pages} {190501} (\bibinfo {year} {2014})}\BibitemShut {NoStop}%
\bibitem [{\citenamefont {Gentile}\ \emph {et~al.}(2021)\citenamefont
  {Gentile}, \citenamefont {Flynn}, \citenamefont {Knauer}, \citenamefont
  {Wiebe}, \citenamefont {Paesani}, \citenamefont {Granade}, \citenamefont
  {Rarity}, \citenamefont {Santagati},\ and\ \citenamefont
  {Laing}}]{gentile021}%
  \BibitemOpen
  \bibfield  {author} {\bibinfo {author} {\bibfnamefont {Antonio~A.}\
  \bibnamefont {Gentile}}, \bibinfo {author} {\bibfnamefont {Brian}\
  \bibnamefont {Flynn}}, \bibinfo {author} {\bibfnamefont {Sebastian}\
  \bibnamefont {Knauer}}, \bibinfo {author} {\bibfnamefont {Nathan}\
  \bibnamefont {Wiebe}}, \bibinfo {author} {\bibfnamefont {Stefano}\
  \bibnamefont {Paesani}}, \bibinfo {author} {\bibfnamefont {Christopher~E.}\
  \bibnamefont {Granade}}, \bibinfo {author} {\bibfnamefont {John~G.}\
  \bibnamefont {Rarity}}, \bibinfo {author} {\bibfnamefont {Raffaele}\
  \bibnamefont {Santagati}}, \ and\ \bibinfo {author} {\bibfnamefont {Anthony}\
  \bibnamefont {Laing}},\ }\bibfield  {title} {\enquote {\bibinfo {title}
  {Learning models of quantum systems from experiments},}\ }\href {\doibase
  10.1038/s41567-021-01201-7} {\bibfield  {journal} {\bibinfo  {journal} {Nat.
  Phys.}\ }\textbf {\bibinfo {volume} {17}},\ \bibinfo {pages} {837--843}
  (\bibinfo {year} {2021})}\BibitemShut {NoStop}%
\bibitem [{\citenamefont {Ghosh}\ \emph {et~al.}(2019)\citenamefont {Ghosh},
  \citenamefont {Opala}, \citenamefont {Matuszewski}, \citenamefont {Paterek},\
  and\ \citenamefont {Liew}}]{ghosh2019}%
  \BibitemOpen
  \bibfield  {author} {\bibinfo {author} {\bibfnamefont {Sanjib}\ \bibnamefont
  {Ghosh}}, \bibinfo {author} {\bibfnamefont {Andrzej}\ \bibnamefont {Opala}},
  \bibinfo {author} {\bibfnamefont {Micha{\l}}\ \bibnamefont {Matuszewski}},
  \bibinfo {author} {\bibfnamefont {Tomasz}\ \bibnamefont {Paterek}}, \ and\
  \bibinfo {author} {\bibfnamefont {Timothy C.~H.}\ \bibnamefont {Liew}},\
  }\bibfield  {title} {\enquote {\bibinfo {title} {Quantum reservoir
  processing},}\ }\href {\doibase 10.1038/s41534-019-0149-8} {\bibfield
  {journal} {\bibinfo  {journal} {npj Quantum Inf}\ }\textbf {\bibinfo {volume}
  {5}},\ \bibinfo {pages} {1--6} (\bibinfo {year} {2019})}\BibitemShut
  {NoStop}%
\bibitem [{\citenamefont {Farhi}\ and\ \citenamefont
  {Neven}(2018)}]{farhi2018}%
  \BibitemOpen
  \bibfield  {author} {\bibinfo {author} {\bibfnamefont {Edward}\ \bibnamefont
  {Farhi}}\ and\ \bibinfo {author} {\bibfnamefont {Hartmut}\ \bibnamefont
  {Neven}},\ }\bibfield  {title} {\enquote {\bibinfo {title} {Classification
  with {{Quantum Neural Networks}} on {{Near Term Processors}}},}\ }\href@noop
  {} {\bibfield  {journal} {\bibinfo  {journal} {arXiv:1802.06002}\ } (\bibinfo
  {year} {2018})}\BibitemShut {NoStop}%
\bibitem [{\citenamefont {Cong}\ \emph {et~al.}(2019)\citenamefont {Cong},
  \citenamefont {Choi},\ and\ \citenamefont {Lukin}}]{cong2019}%
  \BibitemOpen
  \bibfield  {author} {\bibinfo {author} {\bibfnamefont {Iris}\ \bibnamefont
  {Cong}}, \bibinfo {author} {\bibfnamefont {Soonwon}\ \bibnamefont {Choi}}, \
  and\ \bibinfo {author} {\bibfnamefont {Mikhail~D.}\ \bibnamefont {Lukin}},\
  }\bibfield  {title} {\enquote {\bibinfo {title} {Quantum convolutional neural
  networks},}\ }\href {\doibase 10.1038/s41567-019-0648-8} {\bibfield
  {journal} {\bibinfo  {journal} {Nat. Phys.}\ }\textbf {\bibinfo {volume}
  {15}},\ \bibinfo {pages} {1273--1278} (\bibinfo {year} {2019})}\BibitemShut
  {NoStop}%
\bibitem [{\citenamefont {Beer}\ \emph {et~al.}(2020)\citenamefont {Beer},
  \citenamefont {Bondarenko}, \citenamefont {Farrelly}, \citenamefont
  {Osborne}, \citenamefont {Salzmann}, \citenamefont {Scheiermann},\ and\
  \citenamefont {Wolf}}]{beer2020}%
  \BibitemOpen
  \bibfield  {author} {\bibinfo {author} {\bibfnamefont {Kerstin}\ \bibnamefont
  {Beer}}, \bibinfo {author} {\bibfnamefont {Dmytro}\ \bibnamefont
  {Bondarenko}}, \bibinfo {author} {\bibfnamefont {Terry}\ \bibnamefont
  {Farrelly}}, \bibinfo {author} {\bibfnamefont {Tobias~J.}\ \bibnamefont
  {Osborne}}, \bibinfo {author} {\bibfnamefont {Robert}\ \bibnamefont
  {Salzmann}}, \bibinfo {author} {\bibfnamefont {Daniel}\ \bibnamefont
  {Scheiermann}}, \ and\ \bibinfo {author} {\bibfnamefont {Ramona}\
  \bibnamefont {Wolf}},\ }\bibfield  {title} {\enquote {\bibinfo {title}
  {Training deep quantum neural networks},}\ }\href {\doibase
  10.1038/s41467-020-14454-2} {\bibfield  {journal} {\bibinfo  {journal} {Nat.
  Commun.}\ }\textbf {\bibinfo {volume} {11}},\ \bibinfo {pages} {808}
  (\bibinfo {year} {2020})}\BibitemShut {NoStop}%
\bibitem [{\citenamefont {Kottmann}\ \emph {et~al.}(2021)\citenamefont
  {Kottmann}, \citenamefont {Metz}, \citenamefont {Fraxanet},\ and\
  \citenamefont {Baldelli}}]{kottmann2021}%
  \BibitemOpen
  \bibfield  {author} {\bibinfo {author} {\bibfnamefont {Korbinian}\
  \bibnamefont {Kottmann}}, \bibinfo {author} {\bibfnamefont {Friederike}\
  \bibnamefont {Metz}}, \bibinfo {author} {\bibfnamefont {Joana}\ \bibnamefont
  {Fraxanet}}, \ and\ \bibinfo {author} {\bibfnamefont {Niccol{\`o}}\
  \bibnamefont {Baldelli}},\ }\bibfield  {title} {\enquote {\bibinfo {title}
  {Variational quantum anomaly detection: {{Unsupervised}} mapping of phase
  diagrams on a physical quantum computer},}\ }\href {\doibase
  10.1103/PhysRevResearch.3.043184} {\bibfield  {journal} {\bibinfo  {journal}
  {Phys. Rev. Res.}\ }\textbf {\bibinfo {volume} {3}},\ \bibinfo {pages}
  {043184} (\bibinfo {year} {2021})}\BibitemShut {NoStop}%
\bibitem [{\citenamefont {Gong}\ \emph {et~al.}(2023)\citenamefont {Gong},
  \citenamefont {Huang}, \citenamefont {Wang}, \citenamefont {Guo},
  \citenamefont {Li}, \citenamefont {Wu}, \citenamefont {Zhu}, \citenamefont
  {Zhao}, \citenamefont {Guo}, \citenamefont {Qian}, \citenamefont {Ye},
  \citenamefont {Zha}, \citenamefont {Chen}, \citenamefont {Ying},
  \citenamefont {Yu}, \citenamefont {Fan}, \citenamefont {Wu}, \citenamefont
  {Su}, \citenamefont {Deng}, \citenamefont {Rong}, \citenamefont {Zhang},
  \citenamefont {Cao}, \citenamefont {Lin}, \citenamefont {Xu}, \citenamefont
  {Sun}, \citenamefont {Guo}, \citenamefont {Li}, \citenamefont {Liang},
  \citenamefont {Sakurai}, \citenamefont {Nemoto}, \citenamefont {Munro},
  \citenamefont {Huo}, \citenamefont {Lu}, \citenamefont {Peng}, \citenamefont
  {Zhu},\ and\ \citenamefont {Pan}}]{gong2023}%
  \BibitemOpen
  \bibfield  {author} {\bibinfo {author} {\bibfnamefont {Ming}\ \bibnamefont
  {Gong}}, \bibinfo {author} {\bibfnamefont {He-Liang}\ \bibnamefont {Huang}},
  \bibinfo {author} {\bibfnamefont {Shiyu}\ \bibnamefont {Wang}}, \bibinfo
  {author} {\bibfnamefont {Chu}\ \bibnamefont {Guo}}, \bibinfo {author}
  {\bibfnamefont {Shaowei}\ \bibnamefont {Li}}, \bibinfo {author}
  {\bibfnamefont {Yulin}\ \bibnamefont {Wu}}, \bibinfo {author} {\bibfnamefont
  {Qingling}\ \bibnamefont {Zhu}}, \bibinfo {author} {\bibfnamefont {Youwei}\
  \bibnamefont {Zhao}}, \bibinfo {author} {\bibfnamefont {Shaojun}\
  \bibnamefont {Guo}}, \bibinfo {author} {\bibfnamefont {Haoran}\ \bibnamefont
  {Qian}}, \bibinfo {author} {\bibfnamefont {Yangsen}\ \bibnamefont {Ye}},
  \bibinfo {author} {\bibfnamefont {Chen}\ \bibnamefont {Zha}}, \bibinfo
  {author} {\bibfnamefont {Fusheng}\ \bibnamefont {Chen}}, \bibinfo {author}
  {\bibfnamefont {Chong}\ \bibnamefont {Ying}}, \bibinfo {author}
  {\bibfnamefont {Jiale}\ \bibnamefont {Yu}}, \bibinfo {author} {\bibfnamefont
  {Daojin}\ \bibnamefont {Fan}}, \bibinfo {author} {\bibfnamefont {Dachao}\
  \bibnamefont {Wu}}, \bibinfo {author} {\bibfnamefont {Hong}\ \bibnamefont
  {Su}}, \bibinfo {author} {\bibfnamefont {Hui}\ \bibnamefont {Deng}}, \bibinfo
  {author} {\bibfnamefont {Hao}\ \bibnamefont {Rong}}, \bibinfo {author}
  {\bibfnamefont {Kaili}\ \bibnamefont {Zhang}}, \bibinfo {author}
  {\bibfnamefont {Sirui}\ \bibnamefont {Cao}}, \bibinfo {author} {\bibfnamefont
  {Jin}\ \bibnamefont {Lin}}, \bibinfo {author} {\bibfnamefont
  {Yu}~\bibnamefont {Xu}}, \bibinfo {author} {\bibfnamefont {Lihua}\
  \bibnamefont {Sun}}, \bibinfo {author} {\bibfnamefont {Cheng}\ \bibnamefont
  {Guo}}, \bibinfo {author} {\bibfnamefont {Na}~\bibnamefont {Li}}, \bibinfo
  {author} {\bibfnamefont {Futian}\ \bibnamefont {Liang}}, \bibinfo {author}
  {\bibfnamefont {Akitada}\ \bibnamefont {Sakurai}}, \bibinfo {author}
  {\bibfnamefont {Kae}\ \bibnamefont {Nemoto}}, \bibinfo {author}
  {\bibfnamefont {William~J.}\ \bibnamefont {Munro}}, \bibinfo {author}
  {\bibfnamefont {Yong-Heng}\ \bibnamefont {Huo}}, \bibinfo {author}
  {\bibfnamefont {Chao-Yang}\ \bibnamefont {Lu}}, \bibinfo {author}
  {\bibfnamefont {Cheng-Zhi}\ \bibnamefont {Peng}}, \bibinfo {author}
  {\bibfnamefont {Xiaobo}\ \bibnamefont {Zhu}}, \ and\ \bibinfo {author}
  {\bibfnamefont {Jian-Wei}\ \bibnamefont {Pan}},\ }\bibfield  {title}
  {\enquote {\bibinfo {title} {Quantum neuronal sensing of quantum many-body
  states on a 61-qubit programmable superconducting processor},}\ }\href
  {\doibase 10.1016/j.scib.2023.04.003} {\bibfield  {journal} {\bibinfo
  {journal} {Sci. Bull.}\ }\textbf {\bibinfo {volume} {68}},\ \bibinfo {pages}
  {906--912} (\bibinfo {year} {2023})}\BibitemShut {NoStop}%
\bibitem [{\citenamefont {Pollmann}\ \emph {et~al.}(2010)\citenamefont
  {Pollmann}, \citenamefont {Turner}, \citenamefont {Berg},\ and\ \citenamefont
  {Oshikawa}}]{pollmann2010}%
  \BibitemOpen
  \bibfield  {author} {\bibinfo {author} {\bibfnamefont {Frank}\ \bibnamefont
  {Pollmann}}, \bibinfo {author} {\bibfnamefont {Ari~M.}\ \bibnamefont
  {Turner}}, \bibinfo {author} {\bibfnamefont {Erez}\ \bibnamefont {Berg}}, \
  and\ \bibinfo {author} {\bibfnamefont {Masaki}\ \bibnamefont {Oshikawa}},\
  }\bibfield  {title} {\enquote {\bibinfo {title} {Entanglement spectrum of a
  topological phase in one dimension},}\ }\href {\doibase
  10.1103/PhysRevB.81.064439} {\bibfield  {journal} {\bibinfo  {journal} {Phys.
  Rev. B}\ }\textbf {\bibinfo {volume} {81}},\ \bibinfo {pages} {064439}
  (\bibinfo {year} {2010})}\BibitemShut {NoStop}%
\bibitem [{\citenamefont {Chen}\ \emph {et~al.}(2011)\citenamefont {Chen},
  \citenamefont {Gu},\ and\ \citenamefont {Wen}}]{chen2011}%
  \BibitemOpen
  \bibfield  {author} {\bibinfo {author} {\bibfnamefont {Xie}\ \bibnamefont
  {Chen}}, \bibinfo {author} {\bibfnamefont {Zheng-Cheng}\ \bibnamefont {Gu}},
  \ and\ \bibinfo {author} {\bibfnamefont {Xiao-Gang}\ \bibnamefont {Wen}},\
  }\bibfield  {title} {\enquote {\bibinfo {title} {Classification of gapped
  symmetric phases in one-dimensional spin systems},}\ }\href {\doibase
  10.1103/PhysRevB.83.035107} {\bibfield  {journal} {\bibinfo  {journal} {Phys.
  Rev. B}\ }\textbf {\bibinfo {volume} {83}},\ \bibinfo {pages} {035107}
  (\bibinfo {year} {2011})}\BibitemShut {NoStop}%
\bibitem [{\citenamefont {{de L{\'e}s{\'e}leuc}}\ \emph
  {et~al.}(2019)\citenamefont {{de L{\'e}s{\'e}leuc}}, \citenamefont
  {Lienhard}, \citenamefont {Scholl}, \citenamefont {Barredo}, \citenamefont
  {Weber}, \citenamefont {Lang}, \citenamefont {B{\"u}chler}, \citenamefont
  {Lahaye},\ and\ \citenamefont {Browaeys}}]{deleseleuc2019}%
  \BibitemOpen
  \bibfield  {author} {\bibinfo {author} {\bibfnamefont {Sylvain}\ \bibnamefont
  {{de L{\'e}s{\'e}leuc}}}, \bibinfo {author} {\bibfnamefont {Vincent}\
  \bibnamefont {Lienhard}}, \bibinfo {author} {\bibfnamefont {Pascal}\
  \bibnamefont {Scholl}}, \bibinfo {author} {\bibfnamefont {Daniel}\
  \bibnamefont {Barredo}}, \bibinfo {author} {\bibfnamefont {Sebastian}\
  \bibnamefont {Weber}}, \bibinfo {author} {\bibfnamefont {Nicolai}\
  \bibnamefont {Lang}}, \bibinfo {author} {\bibfnamefont {Hans~Peter}\
  \bibnamefont {B{\"u}chler}}, \bibinfo {author} {\bibfnamefont {Thierry}\
  \bibnamefont {Lahaye}}, \ and\ \bibinfo {author} {\bibfnamefont {Antoine}\
  \bibnamefont {Browaeys}},\ }\bibfield  {title} {\enquote {\bibinfo {title}
  {Observation of a symmetry-protected topological phase of interacting bosons
  with {{Rydberg}} atoms},}\ }\href {\doibase 10.1126/science.aav9105}
  {\bibfield  {journal} {\bibinfo  {journal} {Science}\ }\textbf {\bibinfo
  {volume} {365}},\ \bibinfo {pages} {775--780} (\bibinfo {year}
  {2019})}\BibitemShut {NoStop}%
\bibitem [{\citenamefont {Semeghini}\ \emph {et~al.}(2021)\citenamefont
  {Semeghini}, \citenamefont {Levine}, \citenamefont {Keesling}, \citenamefont
  {Ebadi}, \citenamefont {Wang}, \citenamefont {Bluvstein}, \citenamefont
  {Verresen}, \citenamefont {Pichler}, \citenamefont {Kalinowski},
  \citenamefont {Samajdar}, \citenamefont {Omran}, \citenamefont {Sachdev},
  \citenamefont {Vishwanath}, \citenamefont {Greiner}, \citenamefont
  {Vuleti{\'c}},\ and\ \citenamefont {Lukin}}]{semeghini2021}%
  \BibitemOpen
  \bibfield  {author} {\bibinfo {author} {\bibfnamefont {G.}~\bibnamefont
  {Semeghini}}, \bibinfo {author} {\bibfnamefont {H.}~\bibnamefont {Levine}},
  \bibinfo {author} {\bibfnamefont {A.}~\bibnamefont {Keesling}}, \bibinfo
  {author} {\bibfnamefont {S.}~\bibnamefont {Ebadi}}, \bibinfo {author}
  {\bibfnamefont {T.~T.}\ \bibnamefont {Wang}}, \bibinfo {author}
  {\bibfnamefont {D.}~\bibnamefont {Bluvstein}}, \bibinfo {author}
  {\bibfnamefont {R.}~\bibnamefont {Verresen}}, \bibinfo {author}
  {\bibfnamefont {H.}~\bibnamefont {Pichler}}, \bibinfo {author} {\bibfnamefont
  {M.}~\bibnamefont {Kalinowski}}, \bibinfo {author} {\bibfnamefont
  {R.}~\bibnamefont {Samajdar}}, \bibinfo {author} {\bibfnamefont
  {A.}~\bibnamefont {Omran}}, \bibinfo {author} {\bibfnamefont
  {S.}~\bibnamefont {Sachdev}}, \bibinfo {author} {\bibfnamefont
  {A.}~\bibnamefont {Vishwanath}}, \bibinfo {author} {\bibfnamefont
  {M.}~\bibnamefont {Greiner}}, \bibinfo {author} {\bibfnamefont
  {V.}~\bibnamefont {Vuleti{\'c}}}, \ and\ \bibinfo {author} {\bibfnamefont
  {M.~D.}\ \bibnamefont {Lukin}},\ }\bibfield  {title} {\enquote {\bibinfo
  {title} {Probing topological spin liquids on a programmable quantum
  simulator},}\ }\href {\doibase 10.1126/science.abi8794} {\bibfield  {journal}
  {\bibinfo  {journal} {Science}\ }\textbf {\bibinfo {volume} {374}},\ \bibinfo
  {pages} {1242--1247} (\bibinfo {year} {2021})}\BibitemShut {NoStop}%
\bibitem [{\citenamefont {Sachdev}(2011)}]{sachdev2011}%
  \BibitemOpen
  \bibfield  {author} {\bibinfo {author} {\bibfnamefont {Subir}\ \bibnamefont
  {Sachdev}},\ }\href@noop {} {\emph {\bibinfo {title} {Quantum {{Phase
  Transitions}}}}}\ (\bibinfo  {publisher} {Cambridge University Press},\
  \bibinfo {year} {2011})\BibitemShut {NoStop}%
\bibitem [{\citenamefont {Wang}\ \emph {et~al.}(2016)\citenamefont {Wang},
  \citenamefont {Zhang}, \citenamefont {Liu}, \citenamefont {Liu},
  \citenamefont {Tang}, \citenamefont {Song}, \citenamefont {Zhong},
  \citenamefont {Peng}, \citenamefont {Li}, \citenamefont {Nie}, \citenamefont
  {Wang}, \citenamefont {Zhou}, \citenamefont {Ma}, \citenamefont {Xue},\ and\
  \citenamefont {Liu}}]{wang2016}%
  \BibitemOpen
  \bibfield  {author} {\bibinfo {author} {\bibfnamefont {Z.~F.}\ \bibnamefont
  {Wang}}, \bibinfo {author} {\bibfnamefont {Huimin}\ \bibnamefont {Zhang}},
  \bibinfo {author} {\bibfnamefont {Defa}\ \bibnamefont {Liu}}, \bibinfo
  {author} {\bibfnamefont {Chong}\ \bibnamefont {Liu}}, \bibinfo {author}
  {\bibfnamefont {Chenjia}\ \bibnamefont {Tang}}, \bibinfo {author}
  {\bibfnamefont {Canli}\ \bibnamefont {Song}}, \bibinfo {author}
  {\bibfnamefont {Yong}\ \bibnamefont {Zhong}}, \bibinfo {author}
  {\bibfnamefont {Junping}\ \bibnamefont {Peng}}, \bibinfo {author}
  {\bibfnamefont {Fangsen}\ \bibnamefont {Li}}, \bibinfo {author}
  {\bibfnamefont {Caina}\ \bibnamefont {Nie}}, \bibinfo {author} {\bibfnamefont
  {Lili}\ \bibnamefont {Wang}}, \bibinfo {author} {\bibfnamefont {X.~J.}\
  \bibnamefont {Zhou}}, \bibinfo {author} {\bibfnamefont {Xucun}\ \bibnamefont
  {Ma}}, \bibinfo {author} {\bibfnamefont {Q.~K.}\ \bibnamefont {Xue}}, \ and\
  \bibinfo {author} {\bibfnamefont {Feng}\ \bibnamefont {Liu}},\ }\bibfield
  {title} {\enquote {\bibinfo {title} {Topological edge states in a
  high-temperature superconductor {{FeSe}}/{{SrTiO3}}(001) film},}\ }\href
  {\doibase 10.1038/nmat4686} {\bibfield  {journal} {\bibinfo  {journal} {Nat.
  Mat.}\ }\textbf {\bibinfo {volume} {15}},\ \bibinfo {pages} {968--973}
  (\bibinfo {year} {2016})}\BibitemShut {NoStop}%
\bibitem [{\citenamefont {Carrasquilla}\ and\ \citenamefont
  {Melko}(2017)}]{carrasquilla2017}%
  \BibitemOpen
  \bibfield  {author} {\bibinfo {author} {\bibfnamefont {Juan}\ \bibnamefont
  {Carrasquilla}}\ and\ \bibinfo {author} {\bibfnamefont {Roger~G.}\
  \bibnamefont {Melko}},\ }\bibfield  {title} {\enquote {\bibinfo {title}
  {Machine learning phases of matter},}\ }\href {\doibase 10.1038/nphys4035}
  {\bibfield  {journal} {\bibinfo  {journal} {Nat. Phys.}\ }\textbf {\bibinfo
  {volume} {13}},\ \bibinfo {pages} {431--434} (\bibinfo {year}
  {2017})}\BibitemShut {NoStop}%
\bibitem [{\citenamefont {{van Nieuwenburg}}\ \emph {et~al.}(2017)\citenamefont
  {{van Nieuwenburg}}, \citenamefont {Liu},\ and\ \citenamefont
  {Huber}}]{vannieuwenburg2017}%
  \BibitemOpen
  \bibfield  {author} {\bibinfo {author} {\bibfnamefont {Evert~P.~L.}\
  \bibnamefont {{van Nieuwenburg}}}, \bibinfo {author} {\bibfnamefont {Ye-Hua}\
  \bibnamefont {Liu}}, \ and\ \bibinfo {author} {\bibfnamefont {Sebastian~D.}\
  \bibnamefont {Huber}},\ }\bibfield  {title} {\enquote {\bibinfo {title}
  {Learning phase transitions by confusion},}\ }\href {\doibase
  10.1038/nphys4037} {\bibfield  {journal} {\bibinfo  {journal} {Nat. Phys.}\
  }\textbf {\bibinfo {volume} {13}},\ \bibinfo {pages} {435--439} (\bibinfo
  {year} {2017})}\BibitemShut {NoStop}%
\bibitem [{\citenamefont {Greplova}\ \emph {et~al.}(2020)\citenamefont
  {Greplova}, \citenamefont {Valenti}, \citenamefont {Boschung}, \citenamefont
  {Sch{\"a}fer}, \citenamefont {L{\"o}rch},\ and\ \citenamefont
  {Huber}}]{greplova2020}%
  \BibitemOpen
  \bibfield  {author} {\bibinfo {author} {\bibfnamefont {Eliska}\ \bibnamefont
  {Greplova}}, \bibinfo {author} {\bibfnamefont {Agnes}\ \bibnamefont
  {Valenti}}, \bibinfo {author} {\bibfnamefont {Gregor}\ \bibnamefont
  {Boschung}}, \bibinfo {author} {\bibfnamefont {Frank}\ \bibnamefont
  {Sch{\"a}fer}}, \bibinfo {author} {\bibfnamefont {Niels}\ \bibnamefont
  {L{\"o}rch}}, \ and\ \bibinfo {author} {\bibfnamefont {Sebastian~D.}\
  \bibnamefont {Huber}},\ }\bibfield  {title} {\enquote {\bibinfo {title}
  {Unsupervised identification of topological phase transitions using
  predictive models},}\ }\href {\doibase 10.1088/1367-2630/ab7771} {\bibfield
  {journal} {\bibinfo  {journal} {New J. Phys.}\ }\textbf {\bibinfo {volume}
  {22}},\ \bibinfo {pages} {045003} (\bibinfo {year} {2020})}\BibitemShut
  {NoStop}%
\bibitem [{\citenamefont {Rem}\ \emph {et~al.}(2019)\citenamefont {Rem},
  \citenamefont {K{\"a}ming}, \citenamefont {Tarnowski}, \citenamefont
  {Asteria}, \citenamefont {Fl{\"a}schner}, \citenamefont {Becker},
  \citenamefont {Sengstock},\ and\ \citenamefont {Weitenberg}}]{rem2019}%
  \BibitemOpen
  \bibfield  {author} {\bibinfo {author} {\bibfnamefont {Benno~S.}\
  \bibnamefont {Rem}}, \bibinfo {author} {\bibfnamefont {Niklas}\ \bibnamefont
  {K{\"a}ming}}, \bibinfo {author} {\bibfnamefont {Matthias}\ \bibnamefont
  {Tarnowski}}, \bibinfo {author} {\bibfnamefont {Luca}\ \bibnamefont
  {Asteria}}, \bibinfo {author} {\bibfnamefont {Nick}\ \bibnamefont
  {Fl{\"a}schner}}, \bibinfo {author} {\bibfnamefont {Christoph}\ \bibnamefont
  {Becker}}, \bibinfo {author} {\bibfnamefont {Klaus}\ \bibnamefont
  {Sengstock}}, \ and\ \bibinfo {author} {\bibfnamefont {Christof}\
  \bibnamefont {Weitenberg}},\ }\bibfield  {title} {\enquote {\bibinfo {title}
  {Identifying quantum phase transitions using artificial neural networks on
  experimental data},}\ }\href {\doibase 10.1038/s41567-019-0554-0} {\bibfield
  {journal} {\bibinfo  {journal} {Nat. Phys.}\ }\textbf {\bibinfo {volume}
  {15}},\ \bibinfo {pages} {917--920} (\bibinfo {year} {2019})}\BibitemShut
  {NoStop}%
\bibitem [{\citenamefont {Bohrdt}\ \emph {et~al.}(2021)\citenamefont {Bohrdt},
  \citenamefont {Kim}, \citenamefont {Lukin}, \citenamefont {Rispoli},
  \citenamefont {Schittko}, \citenamefont {Knap}, \citenamefont {Greiner},\
  and\ \citenamefont {L{\'e}onard}}]{bohrdt2021}%
  \BibitemOpen
  \bibfield  {author} {\bibinfo {author} {\bibfnamefont {A.}~\bibnamefont
  {Bohrdt}}, \bibinfo {author} {\bibfnamefont {S.}~\bibnamefont {Kim}},
  \bibinfo {author} {\bibfnamefont {A.}~\bibnamefont {Lukin}}, \bibinfo
  {author} {\bibfnamefont {M.}~\bibnamefont {Rispoli}}, \bibinfo {author}
  {\bibfnamefont {R.}~\bibnamefont {Schittko}}, \bibinfo {author}
  {\bibfnamefont {M.}~\bibnamefont {Knap}}, \bibinfo {author} {\bibfnamefont
  {M.}~\bibnamefont {Greiner}}, \ and\ \bibinfo {author} {\bibfnamefont
  {J.}~\bibnamefont {L{\'e}onard}},\ }\bibfield  {title} {\enquote {\bibinfo
  {title} {Analyzing {{Nonequilibrium Quantum States}} through {{Snapshots}}
  with {{Artificial Neural Networks}}},}\ }\href {\doibase
  10.1103/PhysRevLett.127.150504} {\bibfield  {journal} {\bibinfo  {journal}
  {Phys. Rev. Lett.}\ }\textbf {\bibinfo {volume} {127}},\ \bibinfo {pages}
  {150504} (\bibinfo {year} {2021})}\BibitemShut {NoStop}%
\bibitem [{\citenamefont {K{\"a}ming}\ \emph {et~al.}(2021)\citenamefont
  {K{\"a}ming}, \citenamefont {Dawid}, \citenamefont {Kottmann}, \citenamefont
  {Lewenstein}, \citenamefont {Sengstock}, \citenamefont {Dauphin},\ and\
  \citenamefont {Weitenberg}}]{kaming2021}%
  \BibitemOpen
  \bibfield  {author} {\bibinfo {author} {\bibfnamefont {Niklas}\ \bibnamefont
  {K{\"a}ming}}, \bibinfo {author} {\bibfnamefont {Anna}\ \bibnamefont
  {Dawid}}, \bibinfo {author} {\bibfnamefont {Korbinian}\ \bibnamefont
  {Kottmann}}, \bibinfo {author} {\bibfnamefont {Maciej}\ \bibnamefont
  {Lewenstein}}, \bibinfo {author} {\bibfnamefont {Klaus}\ \bibnamefont
  {Sengstock}}, \bibinfo {author} {\bibfnamefont {Alexandre}\ \bibnamefont
  {Dauphin}}, \ and\ \bibinfo {author} {\bibfnamefont {Christof}\ \bibnamefont
  {Weitenberg}},\ }\bibfield  {title} {\enquote {\bibinfo {title} {Unsupervised
  machine learning of topological phase transitions from experimental data},}\
  }\href {\doibase 10.1088/2632-2153/abffe7} {\bibfield  {journal} {\bibinfo
  {journal} {Mach. Learn.: Sci. Technol.}\ }\textbf {\bibinfo {volume} {2}},\
  \bibinfo {pages} {035037} (\bibinfo {year} {2021})}\BibitemShut {NoStop}%
\bibitem [{\citenamefont {Miles}\ \emph {et~al.}(2023)\citenamefont {Miles},
  \citenamefont {Samajdar}, \citenamefont {Ebadi}, \citenamefont {Wang},
  \citenamefont {Pichler}, \citenamefont {Sachdev}, \citenamefont {Lukin},
  \citenamefont {Greiner}, \citenamefont {Weinberger},\ and\ \citenamefont
  {Kim}}]{miles2023}%
  \BibitemOpen
  \bibfield  {author} {\bibinfo {author} {\bibfnamefont {Cole}\ \bibnamefont
  {Miles}}, \bibinfo {author} {\bibfnamefont {Rhine}\ \bibnamefont {Samajdar}},
  \bibinfo {author} {\bibfnamefont {Sepehr}\ \bibnamefont {Ebadi}}, \bibinfo
  {author} {\bibfnamefont {Tout~T.}\ \bibnamefont {Wang}}, \bibinfo {author}
  {\bibfnamefont {Hannes}\ \bibnamefont {Pichler}}, \bibinfo {author}
  {\bibfnamefont {Subir}\ \bibnamefont {Sachdev}}, \bibinfo {author}
  {\bibfnamefont {Mikhail~D.}\ \bibnamefont {Lukin}}, \bibinfo {author}
  {\bibfnamefont {Markus}\ \bibnamefont {Greiner}}, \bibinfo {author}
  {\bibfnamefont {Kilian~Q.}\ \bibnamefont {Weinberger}}, \ and\ \bibinfo
  {author} {\bibfnamefont {Eun-Ah}\ \bibnamefont {Kim}},\ }\bibfield  {title}
  {\enquote {\bibinfo {title} {Machine learning discovery of new phases in
  programmable quantum simulator snapshots},}\ }\href {\doibase
  10.1103/PhysRevResearch.5.013026} {\bibfield  {journal} {\bibinfo  {journal}
  {Phys. Rev. Res.}\ }\textbf {\bibinfo {volume} {5}},\ \bibinfo {pages}
  {013026} (\bibinfo {year} {2023})}\BibitemShut {NoStop}%
\bibitem [{\citenamefont {Smith}\ \emph {et~al.}(2022)\citenamefont {Smith},
  \citenamefont {Jobst}, \citenamefont {Green},\ and\ \citenamefont
  {Pollmann}}]{smith2022}%
  \BibitemOpen
  \bibfield  {author} {\bibinfo {author} {\bibfnamefont {Adam}\ \bibnamefont
  {Smith}}, \bibinfo {author} {\bibfnamefont {Bernhard}\ \bibnamefont {Jobst}},
  \bibinfo {author} {\bibfnamefont {Andrew~G.}\ \bibnamefont {Green}}, \ and\
  \bibinfo {author} {\bibfnamefont {Frank}\ \bibnamefont {Pollmann}},\
  }\bibfield  {title} {\enquote {\bibinfo {title} {Crossing a topological phase
  transition with a quantum computer},}\ }\href {\doibase
  10.1103/PhysRevResearch.4.L022020} {\bibfield  {journal} {\bibinfo  {journal}
  {Phys. Rev. Research}\ }\textbf {\bibinfo {volume} {4}},\ \bibinfo {pages}
  {L022020} (\bibinfo {year} {2022})}\BibitemShut {NoStop}%
\bibitem [{\citenamefont {Satzinger}\ and\ \citenamefont
  {others}(2021)\citenamefont {Satzinger} \emph {et~al.}}]{satzinger2021}%
  \BibitemOpen
  \bibfield  {author} {\bibinfo {author} {\bibfnamefont {K.~J.}\ \bibnamefont
  {Satzinger}} \emph {et~al.},\ }\bibfield  {title} {\enquote {\bibinfo {title}
  {Realizing topologically ordered states on a quantum processor},}\ }\href
  {\doibase 10.1126/science.abi8378} {\bibfield  {journal} {\bibinfo  {journal}
  {Science}\ }\textbf {\bibinfo {volume} {374}},\ \bibinfo {pages} {1237--1241}
  (\bibinfo {year} {2021})}\BibitemShut {NoStop}%
\bibitem [{\citenamefont {Iqbal}\ \emph {et~al.}(2023)\citenamefont {Iqbal},
  \citenamefont {Tantivasadakarn}, \citenamefont {Gatterman}, \citenamefont
  {Gerber}, \citenamefont {Gilmore}, \citenamefont {Gresh}, \citenamefont
  {Hankin}, \citenamefont {Hewitt}, \citenamefont {Horst}, \citenamefont
  {Matheny}, \citenamefont {Mengle}, \citenamefont {Neyenhuis}, \citenamefont
  {Vishwanath}, \citenamefont {{Foss-Feig}}, \citenamefont {Verresen},\ and\
  \citenamefont {Dreyer}}]{iqbal2023}%
  \BibitemOpen
  \bibfield  {author} {\bibinfo {author} {\bibfnamefont {Mohsin}\ \bibnamefont
  {Iqbal}}, \bibinfo {author} {\bibfnamefont {Nathanan}\ \bibnamefont
  {Tantivasadakarn}}, \bibinfo {author} {\bibfnamefont {Thomas~M.}\
  \bibnamefont {Gatterman}}, \bibinfo {author} {\bibfnamefont {Justin~A.}\
  \bibnamefont {Gerber}}, \bibinfo {author} {\bibfnamefont {Kevin}\
  \bibnamefont {Gilmore}}, \bibinfo {author} {\bibfnamefont {Dan}\ \bibnamefont
  {Gresh}}, \bibinfo {author} {\bibfnamefont {Aaron}\ \bibnamefont {Hankin}},
  \bibinfo {author} {\bibfnamefont {Nathan}\ \bibnamefont {Hewitt}}, \bibinfo
  {author} {\bibfnamefont {Chandler~V.}\ \bibnamefont {Horst}}, \bibinfo
  {author} {\bibfnamefont {Mitchell}\ \bibnamefont {Matheny}}, \bibinfo
  {author} {\bibfnamefont {Tanner}\ \bibnamefont {Mengle}}, \bibinfo {author}
  {\bibfnamefont {Brian}\ \bibnamefont {Neyenhuis}}, \bibinfo {author}
  {\bibfnamefont {Ashvin}\ \bibnamefont {Vishwanath}}, \bibinfo {author}
  {\bibfnamefont {Michael}\ \bibnamefont {{Foss-Feig}}}, \bibinfo {author}
  {\bibfnamefont {Ruben}\ \bibnamefont {Verresen}}, \ and\ \bibinfo {author}
  {\bibfnamefont {Henrik}\ \bibnamefont {Dreyer}},\ }\bibfield  {title}
  {\enquote {\bibinfo {title} {Topological {{Order}} from {{Measurements}} and
  {{Feed-Forward}} on a {{Trapped Ion Quantum Computer}}},}\ }\href@noop {}
  {\bibfield  {journal} {\bibinfo  {journal} {arXiv:2302.01917}\ } (\bibinfo
  {year} {2023})}\BibitemShut {NoStop}%
\bibitem [{\citenamefont {Azses}\ \emph {et~al.}(2020)\citenamefont {Azses},
  \citenamefont {Haenel}, \citenamefont {Naveh}, \citenamefont {Raussendorf},
  \citenamefont {Sela},\ and\ \citenamefont {Dalla~Torre}}]{azses2020}%
  \BibitemOpen
  \bibfield  {author} {\bibinfo {author} {\bibfnamefont {Daniel}\ \bibnamefont
  {Azses}}, \bibinfo {author} {\bibfnamefont {Rafael}\ \bibnamefont {Haenel}},
  \bibinfo {author} {\bibfnamefont {Yehuda}\ \bibnamefont {Naveh}}, \bibinfo
  {author} {\bibfnamefont {Robert}\ \bibnamefont {Raussendorf}}, \bibinfo
  {author} {\bibfnamefont {Eran}\ \bibnamefont {Sela}}, \ and\ \bibinfo
  {author} {\bibfnamefont {Emanuele~G.}\ \bibnamefont {Dalla~Torre}},\
  }\bibfield  {title} {\enquote {\bibinfo {title} {Identification of
  {{Symmetry-Protected Topological States}} on {{Noisy Quantum Computers}}},}\
  }\href {\doibase 10.1103/PhysRevLett.125.120502} {\bibfield  {journal}
  {\bibinfo  {journal} {Phys. Rev. Lett.}\ }\textbf {\bibinfo {volume} {125}},\
  \bibinfo {pages} {120502} (\bibinfo {year} {2020})}\BibitemShut {NoStop}%
\bibitem [{\citenamefont {{P{\'e}rez-Garc{\'i}a}}\ \emph
  {et~al.}(2008)\citenamefont {{P{\'e}rez-Garc{\'i}a}}, \citenamefont {Wolf},
  \citenamefont {Sanz}, \citenamefont {Verstraete},\ and\ \citenamefont
  {Cirac}}]{perez-garcia2008}%
  \BibitemOpen
  \bibfield  {author} {\bibinfo {author} {\bibfnamefont {D.}~\bibnamefont
  {{P{\'e}rez-Garc{\'i}a}}}, \bibinfo {author} {\bibfnamefont {M.~M.}\
  \bibnamefont {Wolf}}, \bibinfo {author} {\bibfnamefont {M.}~\bibnamefont
  {Sanz}}, \bibinfo {author} {\bibfnamefont {F.}~\bibnamefont {Verstraete}}, \
  and\ \bibinfo {author} {\bibfnamefont {J.~I.}\ \bibnamefont {Cirac}},\
  }\bibfield  {title} {\enquote {\bibinfo {title} {String {{Order}} and
  {{Symmetries}} in {{Quantum Spin Lattices}}},}\ }\href {\doibase
  10.1103/PhysRevLett.100.167202} {\bibfield  {journal} {\bibinfo  {journal}
  {Phys. Rev. Lett.}\ }\textbf {\bibinfo {volume} {100}},\ \bibinfo {pages}
  {167202} (\bibinfo {year} {2008})}\BibitemShut {NoStop}%
\bibitem [{\citenamefont {Pollmann}\ and\ \citenamefont
  {Turner}(2012)}]{pollmann2012}%
  \BibitemOpen
  \bibfield  {author} {\bibinfo {author} {\bibfnamefont {Frank}\ \bibnamefont
  {Pollmann}}\ and\ \bibinfo {author} {\bibfnamefont {Ari~M.}\ \bibnamefont
  {Turner}},\ }\bibfield  {title} {\enquote {\bibinfo {title} {Detection of
  symmetry-protected topological phases in one dimension},}\ }\href {\doibase
  10.1103/PhysRevB.86.125441} {\bibfield  {journal} {\bibinfo  {journal} {Phys.
  Rev. B}\ }\textbf {\bibinfo {volume} {86}},\ \bibinfo {pages} {125441}
  (\bibinfo {year} {2012})}\BibitemShut {NoStop}%
\bibitem [{\citenamefont {Cong}\ \emph {et~al.}(2024)\citenamefont {Cong},
  \citenamefont {Maskara}, \citenamefont {Tran}, \citenamefont {Pichler},
  \citenamefont {Semeghini}, \citenamefont {Yelin}, \citenamefont {Choi},\ and\
  \citenamefont {Lukin}}]{cong2022}%
  \BibitemOpen
  \bibfield  {author} {\bibinfo {author} {\bibfnamefont {Iris}\ \bibnamefont
  {Cong}}, \bibinfo {author} {\bibfnamefont {Nishad}\ \bibnamefont {Maskara}},
  \bibinfo {author} {\bibfnamefont {Minh~C.}\ \bibnamefont {Tran}}, \bibinfo
  {author} {\bibfnamefont {Hannes}\ \bibnamefont {Pichler}}, \bibinfo {author}
  {\bibfnamefont {Giulia}\ \bibnamefont {Semeghini}}, \bibinfo {author}
  {\bibfnamefont {Susanne~F.}\ \bibnamefont {Yelin}}, \bibinfo {author}
  {\bibfnamefont {Soonwon}\ \bibnamefont {Choi}}, \ and\ \bibinfo {author}
  {\bibfnamefont {Mikhail~D.}\ \bibnamefont {Lukin}},\ }\bibfield  {title}
  {\enquote {\bibinfo {title} {Enhancing detection of topological order by
  local error correction},}\ }\href {\doibase 10.1038/s41467-024-45584-6}
  {\bibfield  {journal} {\bibinfo  {journal} {Nat. Commun.}\ }\textbf {\bibinfo
  {volume} {15}},\ \bibinfo {pages} {1527} (\bibinfo {year}
  {2024})}\BibitemShut {NoStop}%
\bibitem [{\citenamefont {Huang}\ \emph {et~al.}(2022)\citenamefont {Huang},
  \citenamefont {Kueng}, \citenamefont {Torlai}, \citenamefont {Albert},\ and\
  \citenamefont {Preskill}}]{huang2022}%
  \BibitemOpen
  \bibfield  {author} {\bibinfo {author} {\bibfnamefont {Hsin-Yuan}\
  \bibnamefont {Huang}}, \bibinfo {author} {\bibfnamefont {Richard}\
  \bibnamefont {Kueng}}, \bibinfo {author} {\bibfnamefont {Giacomo}\
  \bibnamefont {Torlai}}, \bibinfo {author} {\bibfnamefont {Victor~V.}\
  \bibnamefont {Albert}}, \ and\ \bibinfo {author} {\bibfnamefont {John}\
  \bibnamefont {Preskill}},\ }\bibfield  {title} {\enquote {\bibinfo {title}
  {Provably efficient machine learning for quantum many-body problems},}\
  }\href {\doibase 10.1126/science.abk3333} {\bibfield  {journal} {\bibinfo
  {journal} {Science}\ }\textbf {\bibinfo {volume} {377}},\ \bibinfo {pages}
  {eabk3333} (\bibinfo {year} {2022})}\BibitemShut {NoStop}%
\bibitem [{\citenamefont {Caro}\ \emph {et~al.}(2022)\citenamefont {Caro},
  \citenamefont {Huang}, \citenamefont {Cerezo}, \citenamefont {Sharma},
  \citenamefont {Sornborger}, \citenamefont {Cincio},\ and\ \citenamefont
  {Coles}}]{caro2022}%
  \BibitemOpen
  \bibfield  {author} {\bibinfo {author} {\bibfnamefont {Matthias~C.}\
  \bibnamefont {Caro}}, \bibinfo {author} {\bibfnamefont {Hsin-Yuan}\
  \bibnamefont {Huang}}, \bibinfo {author} {\bibfnamefont {M.}~\bibnamefont
  {Cerezo}}, \bibinfo {author} {\bibfnamefont {Kunal}\ \bibnamefont {Sharma}},
  \bibinfo {author} {\bibfnamefont {Andrew}\ \bibnamefont {Sornborger}},
  \bibinfo {author} {\bibfnamefont {Lukasz}\ \bibnamefont {Cincio}}, \ and\
  \bibinfo {author} {\bibfnamefont {Patrick~J.}\ \bibnamefont {Coles}},\
  }\bibfield  {title} {\enquote {\bibinfo {title} {Generalization in quantum
  machine learning from few training data},}\ }\href {\doibase
  10.1038/s41467-022-32550-3} {\bibfield  {journal} {\bibinfo  {journal} {Nat.
  Commun.}\ }\textbf {\bibinfo {volume} {13}},\ \bibinfo {pages} {4919}
  (\bibinfo {year} {2022})}\BibitemShut {NoStop}%
\bibitem [{\citenamefont {Liu}\ \emph {et~al.}(2023)\citenamefont {Liu},
  \citenamefont {Smith}, \citenamefont {Knap},\ and\ \citenamefont
  {Pollmann}}]{liu2023}%
  \BibitemOpen
  \bibfield  {author} {\bibinfo {author} {\bibfnamefont {Yu-Jie}\ \bibnamefont
  {Liu}}, \bibinfo {author} {\bibfnamefont {Adam}\ \bibnamefont {Smith}},
  \bibinfo {author} {\bibfnamefont {Michael}\ \bibnamefont {Knap}}, \ and\
  \bibinfo {author} {\bibfnamefont {Frank}\ \bibnamefont {Pollmann}},\
  }\bibfield  {title} {\enquote {\bibinfo {title} {Model-{{Independent
  Learning}} of {{Quantum Phases}} of {{Matter}} with {{Quantum Convolutional
  Neural Networks}}},}\ }\href {\doibase 10.1103/PhysRevLett.130.220603}
  {\bibfield  {journal} {\bibinfo  {journal} {Phys. Rev. Lett.}\ }\textbf
  {\bibinfo {volume} {130}},\ \bibinfo {pages} {220603} (\bibinfo {year}
  {2023})}\BibitemShut {NoStop}%
\bibitem [{\citenamefont {Lake}\ \emph {et~al.}(2022)\citenamefont {Lake},
  \citenamefont {Balasubramanian},\ and\ \citenamefont {Choi}}]{lake2022}%
  \BibitemOpen
  \bibfield  {author} {\bibinfo {author} {\bibfnamefont {Ethan}\ \bibnamefont
  {Lake}}, \bibinfo {author} {\bibfnamefont {Shankar}\ \bibnamefont
  {Balasubramanian}}, \ and\ \bibinfo {author} {\bibfnamefont {Soonwon}\
  \bibnamefont {Choi}},\ }\bibfield  {title} {\enquote {\bibinfo {title} {Exact
  {{Quantum Algorithms}} for {{Quantum Phase Recognition}}: {{Renormalization
  Group}} and {{Error Correction}}},}\ }\href@noop {} {\bibfield  {journal}
  {\bibinfo  {journal} {arXiv:2211.09803}\ } (\bibinfo {year}
  {2022})}\BibitemShut {NoStop}%
\bibitem [{\citenamefont {Herrmann}\ \emph {et~al.}(2022)\citenamefont
  {Herrmann}, \citenamefont {Llima}, \citenamefont {Remm}, \citenamefont
  {Zapletal}, \citenamefont {McMahon}, \citenamefont {Scarato}, \citenamefont
  {Swiadek}, \citenamefont {Andersen}, \citenamefont {Hellings}, \citenamefont
  {Krinner}, \citenamefont {Lacroix}, \citenamefont {Lazar}, \citenamefont
  {Kerschbaum}, \citenamefont {Zanuz}, \citenamefont {Norris}, \citenamefont
  {Hartmann}, \citenamefont {Wallraff},\ and\ \citenamefont
  {Eichler}}]{herrmann2022}%
  \BibitemOpen
  \bibfield  {author} {\bibinfo {author} {\bibfnamefont {Johannes}\
  \bibnamefont {Herrmann}}, \bibinfo {author} {\bibfnamefont {Sergi~Masot}\
  \bibnamefont {Llima}}, \bibinfo {author} {\bibfnamefont {Ants}\ \bibnamefont
  {Remm}}, \bibinfo {author} {\bibfnamefont {Petr}\ \bibnamefont {Zapletal}},
  \bibinfo {author} {\bibfnamefont {Nathan~A.}\ \bibnamefont {McMahon}},
  \bibinfo {author} {\bibfnamefont {Colin}\ \bibnamefont {Scarato}}, \bibinfo
  {author} {\bibfnamefont {Fran{\c c}ois}\ \bibnamefont {Swiadek}}, \bibinfo
  {author} {\bibfnamefont {Christian~Kraglund}\ \bibnamefont {Andersen}},
  \bibinfo {author} {\bibfnamefont {Christoph}\ \bibnamefont {Hellings}},
  \bibinfo {author} {\bibfnamefont {Sebastian}\ \bibnamefont {Krinner}},
  \bibinfo {author} {\bibfnamefont {Nathan}\ \bibnamefont {Lacroix}}, \bibinfo
  {author} {\bibfnamefont {Stefania}\ \bibnamefont {Lazar}}, \bibinfo {author}
  {\bibfnamefont {Michael}\ \bibnamefont {Kerschbaum}}, \bibinfo {author}
  {\bibfnamefont {Dante~Colao}\ \bibnamefont {Zanuz}}, \bibinfo {author}
  {\bibfnamefont {Graham~J.}\ \bibnamefont {Norris}}, \bibinfo {author}
  {\bibfnamefont {Michael~J.}\ \bibnamefont {Hartmann}}, \bibinfo {author}
  {\bibfnamefont {Andreas}\ \bibnamefont {Wallraff}}, \ and\ \bibinfo {author}
  {\bibfnamefont {Christopher}\ \bibnamefont {Eichler}},\ }\bibfield  {title}
  {\enquote {\bibinfo {title} {Realizing quantum convolutional neural networks
  on a superconducting quantum processor to recognize quantum phases},}\ }\href
  {\doibase 10.1038/s41467-022-31679-5} {\bibfield  {journal} {\bibinfo
  {journal} {Nat. Commun.}\ }\textbf {\bibinfo {volume} {13}},\ \bibinfo
  {pages} {4144} (\bibinfo {year} {2022})}\BibitemShut {NoStop}%
\bibitem [{\citenamefont {de~Groot}\ \emph {et~al.}(2022)\citenamefont
  {de~Groot}, \citenamefont {Turzillo},\ and\ \citenamefont
  {Schuch}}]{groot2022}%
  \BibitemOpen
  \bibfield  {author} {\bibinfo {author} {\bibfnamefont {Caroline}\
  \bibnamefont {de~Groot}}, \bibinfo {author} {\bibfnamefont {Alex}\
  \bibnamefont {Turzillo}}, \ and\ \bibinfo {author} {\bibfnamefont {Norbert}\
  \bibnamefont {Schuch}},\ }\bibfield  {title} {\enquote {\bibinfo {title}
  {Symmetry {{Protected Topological Order}} in {{Open Quantum Systems}}},}\
  }\href {\doibase 10.22331/q-2022-11-10-856} {\bibfield  {journal} {\bibinfo
  {journal} {Quantum}\ }\textbf {\bibinfo {volume} {6}},\ \bibinfo {pages}
  {856} (\bibinfo {year} {2022})}\BibitemShut {NoStop}%
\bibitem [{\citenamefont {Verresen}\ \emph {et~al.}(2017)\citenamefont
  {Verresen}, \citenamefont {Moessner},\ and\ \citenamefont
  {Pollmann}}]{verresen2017}%
  \BibitemOpen
  \bibfield  {author} {\bibinfo {author} {\bibfnamefont {Ruben}\ \bibnamefont
  {Verresen}}, \bibinfo {author} {\bibfnamefont {Roderich}\ \bibnamefont
  {Moessner}}, \ and\ \bibinfo {author} {\bibfnamefont {Frank}\ \bibnamefont
  {Pollmann}},\ }\bibfield  {title} {\enquote {\bibinfo {title}
  {One-{{Dimensional Symmetry Protected Topological Phases}} and their
  {{Transitions}}},}\ }\href {\doibase 10.1103/PhysRevB.96.165124} {\bibfield
  {journal} {\bibinfo  {journal} {Phys. Rev. B}\ }\textbf {\bibinfo {volume}
  {96}},\ \bibinfo {pages} {165124} (\bibinfo {year} {2017})}\BibitemShut
  {NoStop}%
\bibitem [{\citenamefont {Hauschild}\ and\ \citenamefont
  {Pollmann}(2018)}]{hauschild2018}%
  \BibitemOpen
  \bibfield  {author} {\bibinfo {author} {\bibfnamefont {Johannes}\
  \bibnamefont {Hauschild}}\ and\ \bibinfo {author} {\bibfnamefont {Frank}\
  \bibnamefont {Pollmann}},\ }\bibfield  {title} {\enquote {\bibinfo {title}
  {Efficient numerical simulations with {{Tensor Networks}}: {{Tensor Network
  Python}} ({{TeNPy}})},}\ }\href {\doibase 10.21468/SciPostPhysLectNotes.5}
  {\bibfield  {journal} {\bibinfo  {journal} {SciPost Phys. Lect. Notes}\ ,\
  \bibinfo {pages} {5}} (\bibinfo {year} {2018})}\BibitemShut {NoStop}%
\bibitem [{Note1()}]{Note1}%
  \BibitemOpen
  \bibinfo {note} {We will show in Sec.~\ref {sec:pb} that the error tolerance
  is limited close to phase boundaries as the threshold error probability
  decreases with diverging correlation lengths.}\BibitemShut {Stop}%
\bibitem [{\citenamefont {Eisert}\ \emph {et~al.}(2010)\citenamefont {Eisert},
  \citenamefont {Cramer},\ and\ \citenamefont {Plenio}}]{eisert2010}%
  \BibitemOpen
  \bibfield  {author} {\bibinfo {author} {\bibfnamefont {J.}~\bibnamefont
  {Eisert}}, \bibinfo {author} {\bibfnamefont {M.}~\bibnamefont {Cramer}}, \
  and\ \bibinfo {author} {\bibfnamefont {M.~B.}\ \bibnamefont {Plenio}},\
  }\bibfield  {title} {\enquote {\bibinfo {title} {{\emph{Colloquium}} :
  {{Area}} laws for the entanglement entropy},}\ }\href {\doibase
  10.1103/RevModPhys.82.277} {\bibfield  {journal} {\bibinfo  {journal} {Rev.
  Mod. Phys.}\ }\textbf {\bibinfo {volume} {82}},\ \bibinfo {pages} {277--306}
  (\bibinfo {year} {2010})}\BibitemShut {NoStop}%
\bibitem [{Note2()}]{Note2}%
  \BibitemOpen
  \bibinfo {note} {The correlation length corresponds to a characteristic
  length scale at which quantum correlation functions exponentially decay. For
  matrix product states with the unique largest eigenvalue $|\eta _1| =1$ of
  the corresponding transfer matrix, the correlation length $\xi \propto -
  \protect \frac {1}{\protect \qopname \relax o{log}|\eta _2|}$ is determined
  by the second largest eigenvalue $\eta _2$}\BibitemShut {NoStop}%
\bibitem [{\citenamefont {Kitaev}(2003)}]{kitaev2003}%
  \BibitemOpen
  \bibfield  {author} {\bibinfo {author} {\bibfnamefont {A.~{\relax Yu}.}\
  \bibnamefont {Kitaev}},\ }\bibfield  {title} {\enquote {\bibinfo {title}
  {Fault-tolerant quantum computation by anyons},}\ }\href {\doibase
  10.1016/S0003-4916(02)00018-0} {\bibfield  {journal} {\bibinfo  {journal}
  {Ann. Phys.}\ }\textbf {\bibinfo {volume} {303}},\ \bibinfo {pages} {2--30}
  (\bibinfo {year} {2003})}\BibitemShut {NoStop}%
\bibitem [{\citenamefont {Mesaros}\ and\ \citenamefont
  {Ran}(2013)}]{mesaros2013}%
  \BibitemOpen
  \bibfield  {author} {\bibinfo {author} {\bibfnamefont {Andrej}\ \bibnamefont
  {Mesaros}}\ and\ \bibinfo {author} {\bibfnamefont {Ying}\ \bibnamefont
  {Ran}},\ }\bibfield  {title} {\enquote {\bibinfo {title} {Classification of
  symmetry enriched topological phases with exactly solvable models},}\ }\href
  {\doibase 10.1103/PhysRevB.87.155115} {\bibfield  {journal} {\bibinfo
  {journal} {Phys. Rev. B}\ }\textbf {\bibinfo {volume} {87}},\ \bibinfo
  {pages} {155115} (\bibinfo {year} {2013})}\BibitemShut {NoStop}%
\bibitem [{\citenamefont {Ge}\ \emph {et~al.}(2016)\citenamefont {Ge},
  \citenamefont {Moln{\'a}r},\ and\ \citenamefont {Cirac}}]{ge2016}%
  \BibitemOpen
  \bibfield  {author} {\bibinfo {author} {\bibfnamefont {Yimin}\ \bibnamefont
  {Ge}}, \bibinfo {author} {\bibfnamefont {Andr{\'a}s}\ \bibnamefont
  {Moln{\'a}r}}, \ and\ \bibinfo {author} {\bibfnamefont {J.~Ignacio}\
  \bibnamefont {Cirac}},\ }\bibfield  {title} {\enquote {\bibinfo {title}
  {Rapid {{Adiabatic Preparation}} of {{Injective Projected Entangled Pair
  States}} and {{Gibbs States}}},}\ }\href {\doibase
  10.1103/PhysRevLett.116.080503} {\bibfield  {journal} {\bibinfo  {journal}
  {Phys. Rev. Lett.}\ }\textbf {\bibinfo {volume} {116}},\ \bibinfo {pages}
  {080503} (\bibinfo {year} {2016})}\BibitemShut {NoStop}%
\bibitem [{\citenamefont {Liu}\ \emph {et~al.}(2022)\citenamefont {Liu},
  \citenamefont {Shtengel}, \citenamefont {Smith},\ and\ \citenamefont
  {Pollmann}}]{liu2022}%
  \BibitemOpen
  \bibfield  {author} {\bibinfo {author} {\bibfnamefont {Yu-Jie}\ \bibnamefont
  {Liu}}, \bibinfo {author} {\bibfnamefont {Kirill}\ \bibnamefont {Shtengel}},
  \bibinfo {author} {\bibfnamefont {Adam}\ \bibnamefont {Smith}}, \ and\
  \bibinfo {author} {\bibfnamefont {Frank}\ \bibnamefont {Pollmann}},\
  }\bibfield  {title} {\enquote {\bibinfo {title} {Methods for {{Simulating
  String-Net States}} and {{Anyons}} on a {{Digital Quantum Computer}}},}\
  }\href {\doibase 10.1103/PRXQuantum.3.040315} {\bibfield  {journal} {\bibinfo
   {journal} {PRX Quantum}\ }\textbf {\bibinfo {volume} {3}},\ \bibinfo {pages}
  {040315} (\bibinfo {year} {2022})}\BibitemShut {NoStop}%
\bibitem [{\citenamefont {{Duclos-Cianci}}\ and\ \citenamefont
  {Poulin}(2010)}]{duclos-cianci2010}%
  \BibitemOpen
  \bibfield  {author} {\bibinfo {author} {\bibfnamefont {Guillaume}\
  \bibnamefont {{Duclos-Cianci}}}\ and\ \bibinfo {author} {\bibfnamefont
  {David}\ \bibnamefont {Poulin}},\ }\bibfield  {title} {\enquote {\bibinfo
  {title} {Fast {{Decoders}} for {{Topological Quantum Codes}}},}\ }\href
  {\doibase 10.1103/PhysRevLett.104.050504} {\bibfield  {journal} {\bibinfo
  {journal} {Phys. Rev. Lett.}\ }\textbf {\bibinfo {volume} {104}},\ \bibinfo
  {pages} {050504} (\bibinfo {year} {2010})}\BibitemShut {NoStop}%
\bibitem [{\citenamefont {Feiguin}\ \emph {et~al.}(2007)\citenamefont
  {Feiguin}, \citenamefont {Trebst}, \citenamefont {Ludwig}, \citenamefont
  {Troyer}, \citenamefont {Kitaev}, \citenamefont {Wang},\ and\ \citenamefont
  {Freedman}}]{feiguin2007}%
  \BibitemOpen
  \bibfield  {author} {\bibinfo {author} {\bibfnamefont {Adrian}\ \bibnamefont
  {Feiguin}}, \bibinfo {author} {\bibfnamefont {Simon}\ \bibnamefont {Trebst}},
  \bibinfo {author} {\bibfnamefont {Andreas W.~W.}\ \bibnamefont {Ludwig}},
  \bibinfo {author} {\bibfnamefont {Matthias}\ \bibnamefont {Troyer}}, \bibinfo
  {author} {\bibfnamefont {Alexei}\ \bibnamefont {Kitaev}}, \bibinfo {author}
  {\bibfnamefont {Zhenghan}\ \bibnamefont {Wang}}, \ and\ \bibinfo {author}
  {\bibfnamefont {Michael~H.}\ \bibnamefont {Freedman}},\ }\bibfield  {title}
  {\enquote {\bibinfo {title} {Interacting {{Anyons}} in {{Topological Quantum
  Liquids}}: {{The Golden Chain}}},}\ }\href {\doibase
  10.1103/PhysRevLett.98.160409} {\bibfield  {journal} {\bibinfo  {journal}
  {Phys. Rev. Lett.}\ }\textbf {\bibinfo {volume} {98}},\ \bibinfo {pages}
  {160409} (\bibinfo {year} {2007})}\BibitemShut {NoStop}%
\bibitem [{\citenamefont {Savary}\ and\ \citenamefont
  {Balents}(2017)}]{savary2016}%
  \BibitemOpen
  \bibfield  {author} {\bibinfo {author} {\bibfnamefont {Lucile}\ \bibnamefont
  {Savary}}\ and\ \bibinfo {author} {\bibfnamefont {Leon}\ \bibnamefont
  {Balents}},\ }\bibfield  {title} {\enquote {\bibinfo {title} {Quantum spin
  liquids: A review},}\ }\href {\doibase 10.1088/0034-4885/80/1/016502}
  {\bibfield  {journal} {\bibinfo  {journal} {Rep. Prog. Phys.}\ }\textbf
  {\bibinfo {volume} {80}},\ \bibinfo {pages} {016502} (\bibinfo {year}
  {2017})}\BibitemShut {NoStop}%
\bibitem [{\citenamefont {Pesah}\ \emph {et~al.}(2021)\citenamefont {Pesah},
  \citenamefont {Cerezo}, \citenamefont {Wang}, \citenamefont {Volkoff},
  \citenamefont {Sornborger},\ and\ \citenamefont {Coles}}]{pesah2021}%
  \BibitemOpen
  \bibfield  {author} {\bibinfo {author} {\bibfnamefont {Arthur}\ \bibnamefont
  {Pesah}}, \bibinfo {author} {\bibfnamefont {M.}~\bibnamefont {Cerezo}},
  \bibinfo {author} {\bibfnamefont {Samson}\ \bibnamefont {Wang}}, \bibinfo
  {author} {\bibfnamefont {Tyler}\ \bibnamefont {Volkoff}}, \bibinfo {author}
  {\bibfnamefont {Andrew~T.}\ \bibnamefont {Sornborger}}, \ and\ \bibinfo
  {author} {\bibfnamefont {Patrick~J.}\ \bibnamefont {Coles}},\ }\bibfield
  {title} {\enquote {\bibinfo {title} {Absence of {{Barren Plateaus}} in
  {{Quantum Convolutional Neural Networks}}},}\ }\href {\doibase
  10.1103/PhysRevX.11.041011} {\bibfield  {journal} {\bibinfo  {journal} {Phys.
  Rev. X}\ }\textbf {\bibinfo {volume} {11}},\ \bibinfo {pages} {041011}
  (\bibinfo {year} {2021})}\BibitemShut {NoStop}%
\bibitem [{\citenamefont {Peruzzo}\ \emph {et~al.}(2014)\citenamefont
  {Peruzzo}, \citenamefont {McClean}, \citenamefont {Shadbolt}, \citenamefont
  {Yung}, \citenamefont {Zhou}, \citenamefont {Love}, \citenamefont
  {{Aspuru-Guzik}},\ and\ \citenamefont {O'Brien}}]{peruzzo2014}%
  \BibitemOpen
  \bibfield  {author} {\bibinfo {author} {\bibfnamefont {Alberto}\ \bibnamefont
  {Peruzzo}}, \bibinfo {author} {\bibfnamefont {Jarrod}\ \bibnamefont
  {McClean}}, \bibinfo {author} {\bibfnamefont {Peter}\ \bibnamefont
  {Shadbolt}}, \bibinfo {author} {\bibfnamefont {Man-Hong}\ \bibnamefont
  {Yung}}, \bibinfo {author} {\bibfnamefont {Xiao-Qi}\ \bibnamefont {Zhou}},
  \bibinfo {author} {\bibfnamefont {Peter~J.}\ \bibnamefont {Love}}, \bibinfo
  {author} {\bibfnamefont {Al{\'a}n}\ \bibnamefont {{Aspuru-Guzik}}}, \ and\
  \bibinfo {author} {\bibfnamefont {Jeremy~L.}\ \bibnamefont {O'Brien}},\
  }\bibfield  {title} {\enquote {\bibinfo {title} {A variational eigenvalue
  solver on a photonic quantum processor},}\ }\href {\doibase
  10.1038/ncomms5213} {\bibfield  {journal} {\bibinfo  {journal} {Nat.
  Commun.}\ }\textbf {\bibinfo {volume} {5}},\ \bibinfo {pages} {4213}
  (\bibinfo {year} {2014})}\BibitemShut {NoStop}%
\bibitem [{\citenamefont {McClean}\ \emph {et~al.}(2016)\citenamefont
  {McClean}, \citenamefont {Romero}, \citenamefont {Babbush},\ and\
  \citenamefont {{Aspuru-Guzik}}}]{mcclean2016}%
  \BibitemOpen
  \bibfield  {author} {\bibinfo {author} {\bibfnamefont {Jarrod~R}\
  \bibnamefont {McClean}}, \bibinfo {author} {\bibfnamefont {Jonathan}\
  \bibnamefont {Romero}}, \bibinfo {author} {\bibfnamefont {Ryan}\ \bibnamefont
  {Babbush}}, \ and\ \bibinfo {author} {\bibfnamefont {Al{\'a}n}\ \bibnamefont
  {{Aspuru-Guzik}}},\ }\bibfield  {title} {\enquote {\bibinfo {title} {The
  theory of variational hybrid quantum-classical algorithms},}\ }\href
  {\doibase 10.1088/1367-2630/18/2/023023} {\bibfield  {journal} {\bibinfo
  {journal} {New J. Phys.}\ }\textbf {\bibinfo {volume} {18}},\ \bibinfo
  {pages} {023023} (\bibinfo {year} {2016})}\BibitemShut {NoStop}%
\end{thebibliography}%

\end{document}